\documentclass[12pt,letterpaper]{JHEP3}

\usepackage{amsmath}
\usepackage{setspace}
\usepackage{amssymb, amsthm,graphicx}
\usepackage{caption}
\usepackage{subcaption}
\usepackage{graphicx}

\newcommand{\order}{\mathcal{O}}

\newcommand{\be}{\beta}

\def\beq{\begin{eqnarray}}\def\eeq{\end{eqnarray}}
\def\be{\begin{equation}}\def\ee{\end{equation}}
\def\o{\omega}
\def\g{\gamma}

\def\b{\beta}

\def\hxi{\hat{\xi}}
\def\o{\omega}
\def\g{\gamma}

\def\b{\beta}

\def\r{\rho}

\preprint{TIFR/TH/14-16}

\title{ A Strongly Coupled Anisotropic Fluid From Dilaton Driven Holography}

\author{ Sachin Jain$^{1}$, Nilay Kundu$^{1}$, Kallol Sen$^{2}$,  Aninda Sinha$^{2}$ and Sandip P. Trivedi$^{1}$\\
\it $^{1}$ Department of Theoretical Physics,
\it Tata Institute of Fundamental Research,\\ \it Colaba, Mumbai, 400005, India\\
\it $^{2}$ Centre for High Energy Physics,
\it Indian Institute of Science,\\ \it C.V. Raman Avenue, Bangalore 560012, India.

\vspace{1cm}

E-mail: \email{sachin@theory.tifr.res.in, nilay.tifr@gmail.com, kallol@cts.iisc.ernet.in,  asinha@cts.iisc.ernet.in, trivedi.sp@gmail.com} \\

\vspace{0.5cm}

}

\abstract{
We consider a system consisting of $5$ dimensional gravity with a negative cosmological constant
 coupled to a massless scalar, the dilaton.
We construct a  black brane solution which arises 
 when the dilaton satisfies  linearly varying boundary conditions 
 in the asymptotically  $AdS_5$ region. 
The geometry of this black brane breaks rotational symmetry while preserving 
translational invariance   and corresponds to an anisotropic phase of the system. 
Close to extremality, where the anisotropy is big
 compared to the temperature, some components of the viscosity tensor become parametrically small compared to the entropy density. We study the quasi normal modes in considerable detail and find no instability close to extremality. 
We also obtain the equations for fluid mechanics for an 
anisotropic driven system in general, working upto first order
in the derivative expansion for the stress tensor, and identify additional transport coefficients which appear in the constitutive relation. 
 For the fluid of interest we find that the parametrically small viscosity 
can result in a very small force of friction, when the fluid is enclosed between appropriately 
oriented parallel plates moving with a relative velocity. }

\begin{document}
\maketitle
\section{Introduction}
\label{intro}
The AdS/CFT correspondence has opened the door for an interesting dialogue between 
the study of strongly coupled field  theories, of  interest for example in the study of QCD and
 condensed matter physics, and the study of gravity. See, for example, the reviews 
\cite{Hartnoll:2009sz,Herzog:2009xv,McGreevy:2009xe,Sachdev:2011wg, CasalderreySolana:2011us}
and references therein. 
An important area in which there has already been a  fruitful  exchange  is the study of transport properties,
and the related behavior of fluids, which arise in  strongly coupled systems, see \cite{Son:2007vk} for a review.  
Another interesting area has been  the study of black holes and branes which carry hair or have reduced symmetry. 
The existence of such solutions, which at first sight contradict the conventional lore on 
 no-hair theorems, was motivated by the existence of corresponding phases on the field theory 
side, see \cite{Gubser:2008px,Hartnoll:2008vx,Hartnoll:2008kx,Horowitz:2010gk}, for example. 

Among solutions with reduced symmetry which have been found are black branes which describe 
 homogeneous but anisotropic phases, i.e. phases
which leave  translational invariance, at least in some generalized sense, intact,
 but which break rotational symmetry.
An example of such a phase on the field theory side  
is a spin density wave phase, well known in condensed matter physics. 
Such black brane solutions  were systematically studied in \cite{Iizuka:2012iv,Iizuka:2012pn,Kachru:2013voa} and it was shown that they can be classified using the Bianchi classification developed earlier for understanding homogeneous cosmologies. 
A natural follow up  question  is about the transport properties in such anisotropic 
phases. One would like to know  whether these transport 
properties  reveal some interesting and  qualitatively new types of  
behavior when compared to weakly coupled theories or strongly coupled isotropic situations
which have already been  studied via gravity.
 
In this paper we turn to examining this question in one of the simplest examples of  an homogeneous but anisotropic phase.
This phase is obtained by considering a five dimensional system consisting of gravity coupled to 
a massless scalar, the dilaton, in the presence of a negative cosmological constant.
The anisotropic phase arises when the dilaton is subjected   
 to  a  linearly varying boundary condition along a spatial direction in the asymptotically $AdS_5$ region.  
More precisely,   a non-normalisable mode for the dilaton is turned on in the asymptotically
$AdS_5$ region which is linearly varying along one of the boundary spatial directions. 
The spatial direction along which the dilaton varies breaks isotropy. 
The resulting   solution is  characterized by two parameters, $\rho$, which is proportional to the 
dilaton gradient and is a measure of the breaking of isotropy, and $T$, the temperature.
By taking the dimensionless ratio $\rho /T$ to be  large or small we can consider the highly 
anisotropic case or 
the mildly anisotropic one respectively. 

Next, we   analyze the  transport properties of this system, and studying its stability by both examining the thermodynamics and the spectrum of quasi normal modes in some detail.
We also set up the fluid mechanics which arises in anisotropic phases in some generality and apply 
these general considerations to  study the system at hand.

When the anisotropy is small its effects can be incorporated
systematically, starting with the well known
rotationally invariant black brane in $AdS_5$, in a perturbative expansion in $\rho/T$.
The resulting changes  in thermodynamics and transport properties are small, as expected.
It is the region where the anisotropy is big, with $\rho/T \gg 1$, that is truly interesting, and
where qualitatively new phenomenon or behavior could arise. It is this region of
parameter space
that we mostly explore in this paper.

Before proceeding it is worth mentioning that the linearly varying dilaton actually
 breaks translational invariance along with  rotational invariance. Thus, it would seem that this  system is
 quite different from the ones mentioned above which describe homogeneous but anisotropic phases. 
In fact the similarity  is closer than one might expect  because  for the more standard situation,
where the dilaton takes a constant asymptotic value, it is well known that the  
 thermodynamics of the dual CFT  is independent of this asymptotic value, in the 
gravity approximation. As a result, 
 any corrections to the thermodynamics must be   proportional to the gradient of the dilaton. 
Since this is  a constant for a linearly varying dilaton, the resulting thermodynamics behaves like that of a translationally invariant,
but anisotropic system.

We find that when $T=0$, with fixed $\rho$, the resulting extremal black brane flows to an 
$AdS_4 \times R$ attractor geometry in the near horizon region. This shows that in the dual field theory the anisotropy is 
in fact a relevant perturbation and the theory flows to a new fixed point characterized by the 
symmetries of a $2+1$ dimensional CFT, unlike  the theory in the UV which  is a CFT in $3+1$ dimensions.
Starting from this extremal geometry we then study the effects of turning on a small temperature, so that $\rho/T $ 
continues to be large but not infinite. The  essential features of the thermodynamics can be understood 
analytically from the near-horizon region. We find that the system is thermodynamically stable,   with a positive specific heat.   

We then turn to analyzing the viscosity  of the system in more detail. In a rotationally invariant theory 
there are two independent components of the viscosity, the shear and bulk viscosity. In the conformally 
invariant case  the bulk viscosity vanishes.  Once rotational invariance is broken the viscosity has many more independent components and should be thought of most generally as a fourth rank tensor, with appropriate symmetry properties. In our case, for a dilaton gradient along the $z$ direction, rotational invariance in the $x-y$ plane is preserved and can be used to classify the various components. We find that viscosity, $\eta_\parallel$,
 which corresponds to the spin two component of the metric perturbations continues to saturate the
 famous Kovtun-Son-Starinets (KSS) bound, see \cite{Kovtun:2003wp}, \cite{Kovtun:2004de}, with
 \be
 \label{kssbound}
 \frac{\eta_{\parallel}}{s} = \frac{1}{4\pi},
 \ee
where $s$ is the entropy density. 
But the viscosity, $\eta_\perp$, which   corresponds
 to the spin one component can become much smaller, violating this bound. In fact, 
close to extremality  it goes like,
\be
\label{etaperone}
\frac{\eta_{\bot}}{s} =\frac{8\pi T^2}{3\rho^2}.
\ee 
so that the ratio  $\eta_{\perp} / s$ vanishes as $ T \rightarrow 0$ for fixed $\rho$.

In section \ref{qnmspec} we study the quasi normal modes of the system in considerable detail, see \cite{Kovtun:2005ev} for a general discussion. Our emphasis is on
 the regime
close to extremality where, as mentioned above, components of the viscosity can become very small. 
In the rotationally invariant case, the KSS bound can be violated, see \cite{Kats:2007mq}, \cite{Buchel:2008vz}, \cite{Cremonini:2011iq},
 \cite{Sinha:2009ev}, but attempts to make $\eta/s$ very small 
often lead to pathologies, like causality violation in the boundary theory, see \cite{Brigante:2007nu}, \cite{Brigante:2008gz}.
Here, in the anisotropic case, we find no signs of instability, close to extremality, 
 for the quasi normal modes we study.
 All the frequencies of these  modes we find   lie safely in the lower complex plane. 
While we  do not  study all the  quasi normal modes, we view the absence of any instability in the fairly large
 class we have  studied as an encouraging sign of stability for  the system.   

In section \ref{fluidmech} we turn to a more detailed analysis of the fluid mechanics. We first describe in 
fairly general terms how to set up the fluid mechanics, at least to first order in the derivative expansion 
for the stress tensor, for anisotropic phases. 
 The breaking of rotational invariance is characterized by a four vector, $\xi_{\mu}$, proportional to 
the dilaton gradient which enters in the fluid mechanics. As a result the number of allowed terms  in the constitutive relation for the stress tensor proliferate.  An added complication is that the system we are 
considering actually  breaks translational invariance. While the equilibrium
 thermodynamics
still continues to be   translationally invariant, as was mentioned above, 
 the breaking of translational invariance does  need to be accounted for in out of equilibrium 
situations pertaining to fluid mechanics. We describe how to do so in a systematics manner in the derivative expansion as well.  
Towards the end of this section we consider a simple example of a flow between two plates with a relative velocity between them. This is a canonical situation where the shear viscosity results in a friction force
being exerted by the fluid  on the plates. Due to the breaking of rotational invariance 
 we find that the friction force is different depending on how the plates are oriented. 
 For a suitably chosen  orientation it turns out to be  proportion to $\eta_\perp$, eq.\eqref{txzsol}, and 
can become very small close to extremality, as  $T \rightarrow 0$.

In section \ref{stringemb} we discuss string embeddings of our system. The simplest embedding is the celebrated example of 
 IIB string theory on $AdS_5\times S^5$. However, we find that in this example  there is 
 an instability for the extremal and near-extremal brane which arises from a KK mode on $S^5$. This mode saturates the BF bound in $AdS_5$ but
 lies below the BF bound for the $AdS_4$ geometry which arises in the near horizon limit of the near-extremal black brane.

Let us make some   comments before concluding this introduction. 
First, it is worth pointing out that the equations which govern the fluid mechanics
of an isotropic phase are always rotationally invariant since these equations inherit the symmetries of the underlying equilibrium configuration. The solutions of these equations in contrast of course need not be, and
are often not, rotationally invariant. The   rotational symmetry can be  broken, for example, by
 initial conditions. This is what happens for example in  relativistic heavy ion collision experiments. 
Rotational invariance is not  broken in equilibrium by heating up QCD to temperatures attained at  heavy ion colliders. However the initial conditions for the collisions are anisotropic resulting in anisotropic fluid flows. 
In contrast, the system we are studying has no rotational invariance in equilibrium, and the resulting 
equations of   fluid mechanics themselves break rotational invariance regardless of initial 
 or boundary conditions. 

Second, some references which directly bear on the study being discussed here should be especially mentioned.  The behavior of the gravity-dilaton system,  subjected to a general 
slowly varying dilaton,
 was studied in \cite{BLMNTW}. 
Our results in this paper  extend   this analysis to  the rapidly varying case, after restricting to  the linearly varying profile.
 In fact  the more exotic phenomenon 
arise in the rapidly varying  limit as mentioned above. Anisotropic phases along with  their 
viscosity was discussed  in \cite{Landsteiner:2007bd}, \cite{Azeyanagi:2009pr}, \cite{Natsuume:2010ky}, \cite{Erdmenger:2010xm}, \cite{Basu:2011tt}, \cite{Erdmenger:2011tj}, \cite{Mateos:2011ix}, \cite{Mateos:2011tv}, \cite{Rebhan:2011vd}, \cite{Polchinski:2012nh}, \cite{Iizuka:2012wt}, \cite{Mamo:2012sy}. 
It was found that the shear viscosity can acquire components in some cases which violate the KSS bound \cite{Mateos:2011ix}, \cite{Mateos:2011tv}, \cite{Rebhan:2011vd}, \cite{Polchinski:2012nh}, \cite{Mamo:2012sy}. 
In particular, in \cite{Azeyanagi:2009pr}
  \cite{Mateos:2011ix} and \cite{Mateos:2011tv},   a
 system quite similar to the one studied here was analyzed in some depth. These authors considered
 the gravity system coupled to an axion-dilaton which is  subjected to a linearly varying 
axion instead of dilaton. 
 The related paper by \cite{Rebhan:2011vd}  found that some components of the viscosity tensor 
can become very small violating the KSS  bound eq.\eqref{kssbound}. The main virtue of the 
dilaton system we have studied
 is that it is  somewhat simpler and  this
 allows  the  study  of issues related to stability  to be carried out  in more detail.    
In the paper  \cite{Polchinski:2012nh}  the $F1-NS5$ system, which is  supersymmetric
and thus stable, was studied and the ratio of viscosity to entropy density, for  a component of viscosity,
 was found to go like $T^2$, as in our case.  Gravitational systems where the phase of a complex scalar is linearly varying  have been studied in \cite{Donos:2013eha}.
 The near horizon extremal  $AdS_4 \times R$ solution we obtain, with a linearly varying dilaton, was found earlier in section $6$ of \cite{Freedman:2003ax}, see also \cite{Kanitscheider:2008kd}.

Let us end by mentioning that  besides the contents of various sections summarized above, additional important details
can also be found in appendices \ref{thermofg} - \ref{appqnm}.  
In particular appendix \ref{FMdetail} contains details on how to set up the fluid mechanics in anisotropic situations,
and appendix \ref{kuboapp} and \ref{hydmoap} discuss, from a fluid mechanics perspective,  the response after coupling to a metric and the linearized fluctuations about equilibrium. These allow us to relate the transport coefficients which  appear 
in  fluid mechanics  to  results from the gravity side.  
Appendix \ref{thermofg}  and \ref{appqnm} contain more details of the calculations of the stress energy tensor and quasi  normal modes in the 
gravity theory and appendix \ref{instads} contains a discussion of instabilities in $AdS$ space.
Appendix \ref{instads} analyses instabilities in $AdS_4$ and shows that they  are relatively 
insensitive to the UV completion of the geometry and continue to persist independent of the details of this completion.
 This  is  due to the infinite throat of the $AdS_4$;  
 the argument is also valid for other $AdS_{d+1}$ spaces 
 and should be generalizable for other attractor geometries as well.   
 Finally Appendix \ref{ads3rap} contains an analysis of a similar theory in one lower dimension. In this case the spacetime is  asymptotically $AdS_4$, 
 and  the near horizon extremal  geometry due to the linearly varying dilaton  is $AdS_3\times R$. One finds that a component of the viscosity, compared to the entropy density,  again becomes small. One   advantage in this case  is that some of the analysis of the quasi normal modes can be carried out analytically.   
\section{Anisotropic solution in dilaton gravity system}
\label{gravitysol}
We consider a system consisting of gravity, a massless scalar field, $\phi$, which we call the dilaton, and cosmological constant, $\Lambda$, in $5$ space time dimensions with action,
\be \label{action}
S_{bulk}=\frac{1}{2\kappa^2}\int d^{5}x\sqrt{-g}~\left(R+12\Lambda-\frac{1}{2}(\partial\phi)^2\right).
\ee
Here $2\kappa^2=16\pi G$ is the gravitational coupling with $G$ being Newton's Constant in 5-dimension.
This system is well know to have an $AdS_5$ solution with metric
\be
\label{ads5}
ds^2=L^2\bigg[-u^2 dt^2 +{du^2 \over u^2} + u^2 (dx^2 + dy^2 + dz^2)\bigg]
\ee
and with a constant dilaton. 
$L^2$, the radius of $AdS$ space, is related to $\Lambda$ by 
\be
\label{adssol}
\Lambda ={1 \over L^2}.
\ee
We will also use the constant $N_c$ defined by 
\be
\label{defnc}
{L^3\over G}= {2 N_c^2 \over \pi}.
\ee
In the discussion  which follows we use units  where 
\be
\label{cetlambda}
\Lambda=1,
\ee
so that from eq.(\ref{adssol})
\be
\label{valL}
 L = 1. 
\ee

The system above is well known to arise as a consistent truncation of the bulk system in many cases, e.g., $AdS_5 \times S^5$ solution of IIB supergravity, and more generally several $AdS_5 \times M_5$ solutions where $M_5$ is an Einstein Manifold. 
In the dual field theory, the dilaton is dual to a dimension $4$ scalar operator, e.g., in the $AdS_5 \times S^5$ case dual to the ${\cal N}=4$ SYM theory it is dual to the gauge coupling. 
  
We will be interested in solutions  which break the rotational invariance in the three spatial directions. 
This breaking of rotational invariance will be obtained by turning on the dilaton. 
We will show below   that a set of black brane solutions can be obtained  in which the dilaton varies linearly 
along one of the spatial directions, say $z$, and takes the form:
\be
\label{formdil}
\phi=\rho\, z.
\ee
Note that this will be the form of the dilaton in the bulk, in the solutions we consider \footnote{In a string embedding, e.g., IIB theory 
on $AdS_5 \times S^5$,  as discussed towards the end of this paper, the gravity approximation can break down at large $|z|$ when the dilaton gets very large or small. We have in mind placing the system in a box of size $L_B$ along the $z$ direction to exclude this large $|z|$ region. Mostly our interest
 is in near-extremal situations and our conclusions will follow by taking $1/T <L_B <1/\rho$.}.  In particular the dilaton will be independent of the radial direction $u$. 
The parameter $\rho$ will be an additional scale which will characterizes the breaking of rotational invariance. 

The resulting metric which arises after incorporating the back reaction due to the dilaton 
will preserve rotational invariance in the $x, y$ directions.  In addition, since the 
dilaton stress energy only depends on gradients of the dilaton, it will be translationally invariant in $t,x,y,z$. 
In this way by turning on a   simple linear dilaton profile we will find  a class of black brane solutions 
which correspond to homogeneous but anisotropic phases of the dual field theory. 
 
The  metric which arises after incorporating the dilaton back reaction  can  be written in the form, 
\begin{align}  \label{metans1}
ds^2&=-A(u)dt^2+\frac{du^2}{A(u)}+\b~ B(u)(dx^2+dy^2)+C(u)dz^2.
\end{align}
The coefficient $\b$ on the RHS  above can be clearly  set to unity by rescaling $x,y$, but we will keep it explicitly in the metric, since this  will prove convenient  in some of the following analysis. 

The trace reversed Einstein equations and the equation of motion for the dilaton are given by
\begin{align}
\begin{split} \label{eom1}
R_{\mu\nu}+4\Lambda g_{\mu\nu}-\frac{1}{2}\partial_{\mu}\phi\partial_{\nu}\phi&=0\\
\nabla^2\phi&=0.
\end{split}
\end{align}
In terms of the metric coefficients, the $uu,~tt,~xx,~zz$ components of the trace reversed Einstein equations can be written explicitly as
\begin{align}
\label{eom2}
\begin{split}
\frac{4\Lambda}{A}-\frac{A'B'}{2AB}+\frac{B'^2}{2B^2}-\frac{A'C'}{4AC}+\frac{C'^2}{4C^2}-\frac{A''}{2A}-\frac{B''}{2B}-\frac{C''}{2C}&=0,\\
-4\Lambda A+\frac{AA'B'}{2B}+\frac{AA'C'}{4C}+\frac{1}{2}AA''&=0,\\
4\Lambda B-\frac{1}{2}A'B'-\frac{AB'C'}{4C}-\frac{1}{2}AB''&=0,\\
-\frac{\rho^2}{2}+4\Lambda C-\frac{1}{2}A'C'-\frac{AB'C'}{2B}+\frac{AC'^2}{4C}-\frac{1}{2}AC''&=0.
\end{split}
\end{align} 
It is easy to see that the  linearly varying ansatz, eq.\eqref{formdil}, for the dilaton  satisfies its equation of motion.
Moreover, while the dilaton does not become constant asymptotically, as $u \rightarrow \infty$, its back reaction on the metric can easily be seen to be sub dominant compared to $\Lambda$ in this region. Thus the metric becomes that of 
$AdS_5$, eq.(\ref{ads5}),  as $u \rightarrow \infty$. In particular, we will assume that no non-normalizable deformation for the metric is turned on so that we are considering the dual field theory in flat space.
We also see that the  non-normalizable mode is turned  on for the dilaton, this means that the dual field  theory is subjected to a varying source term, e.g. for the ${\cal N}=4$ case a linearly varying gauge coupling. The varying source term breaks 
rotational invariance and gives rise to the anisotropic phases which we study\footnote{It is worth noting that the source terms in the boundary by themselves do not preserve even translational invariance. However the response of the metric in the
 bulk only cares about gradients of the dilaton and thus is translationally invariant. 
In the boundary theory this 
implies that the resulting
 expectation values of the stress tensor or the operator dual to the dilaton are also translationally 
invariant. 
This  happens from the field theory point of view because  the expectation value of energy density, entropy etc 
in the CFT is independent of the value of the gauge coupling, $g_{YM}$,
 for any constant value of $g_{YM}$ in the gravity approximation. Therefore any deviations from these 
results must be proportional to the gradient of the gauge coupling which preserves translational invariance.}.

We will choose  $\b$ so that  the asymptotic form for the metric is  $AdS_5$, eq.(\ref{ads5}),
  in effect setting the  speed of light \emph{i.e.} $c=1$ at the boundary\footnote{In general the asymptotic behavior of 
 $g_{zz} \rightarrow \gamma \, u^2$, and the additional constant $\gamma$ would also then have to be taken into account in the determining
the thermodynamics. In the analysis below we will see how this constant can be set equal to unity.}. 
With these normalizations the thermodynamics can be easily computed. First, we note that the Euclidean continuation of the above metric in the $t,u$ plane  takes the form 
\be 
ds_E^2 \approx A'(u_H)(u-u_H)dt^2+\frac{du^2}{A'(u_H)(u-u_H)},
\ee 
The requirement of regularity of the metric at the horizon then implies that the inverse of the period of the Euclidean time is identified with the temperature
\be 
\label{defT0}
T=\frac{1}{\delta t_E}=\frac{A'(u_H)}{4\pi},
\ee
The entropy can be calculated as follows. The area element on a $t=\text{const.}, \ u=u_H,$ is given by 
\be
dA_H=\b B(u_H)\sqrt{C(u_H)}dx dy dz.
\ee
The entropy is given by
\be
\label{entropy}
S={A_H\over 4 G}
\ee 
so that the  entropy density per unit volume in the $xyz$-direction is
\be 
\label{entdens}
s=\frac{N_c^2 }{2\pi} \b B(u_H)\sqrt{C(u_H)}.
\ee

\subsection{Gravity solutions}
\label{gravsolsub}
We will be interested in black brane solutions at finite temperature $T$, with a dilaton of the form in \footnote{The work in this section was jointly  done  with Prithvi Narayan.} eq.(\ref{formdil}). 
These solutions have two mass scales, $T$ and $\rho$.
High values of $\rho/T$ will be highly anisotropic, while low values of $\rho/T$ will be nearly isotropic solutions. 

The solution in the regime where $\rho/T \ll 1$ can be constructed by starting with the black brane in $AdS_5$ and incorporating the effects of the varying dilaton perturbatively in $\rho/T$. 
Upto second order in $\rho/T$ one gets 
\begin{align}
\label{lowansol}
\begin{split}
A(u)&=u^{2}-\frac{u_{H}^{4}}{u^{2}}+\rho^{2}a_{1}(u),\\
B(u)&=u^{2}+\rho^{2}b_{1}(u),\\ 
C(u)&=u^{2}+\rho^{2}c_{1}(u),
\end{split}
\end{align}
where the coefficient functions are given by 
\begin{align}
\begin{split}
a_{1}(u)&=\frac{1}{12}\bigg(\frac{u_{H}^{2}}{u^{2}}-1\bigg),\\
b_{1}(u)&=-\frac{u^{2}}{24u_{H}^{2}}\log\bigg[1-\frac{u_{H}^{4}}{u^{4}}\bigg]-\frac{u^{2}}{u_{H}^{2}}\log\bigg[\frac{u^{2}+u_{H}^{2}}{u^{2}-u_{H}^{2}}\bigg],\\
c_{1}(u)&=\frac{u^{2}}{12u_{H}^{2}}\log\bigg[1-\frac{u_{H}^{4}}{u^{4}}\bigg]+2\frac{u^{2}}{u_{H}^{2}}\log\bigg[\frac{u^{2}+u_{H}^{2}}{u^{2}-u_{H}^{2}}\bigg].
\end{split}
\end{align}
This second order solution is a special case of the general discussion in \cite{BLMNTW}. We have also obtained the solution upto fourth order, the resulting expressions are quite complicated and can be found in \cite{HTmath}. 

The solution in the highly anisotropic region, where $\rho/T \rightarrow \infty$ is more interesting. 
It is useful to first consider the extremal case where $T = 0$ with $\rho \ne 0$. 
In this case the full black brane solution is hard to find analytically but  a simple calculation  shows
 that the near horizon region is of the form $AdS_4 \times R$, with metric components,
\be 
\label{extsol1}
A(u)=\frac{4}{3}u^2, \qquad B(u)=u^2, \qquad  C(u)=\frac{\rho^2}{8},
\ee
Note that we have retained  $\beta$ as in eq.\eqref{metans1}, as  a free parameter in the solution. 
In fact, as often happens with attractor geometries, eq.(\ref{extsol1}) is an exact solution of the equations of motion, with the dilaton given by eq.(\ref{formdil}). 

This attractor geometry will play an important role in the subsequent discussion. 
It is worth emphasizing that the linear variation of the dilaton gives rise to extra components in
 the stress tensor which prevent the $z$ direction from shrinking, resulting in the $AdS_4\times R$ geometry. 
Also we note from eq.(\ref{extsol1}) that the $AdS_4$ geometry has a radius smaller than 
that of the asymptotic $AdS_5$,
\be
\label{rad4}
L_4^2=3/4 \,L^2.
\ee
 
Starting with the near horizon geometry in eq.(\ref{extsol1}) we can show that by 
 adding a suitable perturbation which  grows in the UV  the solution matches asymptotically to the $AdS_5$ metric, eq.\eqref{ads5}, 
for a suitable choice of $\beta$. The form of this perturbation is 
\begin{align} \label{pertdef1}
\begin{split}
 A(u) &={4\over 3} u^2 \left(1+\delta A(u)\right),\\
 B(u) &=\beta u^2 \left(1+\delta A(u)\right), \\ 
 C(u) &={\rho^2 \over 8} \left(1+\delta C(u)\right),
 \end{split}
\end{align}
with
\begin{align} \label{pertdef2}
\begin{split}
 \delta A(u) &=  a_1 \ u^{\nu}, \qquad \delta C(u) =  c_1 \ u^{\nu}, \\
 a_1 = \frac{1}{16}&\left(-7+\sqrt{33}\right) c_1, \qquad \nu=\frac{1}{2}
\left(-3+\sqrt{33}\right).
\end{split}
\end{align} 
The analysis showing that after turning on this perturbation one matches with $AdS_5$ at large $u$ was carried out using mathematica. The resulting value of $\beta $ is 
\be
\label{betaval}
\beta={4 \over 3}.
\ee 
By adjusting the coefficient $c_1$ one can ensure that the asymptotic behavior of $C(u)$ in the metric eq.\eqref{extsol1}, eq.\eqref{metans1}
agrees with eq.(\ref{ads5}). 
\begin{figure}
\begin{center}
\includegraphics[width=0.8\textwidth]{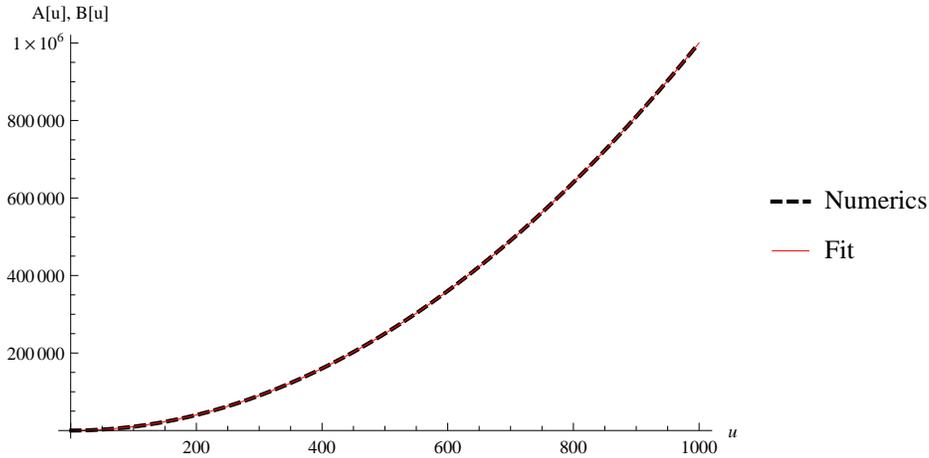}
\caption{Numerical interpolation of near horizon $AdS_4 \times R$ to asymptotic $AdS_5$, for the metric coefficients $A(u),\,B(u)$ with $\rho=1$ and $c_1=2$.
}\label{abvsr}
\end{center}
\end{figure}

\begin{figure}
\begin{center}
\includegraphics[width=0.8\textwidth]{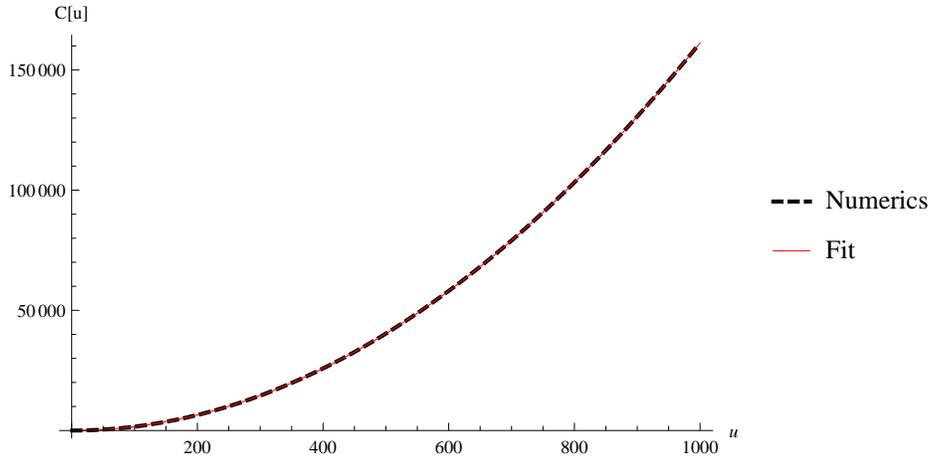}
\caption{Numerical interpolation of near horizon $AdS_4 \times R$ to asymptotic $AdS_5$, for the metric coefficient $C(u)$ with $\rho=1$ and $c_1=2$. 
}\label{cvsr}
\end{center}
\end{figure}
A plot showing the resulting  metric components as a function of $u$
is given in Fig. \ref{abvsr} and Fig. \ref{cvsr}. These plots were obtained by numerical 
interpolation for the case $\rho=1$ and $c_1=2$.  
Asymptotically, as $u \rightarrow \infty$, one finds that
 $A(u), B(u) \rightarrow u^2$, while $C(u)=\gamma\, u^2$,
with $\gamma \ne 1$.

Now that we have understood the extremal solution, one can also turn on a small temperature, while still keeping 
$\rho/T \gg 1$. In the near-horizon $AdS_4 \times R$ region the metric components for the near extremal solution  are given by 

\be 
\label{finTsol1}
A(u)=\frac{4}{3}\bigg(u^2-\frac{u_H^3}{u}\bigg), \qquad B(u)=u^2, \qquad  C(u)=\frac{\rho^2}{8}.
\ee
The horizon radius $u_H$ is related to the temperature by 
\be 
\label{defT}
T=\frac{u_H}{\pi}.
\ee

Note that the effects of the temperature die out at large $u$, thus the perturbation eq.\eqref{pertdef1}, eq.\eqref{pertdef2},  when added to this solution
will flow at large $u$ to $AdS_5$ space, for small enough values of $u_H$. 
Using the definition in eq.\eqref{entdens}, the entropy density for the solution in eq.\eqref{finTsol1} can be obtained as
\be
\label{entdensT}
s= \frac{N_c^2\pi}{3\sqrt{2}}\rho~ T^2,
\ee
where we used the numerical value of the coefficient $\beta$ from eq.\eqref{betaval}. It is fixed by requiring that the metric becomes $AdS_5$ of form eq.(\ref{ads5}). To leading order in $T$ 
the resulting value of $\beta$ is the same as in the extremal case.

\section{Thermodynamics} 
\label{thermo}
The aim of this section is to discuss the thermodynamics of the anisotropic solutions presented in the previous section from both field theory and gravity point of view. First we shall discuss basic facts in general about thermodynamics for the situation when we have anisotropy in one of the spatial directions from purely field theory point of view. Then we shall consider the gravity solutions explicitly and check that the thermodynamics obtained from the study of dual gravity background is consistent with that obtained from the field theory analysis.
\subsection{Thermodynamics of anisotropic phase from boundary analysis}
\label{thermoft}
We consider a relativistically invariant quantum field theory at temperature $T$. It's partition function is given by 
\be
Z[\phi,g_{\mu\nu}]= \int \mathcal{D} \Psi ~e^{-S[\phi,g_{\mu\nu}]}.
\ee
Here $\Psi$ stand for the fields in the theory over which one must integrate. $S$ is the action which depends on the fields,
 $\Psi$, and also on the background metric $g_{\mu\nu}$ and the dilaton $\phi$. 

The metric  has Euclidean signature  and is of the   form
\begin{align} \label{metgf}
\begin{split}
ds^2=& g_{\mu\nu} dx^{\mu}dx^{\nu}\\=&e^{2 \sigma(\vec x)} \left( dt+ a_{i}(\vec x) dx^i \right)^2
+\tilde{g}_{ij}({\vec x}) dx^i dx^j,
\end{split}
\end{align}
where $i=1,2,3$,  and where time $t$ has periodicity, 
\be
t_{E} ~:~ [0,\beta], \qquad \text{with} \qquad \beta={1\over T}.
\ee

The resulting partition function is a functional of $\phi, g_{\mu\nu}$. 
With the system of interest here in mind we would like to consider cases where the dilaton is linearly varying with the 
profile eq.(\ref{formdil}). The system is therefore characterized by two energy scales, $T, \rho$. 
We will take the metric to be varying slowly compared to both $T, \rho$. 

For the systems we are studying this boundary partition function can be obtained by evaluating the dual path 
integral in the bulk. 
As was discussed above in the bulk it is clear that the dependence on $\phi$
  can only arise through gradients of the dilaton. For the dilaton profile under consideration, eq.(\ref{formdil}) this means 
the partition function would depend on 
\be
\label{defchi}
\chi=\xi^2\equiv \xi_\mu \xi^\mu,
\ee
where 
\be
\label{defxi}
\xi_\mu=\partial_\mu\phi.
\ee

 When the metric is varying slowly we can express the dependence of $\log(Z)$ on the metric, $g_{\mu\nu}$, 
in terms of a derivative expansion. It will be enough for our case to set $a_i=0$ in the metric in eq.\eqref{metgf}. 
This gives\footnote{Note that the dependence of partition function on $\chi$ and the subsequent discussions are similar in structure to the superfluid case studied in \cite{Bhattacharyya:2012xi}.}, 
\begin{equation}\label{zopf} \begin{split}
 W=& \log Z= \int d^{3}x~\sqrt{\tilde{g}}~\frac{1}{{\hat T}} ~P({\hat T},\chi)+\cdots,\\
 \sqrt{\tilde{g}} =&\text{Det}[\tilde{g}_{ij}], \qquad {\hat T}= T e^{-\sigma}+\cdots,
\end{split}
\end{equation}
where the terms in the ellipses stand for terms with one or more derivatives of the metric.  
It will be enough for our purposes to consider the leading term on the RHS above.

In the flat space limit, the stress tensor is defined in terms of the variation of $\log Z$  with respect to the 
 metric $g_{\mu\nu}$
\be 
\label{stdef}
\langle T^{\mu\nu} \rangle = -{2\over \sqrt{g}} {\delta W \over \delta g_{\mu\nu}} \bigg \rvert_{g_{\mu\nu}=\delta_{\mu\nu}},
\ee
where $\sqrt{g}= \text{Det}[g_{\mu\nu}]$. 
With the metric in eq.(\ref{metgf}) and $a_i=0$ we get 
\be
\label{stdef1}
T^{\mu\nu} = {2 \hat{T}\over \sqrt{\tilde{g}}} {\delta \log Z \over \delta g_{\mu\nu}} \bigg \rvert_{\sigma=0,
~\tilde{g}_{ij}=\delta_{ij}}.
\ee
The $\mu=0,~ \nu=0$ component of the stress tensor can be obtained from eq.\eqref{stdef1} as,
\be
\label{st00}
T_{00} = e^{2\sigma} \bigg[-P+\hat{T} {\partial P \over \partial \hat{T}}\bigg]\bigg \rvert_{\sigma=1}=-P+T {\partial P \over \partial T}.
\ee
With the following definitions of the energy density
\be
\label{relep}
T_{00} = \epsilon
\ee
and the entropy density
\be
\label{releps}
s= {\partial P \over \partial T}\bigg \rvert_{ \chi},
\ee
one readily obtains the thermodynamic relation,
\be
\label{threl1}
\epsilon+P=sT.
\ee

Next,  we consider the $\mu=i,~ \nu=j$ component of the stress tensor
\be
\label{stij}
T^{ij} = \frac{-2\hat{T}}{ \sqrt{\tilde{g}}} g^{ik}g^{jl} \frac{\delta \log Z}{\delta g^{kl}} 
        = -2 g^{ik}g^{jl} \left[ -{1\over2} g_{kl} P +  \frac{\partial P}{\partial \chi} \frac{\partial \chi}{\partial g^{kl}} \right] 
        = P g^{ij} + f \xi^i \xi^j,
\ee
where we have defined
\be
\label{deff}
 f= -2\frac{\partial P}{\partial \chi} \bigg\rvert_{T}.
\ee
Using the definitions of entropy density in eq.\eqref{releps} and of the function $f$ from eq.\eqref{deff} one obtains the relation
\be \label{relpsch}
dP= s~ dT-\frac{1}{2}f~ d\chi.
\ee
From eq.\eqref{stij} we obtain 
\be\begin{split}
\label{stxyz}
T^{xx}&=T^{yy}=P,\\
T^{zz}&=P_z=P+ f (\xi^z)^2.
\end{split}\ee
The  last equation in eq.(\ref{stxyz})  implies
\be\label{consis1}
P_z=P-2\frac{\partial P}{\partial \chi}(\xi^z)^2.
\ee
Defining the extensive quantities \emph{i.e.} total energy $E=\epsilon~V$ and total entropy $S=s~V$ in terms of the energy density, entropy density and the volume $V$, one can further obtain a thermodynamic relation from eq.\eqref{threl1}, 
 and eq.\eqref{relpsch}, 
\be
\label{threl2}
dE+P~dV+\frac{1}{2}f~V~ d\chi =T~dS.
\ee
This is the first law of thermodynamics but  now also including  a  term which arises due to the presence 
of the dilaton gradient.

A few comments are now in order.  
First, we note from eq.(\ref{threl2}) and eq.\eqref{stxyz} that  the  pressure $P$ which is  conjugate to the volume $V$ is 
$\{\mu=x,~ \nu=x\}$ or $\{\mu=y,~ \nu=y\}$ component of the stress tensor $T^{\mu\nu}$. 
Second, a useful analogy to keep in mind for comparison is that of a system in a magnetic field, $H$.
For such a system the first law takes the form, 
\be
dE+P~dV+M~ dH =T~dS,
\ee
with $M$ being the magnetization  of the system. By comparing we see that $\chi$,  which is determined by the external dilaton's
gradient, plays the role of the external magnetic field and  ${1\over 2} f V$ the role of $M$. 

The various components of  the stress tensor  in eq.\eqref{st00}, eq.\eqref{stij}  can be combined in the following simple form
\be \label{lto}
 T^{\mu\nu} = (\epsilon+ P){\hat u}^{\mu}{\hat u}^{\nu}+ P g^{\mu\nu}+ 
f \xi^{\mu}\xi^{\nu},
\ee
where ${\hat u}^\mu$ the velocity four-vector is given by 
\be
\label{defhatu}
{\hat u}^{\mu}= (1,0,0,0).
\ee

With this general analysis of the thermodynamics in hand we are now ready to turn   to the  system at hand. 
We will consider first the low anisotropy region and then the high anisotropy region below.

\subsection{More on thermodynamics of the highly anisotropic regime: $\rho/T \gg 1$}
\label{mohianth}
We are now ready to  consider some aspects of the highly anisotropic regime. 

We begin with the $T\rightarrow 0$ extremal limit,  where, $\rho/T \rightarrow \infty$.  
In this limit the thermodynamic identity eq.(\ref{threl1}) becomes 
\be
\label{relextep}
\epsilon_{0} +P_{0} = 0.
\ee
where the subscript $0$ indicates the $T=0$ values. 
Notice that it follows from eq.(\ref{relextep}) that if the
 energy density $\epsilon_0>0$ the pressure $P_0<0$. 
 We will comment on this feature further below.  Finally,  since the near horizon geometry is 
$AdS_4\times R$ the entropy vanishes in the extremal case. 

Close to extremality, for  ${T \over \rho} \ll 1$, the  entropy density no longer vanishes. To leading order we see that 
\be
\label{scales}
s = c_1\rho ~T^2,
\ee
where $c_1>0$. We can express the other thermodynamic variables as  corrections about the extremal values,
\be
\label{varcorr}
\epsilon=\epsilon_0+\delta \epsilon,~~P=P_0+\delta P,~~~f=f_0+\delta f.
\ee

Using the identity eq.\eqref{releps} one obtains
\be
\label{reldelp}
\delta P= \frac{c_1 }{3} \rho T^{3}.
\ee
Further using the relation eq.\eqref{threl1} along with eq.\eqref{relextep} one obtains,
\be
\label{reldele}
\delta \epsilon = \frac{2c_1 }{3}\rho T^{3}.
\ee

Next,  can also use the relation eq.\eqref{deff} and eq.\eqref{reldelp} to get
\be
\label{reldelf}
\delta f = - {1 \over \rho^2}  \delta P.
\ee
This in turn gives  from eq.(\ref{stxyz}) that 
\be
\label{reldelpz}
\delta P_z = 0.
\ee

Let us end this subsection with two comments.
First, we will see from the gravity analysis that the trace anomaly in this system takes the form
\be
\label{tracean2}
T^\mu_\mu={N_c^2\rho^4 \over 192 \pi^2}.
\ee
In particular the RHS above is independent of $T$. This means that the corrections to the energy and pressures must satisfy the relation 
\be
\label{fintthrel}
Tr[\delta T_{\mu\nu}] = \delta T^{\mu}_{\mu}=-\delta \epsilon + 2~ \delta P +\delta P_z =0.
\ee
We see from the expressions obtained above that this relation is indeed met. 
We have also seen that $\delta P_z$ vanishes at small temperature. This can be understood in terms of the 
$AdS_4 \times R$ solution which describes the near horizon geometry. The $AdS_4 \times R$ solution should be dual to a CFT 
which is Lorentz invariant in $2+1$ dim. Small temperature or energy excitations should be states in this CFT.
The trace anomaly in this CFT  takes the form
\be
\label{tan3}
-\delta \epsilon_0+  \delta P_x + \delta P_y=0.
\ee
On the other hand the full asymptotically $AdS_5$ solution is dual to a Lorentz invariant $3+1$ dimensional theory (with a source turned on)  with the trace anomaly eq.(\ref{tracean2}). 
This implies that eq.(\ref{fintthrel}) must also be valid. Both these constraints then necessarily lead to 
vanishing of $\delta P_z$. 

Second, we had mentioned  above that the pressure at extremality $P_0$ can be negative. 
  This might lead to the worry that the system is unstable. However, we see that the small temperature excitations above extremality have positive $\delta P$, and also 
$\delta P_z$ is not negative. The stress energy of small excitations there does not show any instability. 
 In addition we see from eq.(\ref{scales}) that the entropy density and therefore specific heat is positive, 
which shows that the system is also thermodynamically stable. An instability would be present if the compressibility, ${\partial P\over \partial V}>0$. But this is not true for our system. 
In fact a  negative pressure at extremality is akin to what happens in a system with a positive cosmological constant
and does not by itself signal an instability. 
Let us also mention that the parameter $\rho$ determines the magnitude of the external forcing function provided by the dilaton, it is the analogue of an external magnetic field $H$. Thermodynamic stability does not require
${\partial^2 F \over \partial \rho^2}$ to necessarily have a definite sign, where $F$ denotes the Free energy. 
This is analogous to what happens in an external magnetic field. The susceptibility of a system coupled to a magnetic field can be positive or negative 
depending on whether the system is para or diamagnetic and stable systems of both kinds can arise.

\subsection{Thermodynamics from gravity}
\label{thermogr}
To study the thermodynamics from gravity we need the near boundary behavior of the metric and the dilaton. From this behavior 
one can extract the stress energy tensor $\langle T^{\mu \nu} \rangle$ after suitably subtracting divergences using the procedure of holographic renormalization, see \cite{Balasubramanian:1999re}, \cite{Skenderis:2002wp}. 
For this purpose one needs to work with a total action of the form, 
\be
\label{stot}
S_{tot}=S_{bulk}+S_{GH}+S_{ct},
\ee
where $S_{bulk}$ is the bulk action given in eq.\eqref{action}, $S_{GH}$ is the Gibbons Hawking boundary term, and $S_{ct}$ is the counter term action needed to subtract the divergences.  

 It is convenient to  work in Fefferman-Graham coordinates in which the near boundary metric and dilaton take the form
\be \begin{split}\label{FG1}
ds^2&=\frac{dv^2}{v^2}+\g_{\mu\nu}(x,v)dx^\mu dx^\nu,\\
\g_{\mu\nu}&=\frac{1}{v^2}\big[g^{(0)}_{\mu\nu}+v^2 g^{(2)}_{\mu\nu}+v^4(g^{(4)}_{\mu\nu}+2\log v~ \tilde{g}^{(4)}_{\mu\nu})+O(v^6)\big],\\
\phi&=\phi^{(0)}+v^2 \phi^{(2)}+v^4(\phi^{(4)}+2\log v~ \tilde{\phi}^{(4)})+O(v^6).
\end{split}
\ee
Here $v$ is a coordinate which vanishes at the boundary and it is related to the coordinate $u$ by 
\be
\label{defvu}
v = {1\over u} + \cdots,
\ee
where the ellipses stand for corrections which vanish near the boundary where $v \rightarrow 0$. 

We will follow the analysis in \cite{Papadimitriou:2011qb}.
The counter term action is given in eq. B.12 of \cite{Papadimitriou:2011qb}. In our case the axion $\chi=0$ and the boundary metric is flat,
$\gamma_{ij}=\delta_{ij}$. 
Since there is no normalizable mode for the dilaton turned on we have
\be
\label{opdual}
\langle O \rangle=0,
\ee
where $O$ is the operator dual to the dilaton in the boundary. 
After subtracting the divergences, the stress tensor has an expectation value
\begin{align}
 \label{STnsor}
 \begin{split}
 \langle T_{\mu\nu} \rangle=  {1\over 8 \pi G}\bigg[& 2\bigg(g^{(4)}_{\mu\nu}-\text{Tr}[g^{(4)}]~g^{(0)}_{\mu\nu} \bigg)+ \bigg(\tilde{g}^{(4)}_{\mu\nu} -\text{Tr}[\tilde{g}^{(4)}]~ g^{(0)}_{\mu\nu} \bigg)\\ &-\bigg(\text{Tr}[g^{(2)}]~g^{(2)}_{\mu\nu}-\text{Tr}[g_{(2)}^{2}]~g^{(0)}_{\mu\nu}\bigg) + \frac{\rho^{2}}{4}\bigg(g^{(2)}_{\mu\nu}-g^{(2)zz}g^{(0)}_{\mu\nu}\bigg)\bigg],
 \end{split}
 \end{align}
From eq.(B.8), eq.(B.11), eq.(B.16) of \cite{Papadimitriou:2011qb} we have that\footnote{It is important to note that the dilaton kinetic term in the action, eq.(2.1) of \cite{Papadimitriou:2011qb} has a relative factor of $2$ with the dilaton kinetic term in our action eq.\eqref{action}.}  
\be
\label{combex}
{\tilde g}^{(4)}_{\mu\nu} - Tr[{\tilde g}^{(4)}]~ g^{(0)}_{\mu\nu}= -{1\over 24} {\rho^4\over 4}\bigg( 2 \delta_{z\mu} \delta_{z\nu}-{1\over 2} \delta_{\mu\nu}\bigg).
\ee
Note in particular that the RHS is independent of  temperature and only dependent on $\rho$. 

Also, from eq.(\ref{STnsor}) we get that trace
\be
\label{tranom}
\langle T^\mu_\mu\rangle=-\epsilon +2P+P_z=\frac{N_c^2 \rho^4}{192\pi^2},
\ee
where the last equality follows from eq.\eqref{tracean2} and we have also used the relation between $G$ and $N_c$ given in eq.\eqref{defnc} with $L=1$. 

Note we see from eq.(B.21), eq.(B.22) of \cite{Papadimitriou:2011qb} that for the system at hand, the trace anomaly  is given by 
\be
\label{traceano}
 T^\mu_\mu ={L^3\over 384 \pi G_{N}} ((\partial \phi)^2)^2.
\ee
This agrees with eq.(\ref{tranom}) after using eq.(\ref{defnc}), eq.(\ref{formdil}).

Let us end this subsection with one comment. The choice of counter-terms one makes does affect the 
 the final result for the stress tensor. However important physical  
 consequences are independent of  this choice.  
For example, the choice of counter terms will affect the value of the energy density, $\epsilon$, 
$P$, etc, at extremality,
as calculated below in eq.(\ref{EPhian}), but it will not change the additional finite temperature corrections,
$\delta \epsilon, \delta P$, etc which determine the physical consequences. 

\subsubsection{Low anisotropy regime: $\rho / T \ll 1$}
\label{loanth}

We are now ready to study the low anisotropy region. 
The asymptotic form of the metric is given in eq.(\ref{hiansfg}), eq.(\ref{hiansfg1}) appendix \ref{apthloan}. 
 The resulting values for the stress energy tensor in terms of $T, u_H$ are given in 
eq.(\ref{hiansth}) of appendix \ref{apthloan}.
The thermodynamic quantities can be obtained, as is explained in appendix \ref{apthloan}, as
\begin{align}
\label{Eloan}
\epsilon=&\frac{3}{8}N_{c}^{2}\pi^{2}T^{4}+\frac{1}{32}N_{c}^{2}\rho^{2}T^{2}-\frac{N_{c}^{2}(13-8\log[2])}{1536\pi^{2}}\rho^{4}+O(\rho^{6}),\\ \label{Pxloan}
 P=&\frac{1}{8}N_{c}^{2}\pi^{2}T^{4}+\frac{1}{32}N_{c}^{2}\rho^{2}T^{2}+\frac{N_{c}^{2}(15-24\log[2])}{4608\pi^{2}}\rho^{4}+O(\rho^{6}),\\ \label{Pzloan}
P_{z}=&\frac{1}{8}N_{c}^{2}\pi^{2}T^{4}-\frac{1}{32}N_{c}^{2}\rho^{2}T^{2}+\frac{N_{c}^{2}(-45+72\log[2])}{4608\pi^{2}}\rho^{4}+O(\rho^{6}),
\end{align}   
The resulting entropy density becomes 
\be 
\label{Sloan}
s=\frac{1}{2}N_{c}^{2}\pi^{2}T^{3}+\frac{1}{16}N_{c}^{2}\rho^{2}T-\frac{N_{c}^{2}\rho^{4}}{192\pi^{2}T}+O(\rho^{6}),
\ee
Using  eq.\eqref{Eloan}, eq.\eqref{Pxloan} and eq.\eqref{Pzloan} we can also obtain the value of the trace anomaly
\be
\label{tranomap}
\langle T^\mu_\mu\rangle=-\epsilon\,+2P+P_z=\frac{N_c^2 \rho^4}{192\pi^2},
\ee
which agrees with eq.(\ref{tranom}). 

It is also worth pointing out  from eq.(\ref{hiansfg1}) and eq.\eqref{FG1} that 
\be
\label{combex2}
({\tilde g}_{4\,\mu\nu} - Tr({\tilde g}_4) g_{0\,\mu\nu})=-{1\over 24} {\rho^4 \over 4}[ 2 \delta_{z\mu} \delta_{z\nu}-{1\over 2} \delta_{\mu\nu}],
\ee
which agrees with eq.(\ref{combex}). In fact $Tr({\tilde g}_4)$ vanishes.

Let us mention that the low anisotropy regime was also studied in \cite{BLMNTW} in some generality. 
\subsubsection{High anisotropy regime: $\rho / T\gg 1$ }
\label{hianth}
In this case the full solution cannot be obtained analytically. Instead one can understand the near horizon region analytically
and then numerically interpolate to go to the asymptotically $AdS_5$ region. 

We start with the extremal case, with $T=0$. 
The full solution preserves Lorentz invariance in the $t,x,y$ directions. As a result the metric components in the asymptotic,
$u\rightarrow \infty$ region take the form
\begin{align}
\begin{split}\label{fghan1}
g_{tt}&=-1+\frac{\rho ^2 v^2}{24}-v^4 a_4(\rho )+\frac{1}{96} \rho ^4 v^4 \log (\rho  v),\\
g_{xx}&=g_{yy}=1-\frac{\rho ^2 v^2}{24}+v^4 a_4(\rho )-\frac{1}{96} \rho ^4 v^4 \log (\rho  v),\\
g_{zz}&=1+\frac{5 \rho ^2 v^2}{24}+v^4 c_4(\rho)+\frac{1}{32} \rho ^4 v^4 \log (\rho  v),
\end{split}
\end{align}
where $a_4(\rho), c_4(\rho)$ are unknown functions of $\rho$.
Here we are using the Fefferman -Graham coordinate defined in eq.\eqref{defvu}.

Note that the coefficient of the $\log v$ term is fixed by the anomaly 
 and coefficient of $v^2$ is fixed in terms of the
dilaton in the FG expansion (see \cite{Papadimitriou:2011qb}).

Using eq.(\ref{fghan1}) in eq.(\ref{STnsor}) gives 
\be
\label{lanth1}
\begin{split}
\epsilon&=\frac{N_c^2 \left(768 a_4(\rho )+384 c_4(\rho )-5 \rho ^4\right)}{768 \pi ^2},\\
P&=\frac{N_c^2 \left(-768 a_4(\rho )-384 c_4(\rho )+5 \rho ^4\right)}{768 \pi ^2},\\
P_z&=\frac{N_c^2 \left(\rho ^4-384 a_4(\rho )\right)}{256 \pi ^2}.
\end{split}
\ee
The anomaly constraint eq.(\ref{tranom}) allows us to  solve for $a_4(\rho)$ in terms of $c_4(\rho)$ and yields, 
\be
\label{a4eq}
a_4(\rho) = \frac{7}{1728}\rho^4-\frac{1}{3}c_4(\rho).
\ee
From eq.(\ref{a4eq}), eq.(\ref{lanth1}) and eq.(\ref{consis1}) we get a differential equation for $c_4$ which yields,
\be
c_4(\rho)=c_1 \rho^4+\frac{1}{32}\rho^4 \log(\rho),
\ee
where $c_1$ is an integration constant. 
$c_1$ can be absorbed by introducing a  suitable scale $\mu$ giving,
\be
\label{a4c4hian}
a_4(\rho)=\frac{7}{1728}\rho^4 -\frac{1}{96}\rho^4 \log(\frac{\rho}{\mu}), \qquad c_4(\rho)=\frac{1}{32}\rho^4 \log(\frac{\rho}{\mu}).
\ee

Finally, inserting eq.\eqref{a4c4hian} back in eq.\eqref{lanth1} gives 
\be
\label{EPhian}
\begin{split}
\epsilon&=\frac{N_c^2 \rho^4\left(-17+36 \ln(\frac{\rho}{\mu})\right)}{6912 \pi ^2},\\
P&=-\frac{N_c^2 \rho^4\left(-17+36 \ln(\frac{\rho}{\mu})\right)}{6912 \pi ^2},\\
P_z&=\frac{N_c^2 \rho^4\left(-5+36 \ln(\frac{\rho}{\mu})\right)}{2304 \pi ^2}.
\end{split}
\ee

Note that at the leading order $\epsilon=-P$ in eq.\eqref{lanth1} and also in eq.\eqref{EPhian}, hence  from eq.\eqref{threl1} the entropy is zero.
It is worth noting that an additional  scale $\mu$ has appeared in the expressions above. The value of $\mu$
 can be determined from a numerical analysis where the full solution which interpolates between the 
near horizon region and asymptotic AdS space is constructed. 
Note also that  eq.(\ref{a4c4hian}) determines all   terms in the metric eq.(\ref{fghan1}) which go like $v^4 \log(v)$. One can verify that these terms satisfy the relation eq.\eqref{combex}.
 
The behavior at small temperatures above extremality was discussed in  subsection \ref{mohianth} above. 
Numerically we find that the coefficient $c_1$ in eq.\eqref{scales} takes the value 
\be
\label{valc11}
c_1= {N_c^2 \pi \over 3\sqrt{2}}.
\ee
The resulting values of thermodynamics quantities can then be obtained from the formula in subsection \ref{mohianth}.

\section{Computation of the viscosity from gravity}
\label{grvis}
In isotropic situations it is well known  that the shear viscosity for  any system having a  gravity dual is given by 
\be
\label{shvisa}
{\eta \over s}={1\over 4 \pi},
\ee
where $s$ is the entropy density. 
This result holds as long as Einstein's two derivative theory is a good approximation on the gravity side. 
The only other independent component is the bulk  viscosity which vanishes for a conformally invariant theory. 

More generally, for anisotropic situations, the viscosity should be thought of as a tensor $\eta_{ij,kl}$ where the indices,
$i,j,k,l,$ take values along the spatial directions. 
A Kubo formula can be written down relating the viscosity to the two point function of the stress tensor and takes the form:
\begin{equation}
\label{kuboa}
\eta_{ij,kl} = -\lim_{\omega\rightarrow 0}\frac{1}{\omega}\, Im \big[ G^{R}_{ij,kl}(\omega)\big],
\end{equation}
where
\begin{equation}
\label{defGR}
 G^{R}_{ij,kl}(\omega,0) = \int dt d{\bf x}\, e^{i \omega t}\, \theta(t)\, \langle[T_{ij}(t,{\bf x}),T_{kl}(0,0)]\rangle
\end{equation}
 is the retarded Green's function, and $Im$ denotes its imaginary part. 
Note that  we are interested here in the  viscosity at vanishing spatial momentum. It is clear 
from eq.(\ref{kuboa}) that the viscosity is a fourth rank tensor, $\eta_{ij,kl}$, symmetric in $i,j$ and $k,l$
and also symmetric with respect to the exchange $(ij)\leftrightarrow (kl)$. This means it has $21$ independent components. 

In the gravity theory, this two point function can be calculated by studying the behavior of metric perturbations 
in the corresponding black brane solution. 
The solution we are interested in preserves partial rotational invariance in the $x-y$ plane. We can use this unbroken
 $SO(2)$ subgroup, ${\bf R}$,  to classify the perturbations. 
We write the metric as 
\be
\label{metpert}
ds^2=g_{MN}\,dx^Mdx^N + h_{\mu \nu}\,dx^{\mu} dx^{\nu}.
\ee
Here $g_{MN}$, with $M,N=t,u,x,y,z,$ stands for the background black brane metric  
while, $h_{\mu\nu}$, with $ \mu, \nu =t,x,y,z,$ correspond to general metric perturbations. 
We have made a gauge choice to set $h_{uu}=h_{u\mu}=0$. Let us note before proceeding that for the brane solution we consider
the black brane metric is  given in eq.(\ref{metans1}), in the following discussion we will take it to  be more generally of the form 
\be
\label{metricbb}
ds^2=-g_{tt} dt^2 + g_{uu}du^2 + g_{xx} dx^2 + g_{yy}dy^2 + g_{zz}dz^2.
\ee

The two-point functions of the spatial components of the stress tensor, which appear in eq.(\ref{defGR}) require us to study the 
behavior of the metric perturbations $h_{\mu\nu}$ where $\mu,\nu=x,y,z$. There are six independent components of this type. 
Two of these, $(h_{xx}- h_{yy}),\, h_{xy},$ carry spin $2$ with respect to ${\bf R}$. Two more, $h_{xz}, h_{yz},$ carry spin $1$.
The remaining two,  $(h_{xx}+h_{yy})$ and $h_{zz}$, are of spin $0$. Correspondingly, we see that the viscosity tensor 
will have $5$ independent components \footnote{There are three independent two point functions among the two spin $0$
perturbation.}. Actually,   the analysis of the spin $0$ sector is
  more complicated due to  mixing with the 
dilaton perturbation. We will comment on this more below. 

The spin $2$ and spin $1$ perturbations, as we will see below, satisfy an equation of the form, 
\begin{equation}\label{eqnscal}
 \partial_{u}\left(\sqrt{-g}P(u)g^{uu}\partial_{u}\phi(u)\right)-\omega^2 N(u)g^{tt} \phi(u)=0,
\end{equation} 
where $\phi$ is a scalar field, whose precise relation to the metric perturbation will be discussed shortly, and 
the  functions $P(u), N(u)$ are determined in terms of background metric.
The expectation value of the dual operator is determined, as per the standard AdS dictionary in terms of the canonical momentum
\begin{equation}\label{CNM}
\Pi(u,\omega)=\frac{1}{16\pi G}\sqrt{-g}P(u)g^{uu}\partial_{u}\phi(u).
\end{equation}
It then follows that the retarded Green's function takes the form 
\begin{equation}
 G^{\rm{ret}} = -\frac{\Pi(u,\omega)}{\phi(u,\omega)} \Bigg{|}_{u\rightarrow \infty}~~~\label{retardedg}.
\end{equation}
And the response function which will enter in the definition of the viscosity eq.(\ref{kuboa})  is given by 
\begin{equation}
\label{responsefun}
 \chi = \lim_{\omega\rightarrow 0}\frac{\Pi(u,\omega)}{i \omega \phi(u,\omega)}\Bigg{|}_{u \rightarrow \infty}.\end{equation}

An important fact, see \cite{Iqbal:2008by}, is that the RHS of eq.(\ref{responsefun})
can equally well be  evaluated very close to  the horizon, $u=u_H$,
instead of at the boundary of AdS space, $u\rightarrow \infty$. 
This follows from  noting that to evaluate the  RHS  we are interested in the behavior of the ratio  $({\Pi(u)\, / \,\phi(u)})$ only upto $O(\omega)^2$ correction,  as $\omega \rightarrow 0$. 
Now from the equation of motion, eq.(\ref{eqnscal}), we see that unless we are very close to the horizon,  where $g^{tt}$ 
diverges, the second term can be neglected since it is proportional to $\omega^2$. 
Thus 
\be
\label{condpi}
\partial_u \Pi=0
\ee
upto $O(\omega)^2$.
From eq.(\ref{condpi}) it follows that 
\be
\label{condphi}
\Pi = C,
\ee
where $C$ is a  constant, independent of $u$. 
At $u \rightarrow \infty$ it will turn out that $\phi$  goes to a constant upto $O(\omega^2)$. 
Thus it will turn out that $C$ in eq.(\ref{condphi}) vanishes upto $O(\omega^2)$ and therefore $\phi$ itself is a constant.

From the discussion above it follows that $\partial_u({\Pi\over \phi})=0$  upto $O(\omega^2)$  leading\footnote{More
 correctly the vanishing of $\partial_u({\Pi\over \phi})=0$ holds as long as we are not too close to the horizon,  more properly the analysis involves a matched asymptotic expansion for small $\omega$.} to
\be
\label{membrane}
\chi=\lim_{\omega\rightarrow 0}\,\frac{\Pi(u,\omega)}{i \, \omega \, \phi(u,\omega)}\Bigg{|}_{u \rightarrow u_H}.
 \ee

Now the solution one considers must be regular at the future horizon. This means that quite generally near the horizon,  
\be
\label{clh}
\phi \sim e^{-i \omega (t+r_*)},
\ee
where the tortoise coordinate,
\be
\label{tcoord}
r_*=\int \sqrt{g_{uu}\over g_{tt}} \, du.
\ee

It is then simple to show from eq.(\ref{membrane}) that 
\be
\label{response11}
\chi=-\frac{1}{16\pi G}P(u_H)\sqrt{\frac{-g}{g_{tt} g_{uu}}}\Bigg{|}_{u \rightarrow u_H}.
\ee
The spin $2$ and spin $1$ components of the viscosity will  then follow from the appropriate response   $\chi$.

It will turn out, interestingly, that whereas the spin $2$ component still satisfies the viscosity bound, eq.\eqref{shvisa}, 
the spin $1$ component does not and can become much smaller. In fact,  we will find that  at extremality, 
when $T \rightarrow 0$, the spin $1$ component will vanish, while the entropy as discussed above stays finite in this limit. 
We will denote the spin $2$ component of the viscosity as $\eta_{\parallel}$ and the spin $1$ component as $\eta_\bot$. 

Before proceeding let us mention that our system is quite analogous to that studied in \cite{Mateos:2011ix}, \cite{Mateos:2011tv} and subsequently in \footnote{We should alert the reader that in \cite{Mateos:2011ix}, \cite{Mateos:2011tv}, \cite{Rebhan:2011vd}, the notation for $\eta_\parallel, \eta_\perp$ is reversed compared to ours.} \cite{Rebhan:2011vd}, where it was found that  the $\eta_\perp$  component of the viscosity  scales with $T$ as in eq.\eqref{etaperone}. This was also found in \cite{Polchinski:2012nh} which studied the D1-D5 system.

\subsection{Computation of the spin $1$ component of the viscosity, $\eta_{\bot}/s$}
\label{grvissub1}
 The spin $1$ component of viscosity can be written, using eq.\eqref{kuboa}, as
\be
\label{defetperp1}
\eta_{\perp}=\eta_{xz,xz} = -\lim_{\omega\rightarrow 0}\frac{1}{\omega}\, Im \big[ G^{R}_{xz,xz}(\omega)\big].
\ee
To calculate the spin $1$ component it is enough to  consider the  $h_{xz}$ component of a metric perturbation, 
eq.\eqref{metpert}, so that the full metric is of the   form
\begin{equation}
\begin{split}
 ds^2=-g_{tt}(u) dt^2+g_{uu}(u)du^2+ &g_{xx}(u)\sum_{i=1}^{d-2}dx_{i}^{2}+g_{zz}(u)dz^2 \\ &+2e^{-i\omega t}Z(u)g_{xx}(u)dx dz,
 \end{split}
\end{equation}
where $Z(u)$ is the perturbation that we need to study.  
One can easily show that the other modes decouples from 
$Z(u)$  and so  can be consistently set to  zero.

The equation that one obtains for this mode $Z(u)$ is 
\begin{equation}\label{eq1}
 \partial_{u}(\sqrt{g}g^{zz}g^{uu}g_{xx}\partial_{u}Z(u))-\omega^2 Z(u)\sqrt{g}\,g^{zz}g^{tt}g_{xx}=0.
\end{equation} 
Comparing with \eqref{eqnscal} we see  that $P(u)=g^{zz}g_{xx}$, and 
\be
\label{valphi}
\phi=Z(u).
\ee  
It also follows from the action eq.\eqref{action} that the canonical momentum is given by eq.(\ref{CNM}) with $\phi$ given by eq.(\ref{valphi}).
 
The expression for the entropy density is given by
\begin{equation}
 s=\frac{1}{4 G} \frac{\sqrt{-g}}{\sqrt{-g_{uu}g_{tt}}}\Bigg{|}_{u_{H}}.
\end{equation} 
Further using \eqref{response11}, eq.(\ref{retardedg}) with eq.(\ref{defetperp1}) we  obtain 
\begin{equation}
\label{finaletaperp}
 \frac{\eta_{\bot}}{s}=\frac{1}{4\pi}\frac{g_{xx}}{g_{zz}}\Bigg{|}_{u_{H}}.
\end{equation}
\subsubsection{Low anisotropy regime: $\rho/T \ll 1$}
Using the form of the metric in this regime given in \eqref{lowansol} (see also appendix \ref{apthloan}) near the horizon, the ratio turns out to be 
\be 
\frac{\eta_{\bot}}{s}=\frac{1}{4\pi}-\frac{\rho^2 \log 2}{16 \pi^3 T^2}+\frac{(6-\pi^2+54 (\log 2)^2)\rho^4}{2304\pi^5 T^4}+O\bigg[\bigg(\frac{\rho}{T}\bigg)^6\bigg].
\ee
As $\rho/T\rightarrow 0$ the ratio approaches the universal value
\be \frac{\eta_{\bot}}{s}\rightarrow\frac{1}{4\pi}.\ee
Interestingly, the effect of anisotropy is to reduce the value of this ratio from this universal value.  
\subsubsection{High anisotropy regime: $\rho/T \gg 1$}
In this region of parameter space the  near  horizon region is given by the metric of a black brane in 
 $AdS_4\times R$ geometry, with components, eq.(\ref{finTsol1}). From eq.(\ref{finaletaperp}), with eq.(\ref{defT}) and eq.(\ref{betaval}) we then get, 
\be
\label{etaperpint}
\frac{\eta_{\bot}}{s} =\frac{8\pi T^2}{3\rho^2}.
\ee 
We see that the ratio can be made arbitrarily small, with  
$ \frac{\eta_{\bot}}{s}\rightarrow 0,$ as $T \rightarrow 0$ keeping $\rho$ fixed. 
Eq.(\ref{etaperpint}) is also reproduced in the introduction, eq.(\ref{etaperone}), since it is one of our main results. 

\subsection{Computation of the spin $2$ component of the viscosity,   $\eta_{\parallel}/s$}
The spin $2$ component of the viscosity can be written, using eq.\eqref{kuboa}, as
\be
\label{defetprl1}
\eta_{\parallel}=\eta_{xy,xy} = -\lim_{\omega\rightarrow 0}\frac{1}{\omega}\, Im \big[ G^{R}_{xy,xy}(\omega)\big].
\ee
To compute the spin $2$ component it is enough to consider a metric perturbation with $h_{xy}$ non-zero.
The full metric then takes the form,
\begin{equation}
\begin{split}
 ds^2=-g_{tt}(u) dt^2+g_{uu}(u)du^2&+g_{xx}(u)\sum_{i=1}^{d-2}dx_{i}^{2}+g_{zz}(u)dz^2 \\ & +2e^{-i\omega t}Y(u)g_{xx}(u)dx dy,
 \end{split}
\end{equation}
where $Y(u)$ is the perturbation that determines $\eta_{\parallel}$. 
One can easily show that the other modes decouples from 
$Y(u)$ and therefore can be consistently  set them to zero.

The equation that one obtains for the mode $Y(u)$ is of the form
\begin{equation}\label{eq11}
 \partial_{u}(\sqrt{g}g^{uu}\partial_{u}Y(u))-\omega^2 Y(u)\sqrt{g}g^{tt}=0.
\end{equation}
The analysis which follows is similar to the one above, so we will be brief. 
Comparing with \eqref{eqnscal} we immediately conclude that $P(u)=1.$
The ratio $\eta_{\parallel}/s$ is then  easily computable and is given by
\begin{equation}
 \frac{\eta_{\parallel}}{s}=\frac{1}{4\pi}.\label{etaprll1}
\end{equation}
We see that this is the same as in the isotropic case. 
Thus the spin $2$ component of the viscosity takes the same value as eq.(\ref{shvisa}) in the isotropic case. 

\subsection{Comments on the spin $0$ case } 

As mentioned above the analysis in the spin $0$ sector is more complicated since the two metric 
perturbations, $(h_{xx}+h_{yy})$ and $h_{zz}$, can mix also with the dilaton perturbation. 
We will not discuss this case in further detail here. However, when we analyze the quasi normal modes we
see how the imaginary parts of the hydrodynamical modes   
determine or constrain
transport coefficients in the spin zero sector.

\section{The spectrum of quasi normal modes}
\label{qnmspec}
In this section we analyze the spectrum of quasi normal modes in some detail with a view to studying the stability of the brane solutions 
presented above. Previous attempts, in the isotropic case, to violate the viscosity bound, eq.\eqref{shvisa}, have sometimes 
lead to inconsistencies, see  \cite{Brigante:2007nu}, \cite{Brigante:2008gz}. 
Since the anisotropic solutions we have found violate the bound 
for some components of the viscosity, as was discussed in section \ref{grvissub1}, and in fact do so dramatically close to extremality, it is 
natural to ask about the stability of these solutions. 

The analysis of the quasi normal modes is quite non-trivial since the solutions of interest are not known analytically, for general values
of the anisotropy parameter $\rho/T$. When $\rho /T \ll 1$ the solution is close to a Schwarzschild black brane and one does not expect any instability. The imaginary parts of the quasi normal modes should be safely in the lower half of the complex plane,  for all modes. The 
only possible exceptions are the hydrodynamic modes, but these are Goldstone modes and protected by symmetries.
The opposite regime close to extremality, with $\rho/T \gg 1$, is the interesting one  where potential instabilities could be present. 
As mentioned above, this is also the regime where the violations of the viscosity bound are big. 

Close to extremality the solution interpolates between asymptotically $AdS_5$ space in the UV and an $AdS_4 \times R$ attractor in the IR, 
more correctly its finite temperature deformation. Quasi normal modes of very high frequency, $|\omega| \gg \rho$,
would be localized near the boundary and one expects them to be very similar to those in $AdS_5$ and thus to be stable. 
It is  the  low frequency   modes with 
\be
\label{limomega}
|\omega| \ll \rho,
\ee
for which  instabilities might appear. One expects these modes to be  localized in the near horizon $AdS_4 \times R$ region,  which is 
quite different from $AdS_5$.  

Fortunately, as often happens in the study of near extremal solutions, the black brane in $AdS_4 \times R$ is a solution of the equations in motion in its own right. To study the possible instabilities in the frequency regime, eq.(\ref{limomega}), we therefore study
quasi normal modes about the $AdS_4 \times R$ black brane solution below. 
As we will see,  while the analysis is not fully exhaustive, 
 in the many channels we study we do not find any instability. The analysis therefore provides considerable evidence that the black brane solutions studied in this paper are stable.  

We should add that  analysis we have carried out
  does not apply to modes with frequencies $|\omega| \sim \rho$, which must be analyzed in the full 
interpolating geometry. It seems unlikely to us on physical grounds that an unstable mode appears in this regime, but we have not ruled this out. Doing so would require a numerical analysis about the background solution which is also known only numerically.   
We leave such an analysis for the future. 

\subsection{General strategy for finding quasi normal modes}
Let us now turn to discussing the perturbations in more detail. The metric and dilaton will be denoted by 
\begin{align}
g_{MN}& = g_{MN}^{(0)} + h_{MN}, \qquad \text{where} \, \, M, N=t,x,y,z, u, \label{metperta} \\
\phi & = \phi^{(0)}  + \psi. \label{dilperta}
\end{align}
Here $g_{MN}$  denotes the background metric for a black brane in  $AdS_4 \times R$,  and $\phi^{(0)}$ the background value for the dilaton
given in eq.(\ref{formdil}).
We will work in coordinates where the background metric is given by 
\be
\label{metbackgrnd}
 ds^2 = -A(u) dt^2 + {du^2 \over A(u)} + B(u) (dx^2+dy^2) + C(u) dz^2,
\ee
with $A(u), B(u), C(u)$ given in eq.(\ref{extsol1}). Note that since we will be working in the $AdS_4 \times R$ region we have set the parameter
$\beta$ which appears in eq.(\ref{metans1}) to unity. 
Indices will be raised and lowered in the following discussion using this background metric. 

 In the analysis below we choose the  gauge
\begin{align}
h_{u \mu} & =  0, \qquad \text{where} \, \, \mu = x,\,y,\,z,\,t, \label{gaugepa} \\
\text{and} \, \,h_{uu} & =  0. \label{gaugepb}
\end{align}
The metric perturbation then has $10$ components, and the dilaton perturbation has one component, making for a total of $11$ independent perturbations.

A perturbation will be partially specified by its momentum $\vec{q} = (q_x, q_y, q_z)$ and frequency $\omega$ with dependence
on $(x,y,z,t)$ given by 
\be
\label{pertdep}
\delta \phi \sim e^{i \, \vec{q} \cdot \, \vec{x} - i \, \omega  \,t}.
\ee 
The perturbations  satisfy the Einstein equations and also the equation of motion for the dilaton.
Once the dependence on $(x,y,z,t)$ is specified these determine the radial dependence of the perturbations. 
There are $15$ components of the Einstein equations, of these $5$ are constraints which only involve first derivatives with respect 
to the radial variable.  These $5$ constraints reduce the number of independent perturbations from $11$ to $6$. 
We will study the behavior of these $6$ independent perturbations when we analyze the quasi normal mode spectrum.  

Quasi normal modes are perturbations which satisfy   the following two boundary conditions:
\begin{itemize}
\item The modes are in going at the future horizon
\item Only normalizable components are  turned on at the boundary so that no source terms are activated in the dual field theory. 
\end{itemize}

These conditions can be met only if the frequencies take ``quantized'' values,  which are in general complex. 
A quasi normal mode with  a  frequency lying in the upper half of the complex plane  will grow  exponentially in time  signaling
 an instability.

We will study the spectrum of quasi normal modes about the black brane in $AdS_4 \times R$. This solution was given in eq.\eqref{finTsol1} with the
radial coordinate being denoted as $u$. It will be convenient in the subsequent analysis to work with the coordinate $v$, 
\be 
\label{defvcord}
v=\frac{u_{H}}{u}.
\ee
which takes value $v=1$ at the horizon, where $u=u_H$ and $v=0$ at the boundary, $u \rightarrow \infty$. In what follows, throughout this section, we will work without any loss of generality in units where $u_H=1$ or equivalently with $T=1 / \pi$. We will reinsert them whenever needed through dimensional analysis.

Our strategy will be as follows. 
As mentioned in eq.(\ref{pertdep}) a   perturbation has  momentum $\vec{q} = (q_x, q_y, q_z)$ and frequency $\omega$.
For any choice of $\vec{q}, \omega$, we  first identify suitable combinations of perturbations 
for which the linearized  equations  decouple. 
Denoting a generic perturbation of this type  by $Z(v)$,  we will find that such a perturbation  satisfies an equation of the type
\be
\label{diffeqz} 
a(v)Z''(v)+b(v)Z'(v)+c(v)Z(v)=0.
\ee

The two boundary conditions mentioned above, then give rise to the conditions, 
\be 
\label{limZbh}
Z(v)|_{v\rightarrow 1}\sim (1-v^3)^{-\frac{i\o}{4}},  \qquad Z(v)|_{v\rightarrow 0}\sim v^{\gamma},
\ee where $\gamma > 0$.

We will then use  use Leaver's method, \cite{Leaver:1990zz}, to numerically compute the frequencies of quasi normal modes  
for  the $Z(v)$ perturbation\footnote{We thank Sean Hartnoll for 
 generously  sharing  a  mathematica notebook where a similar analysis of quasi normal modes had been performed.}  \cite{Denef:2009yy, Denef:2009kn}.
In this method we expand $Z(v)$  about the midpoint $v=1/2$ as 
\be 
\label{expZser}
Z(v)=\left(1-v^3\right)^{-\frac{i\o}{4}}\,v^\gamma \sum_{n=0}^{M}d_{n}\,\left(v-\frac{1}{2}\right)^n. 
\ee
Inserting  in eq.\eqref{diffeqz}  and also expanding the coefficients $a(v), b(v), c(v)$ in a Taylor series about $v=1/2$,
 we then collect terms of the same order in $(v-1/2)$ to obtain $M+1$ linear relations among the $M+1$ coefficients $d_n$. 
These can be summarized as a matrix equation
\be 
\label{eqndn}
\sum_{n=0}^{M} A_{mn}(z)\,d_{n}=0.
\ee
The matrix elements $A_{mn}$ depend on $\vec{q}, \omega$. For a non-trivial solution  we get the condition,
\be
\label{condA}
det[A_{mn}(\vec{q}, \omega)]=0,
\ee
which determines $\omega$ in terms of the momentum $\vec{q}$. 
As mentioned above the solutions for $\omega$ will be complex in general.
With $M$ big enough one can get reasonable numerical accuracy for $\omega$. 

\subsection{Modes with $q_z\ne 0$}
\label{momqz}
We first consider modes where the momentum has vanishing component along the $x-y$ direction, $\vec{q}=(0,0,q_z)$. 
In the subsequent discussion of this subsection we will use the notation
\be
\label{defq}
q=q_z
\ee
to denote the only non-zero component of the momentum. 
Additional details pertaining to this subsection can be found in appendix \ref{qnmzap}. 

We can use the rotational invariance in the $x-y$ plane to classify the perturbations. 
The $11$ perturbations (without using the constraint equations) split up as follows:
 \be
 \label{qzdifch}
 \begin{array}{cc}
  \text{Spin} \, \, 2: & h_{y}^{x},\, (h_{x}^{x} - h_{y}^{y} )/2\,\, \\
  \text{Spin} \, \,  1:& h_{t}^{x},\,h_{z}^{x},\, h_{t}^{y},\, h_{z}^{y} \\
  \text{Spin} \, \,  0:& ( h_{xx} + h_{yy} )/2\,\, , h_{zz}\,\, , h_{tt}\,\, , h_{tz}\,\, ,\psi.
\end{array}
 \ee

As we will see below, the $5$ constraint equations  among the Einstein equations will cut down the number of 
independent perturbations  to $6$. We will be able to find linear combinations among these for which the equations decouple and then use Leaver's method for obtaining the quasi normal modes.
\subsubsection{Spin 2} 
\label{sp2momz}
It turns out that the equations for $h_{y}^{x}$ and $(h_{x}^{x}-h_{y}^{y})/2$ automatically decouple and each is of the form, eq.\eqref{diffeqz} withe the coefficients, $a(v), b(v), c(v)$ given in eq.\eqref{abcsp2qz} in appendix \ref{qnmzapsp2}.  
\begin{figure}
\begin{center}
\includegraphics[width=0.8\textwidth]{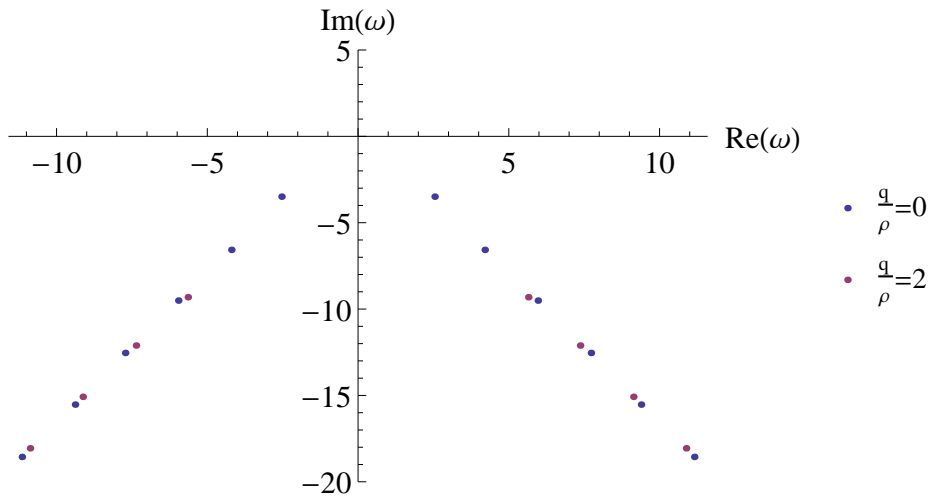}
\caption{QNM plot for spin 2 mode for $\frac{q}{\rho}=0,\,2$, with $\vec{q}$ along $z$ direction.}\label{sp2z}
\end{center}
\end{figure}
The resulting spectrum of quasi normal modes is shown in Fig \ref{sp2z}. There are two plots in this figure, the blue dots correspond to $q/\rho = 0$, and the red dots to $q/\rho=2$.
We see that all the allowed values of $\omega$ are in the lower half plane, showing that there are no instabilities in this channel. 
Changing the value of $q/\rho$ results in qualitatively similar plots. 
From the fact that there is no mode with $\omega \rightarrow 0$ as $q \rightarrow 0$ we see that there is no hydrodynamic mode in this 
channel.

\subsubsection{Spin 1}
\label{sp2momzsp1}

In the spin $1$ case the equations are coupled together. To decouple them it is useful to consider combinations of perturbations which are invariant under a gauge transformation
\be
\label{defgic}
h_{\mu\nu} \rightarrow h_{\mu\nu} + \nabla_\mu \xi_\nu + \nabla_\nu \xi_\mu.
\ee
where the gauge transformation parameter is of the form
\be
\label{gtpform}
\xi_M=f_M (v)\, e^{i\, q\, z -i\, \omega\, t},\qquad \text{with} \, \, M=t,x,y,z,v,
\ee
and $f_M(v)$ is a general function of $v$. 

Two gauge invariant combinations are 
\begin{align}
Z_1^{(1)}=&q\, h^x_t+\omega \,h^x_z \label{gica1}, \\
Z_1^{(2)}=&q\, h^y_t+ \omega \,h^y_z \label{gica2}.
\end{align}
These satisfy equations of the form, eq.\eqref{diffeqz} with  the coefficient functions given in eq.\eqref{abcsp1qz} of appendix \ref{qnmzapsp1}. 
\begin{figure}
\begin{center}
\includegraphics[width=0.8\textwidth]{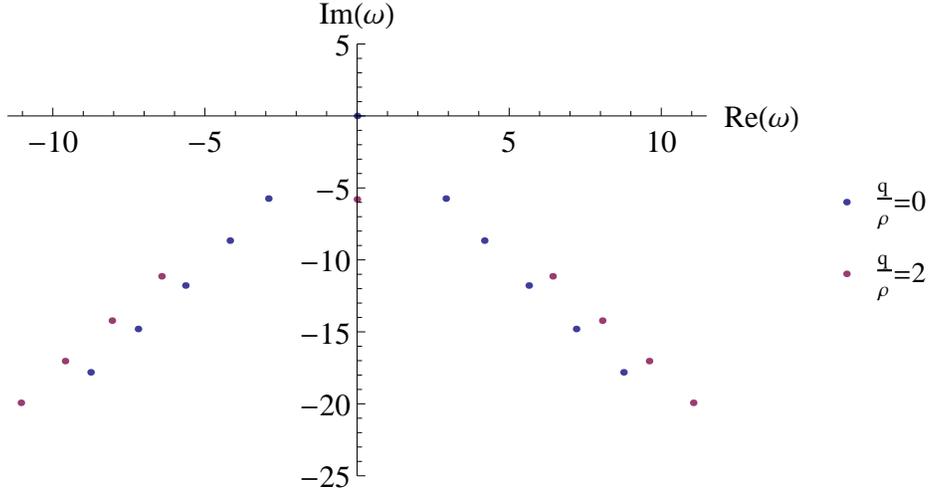}
\caption{QNM plots for spin 1 mode with $\frac{q}{\rho}=0,2$ with $\vec{q}$ along $z$ direction}\label{sp1z}
\end{center}
\end{figure}

The resulting spectrum of quasi normal modes is shown in Fig \ref{sp1z}. 
The blue dots in this figure correspond to $q/\rho=0$ and the red dots to $q/\rho=2$. 
Once again we see that all values of $\omega$ have $Im(\omega)<0$. 
There is a  mode, for $Z_1^{(1)}$ and $Z_{1}^{(2)}$,  with $\omega\rightarrow 0$ as $q\rightarrow 0$;
these are hydrodynamic modes. In appendix \ref{qnmzapsp1} we analyze this limit in more detail. The resulting dispersion relation, in this limit, we find is given by 
\be 
\label{sp1disz}
\omega=-i\frac{8\, \pi\,T\,q^2}{3\rho^2}.
\ee
This gives rise to a viscosity coefficient $\eta_\perp$ that agrees with the result obtained  using the Kubo formula in section \ref{grvissub1}, eq.(\ref{etaperpint}). 
Note that, eq.\eqref{gica1}, eq.\eqref{gica2}  suggests that in the fluid dynamics side the fluid flow has nontrivial solutions for fluctuation for 
velocity $\delta u_{x}(t,z)$ 
or $\delta u_{y}(t,z)$ but the temperature is fixed to its equilibrium value. This is precisely what we found in subsection \ref{apfltz} 
eq.\eqref{apfltz1}. 

There are two additional perturbations in the spin $1$ channel besides $Z_1^{(1)}, Z_1^{(2)}$. The constraint equations among the Einstein equations, which were mentioned above,  give rise to first order equations for these perturbations. 
For example, choosing one of these two remaining perturbations to be $h_{t}^{x}$ we get the equation
\be
\label{perthtx}
{h_{t}^{x}}'(v)=\frac{32 q (1-v^3)}{32q^2 (1-v^3)-3 v^2 \rho^2\omega^2}{Z_1^{(1)}}'(v),
\ee
here $Z_1^{(1)}$, which is a known function solving eq.\eqref{diffeqz} for any given $(q, \,\omega)$, can be thought of as a 
source term in this equation. The general solution to eq.(\ref{perthtx})  can be written as 
\be
\label{gensol}
h_{t}^{x}=(h_{t}^{x})^{hom} + (h_{t}^{x})^{P},
\ee
where $(h_{t}^{x})^{hom}$ solves the homogeneous equation
\be
{h_{t}^{x}}'(v)=0
\ee
and $(h_{t}^{x})^{P}$ is a particular solution sourced by $Z_1^{(1)}$. Now $(h_{t}^{x})^{P}$ will have the same frequency dependence as $Z_1^{(1)}$.
So the only additional possibility for  independent  modes is  from $(h_{t}^{x})^{hom}$. 
It is easy to see though that eq.\ref{perthtx} does not admit any solutions which  meet the required two boundary conditions at 
the horizon and infinity. Thus the full spectrum of allowed frequencies in the spin $1$  case 
is obtained from those for $Z_1^{(1)}, Z_1^{(2)}$.   
\subsubsection{Spin 0}
There are five perturbations in this channel. Two gauge invariant combinations are  denoted by $Z_0^{(1)}$ and $Z_0^{(2)}$. They are  given by 
\begin{align}
\label{sp0defz}
\begin{split}
Z_0^{(1)}(v)=\frac{v^2\rho ^2 \left(8  v+3 \omega ^2\right)}{8 \left(3 q^2+\rho^2\right)}h_{z}^{z} &-\frac{8 i q^2 \rho  v^3-i \rho^3v^2 \omega ^2}{12 q^3+4 \rho ^2 q}\psi-\frac{8 \left(v^3-1\right) \omega }{3 q}h^{z}_{t} \\ &
-\frac{4  \left(v^3-1\right)}{3}h^{t}_{t} +\frac{v^3+2}{3} h,
\end{split}
\end{align}
\begin{align}
\label{sp0defzt}
Z_0^{(2)}(v)=\frac{1}{6} \rho ^2 \left(v^2+\frac{2}{v}\right)h^{z}_{z}-\frac{1}{3} i q \rho  \left(v^2+\frac{2}{v}\right)\psi,
\end{align}
where $h=h_{x}^{x}+h_{y}^{y}$.
These satisfy the decoupled equations 
\begin{align} 
\label{sp0eqnZ}
a_1(v){Z_0^{(1)}}''(v)+b_1(v){Z_0^{(1)}}'(v)+c_1(v)Z_0^{(1)}(v)&=0, \\ \label{sp0eqnZt}
a_2(v){Z_0^{(2)}}''(v)+b_2(v){Z_0^{(2)}}'(v)+c_2(v)Z_0^{(2)}(v)&=0.
\end{align}
where the coefficients $a_1,\, a_2,\,  b_1,\, b_2,\, c_1,\, c_2$ are given in eq.\eqref{abcsp0qz}, eq.\eqref{abcsp0qz1}  appendix \ref{qnmzapsp0}. 
\begin{figure}
\begin{center}
\includegraphics[width=0.8\textwidth]{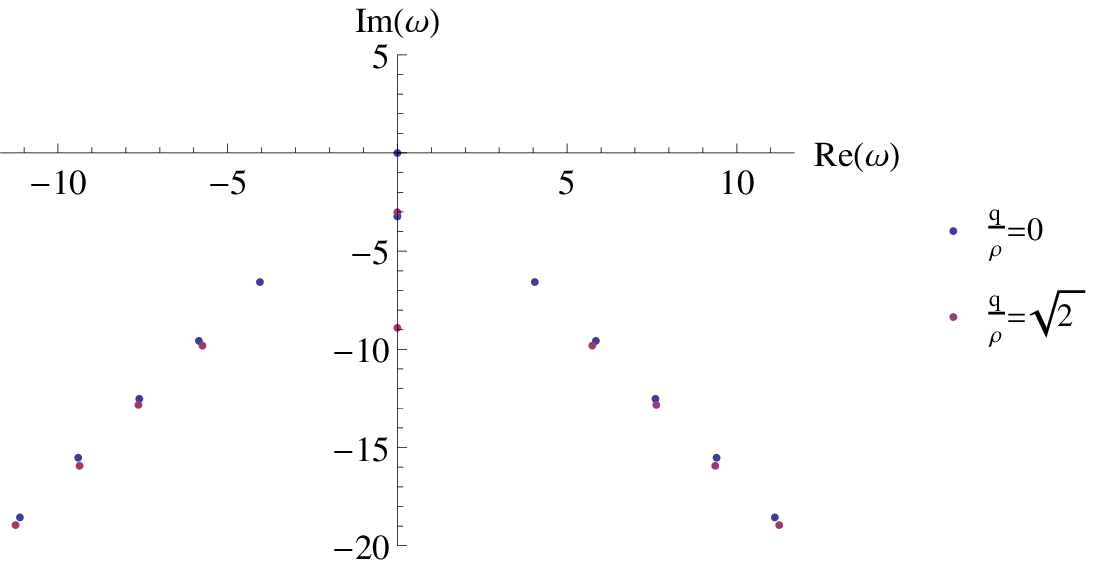}
\caption{QNM plot for spin 0 mode with $\frac{q}{\rho}=0,\,\sqrt{2}$~ for $Z_0^{(1)}(v)$ with $\vec{q}$ along $z$ direction}\label{sp0Z}
\end{center}
\end{figure}

\begin{figure}
\begin{center}
\includegraphics[width=0.8\textwidth]{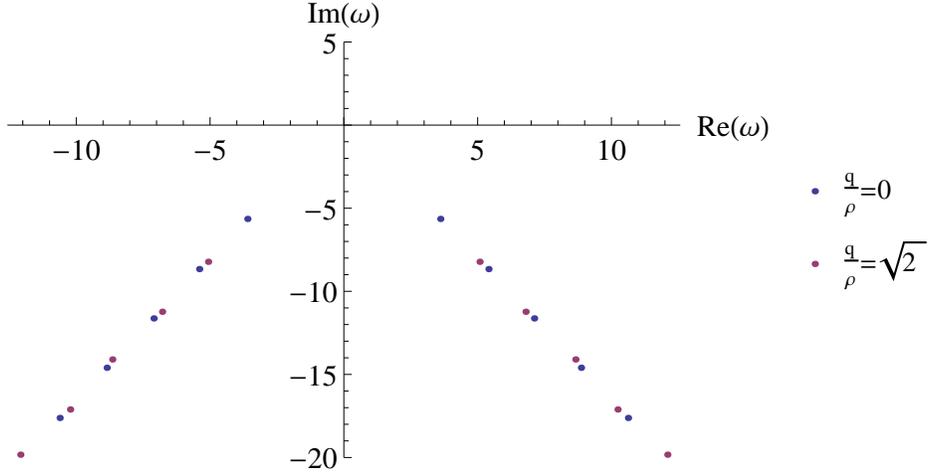}
\caption{QNM plot for spin 0 mode with $\frac{q}{\rho}=0,\,\sqrt{2}$ for $Z_0^{(2)}(v)$ with $\vec{q}$ along $z$ direction}\label{sp0Zt}
\end{center}
\end{figure}
 The resulting spectrum of quasi normal modes for $Z_0^{(1)}(v)$ and $Z_0^{(2)}(v)$  are  shown in Fig. \ref{sp0Z} and \ref{sp0Zt} respectively with the blue dots for $q/\rho=0$ and red dots for $q/\rho=\sqrt{2}$.
We see that all modes have frequencies lying in the lower half plane. 
From the plot for  the $Z_0^{(1)}(v)$ perturbation we see that there is a hydrodynamic mode. As discussed in appendix \ref{qnmzapsp0} it has a dispersion relation
\be 
\label{sp0disz}
\omega=-3i \frac{\pi\,T\,q^2}{\rho^2}.
\ee
Note that, $Z_0^{(1)}(v)$ in eq.\eqref{sp0defz}  suggests that in the fluid dynamics side the fluid flow has nontrivial solutions for 
fluctuation for velocity
$\delta u_{z}(t,z)$ and  for temperature $\delta T(t,z)$. This is precisely what we found in subsection \ref{apfltz} 
eq.\eqref{apfltz2}. 

The analysis of linearized perturbation in fluid mechanics is carried out in appendix \ref{hydmoap}. The mode
which corresponds to the hydrodynamic mode above is discussed in section \ref{apfltz}, eq.(\ref{rela123}).
Comparing we see that eq.(\ref{sp0disz}) puts constraints on a combination of the transport coefficients 
$\zeta^b_1, \zeta^b_2$ and the coefficients $n_{5_1}, n_{5_2}, n_{6_3}$ which appear in the derivative expansion of $\langle O \rangle$, eq.(\ref{tmrel3}). Assuming all six terms on RHS of eq.(\ref{rela123}) are comparable, barring cancellations,  we find that 
the transport coefficients $\zeta_1^b, \zeta_2^b$ scale with $T$ as 
\begin{eqnarray}
{\zeta_1^b \over s} &  \sim &  {1/T}, \label{scalez1} \\
{\zeta_2^b\over s} & \sim & T .\label{scalez2}
\end{eqnarray}
As a result $\zeta_1^b \gg \eta_\parallel$  while $\eta_\parallel \gg \zeta_2^b \gg \eta\perp$
at small $T$. 
Note that in deriving  eq.(\ref{scalez1}), eq.(\ref{scalez2})  we made  use of the relations $P_0 +\epsilon_0 + f\rho^2 \sim \rho^4 \sim O(T)^0$,
and $(\partial_T\epsilon)_0 \sim T^2$.

As in the spin $1$ case, the constraint equations can be used to obtain first order equations for the 
remaining three spin $0$ perturbations in terms of $Z_0^{(1)}, Z_0^{(2)}$. These equations do not give rise to any additional modes.
 
 Let us end this subsection by noting that the  independent modes  from which the quasi normal modes arose, were,
$h_{y}^{x},\, (h_{x}^x-h_{y}^y)/2,\, Z_1^{(1)},\, Z_1^{(2)},\, Z_0^{(1)},\, Z_0^{(2)}$, which is indeed $6$ in number as mentioned above. 
 
\subsection{Modes with $q_y \ne 0$}
\label{momqxy}
Next we consider modes with $q_z=0$. By using the rotational symmetry in the $x-y$ plane we can take the non-zero momentum to be along the $y$ direction and denote it by $q$ below. The perturbations then have a dependence going like $\delta \phi \sim e^{i q y - i \omega t}$. 
Since the analysis  parallels that of the previous subsection we will be more brief in the following discussion. 
Also, additional details can be found in appendix \ref{qnmxyap}. 

There are $11$ perturbations to begin with, of which only $6$ remain after using the constraints. We describe these modes and their quasinormal modes below.
The $6$ can be combined into the following combinations
\begin{align}
\label{gicmodeqz}
 \begin{split}
Z_1\,=&\,q \,h^{x}_{t}+\omega \, h^{x}_{y}, \\
Z_2\,=&\,h^x_z, \\
Z_3\,=&\,v^3 h_{z}^{z},  \\
Z_4\,=&\,\frac{4}{3} q^2 \left(v^3-1\right)\,h_{t}^{t}+2 q \omega  h_{t}^{y}+\omega^2\, h_{y}^{y} \\ &+\frac{1}{3} \left(2 q^2 \left(v^3+2\right)-3 \omega ^2\right)h^x_x+q^2 v^3\,h_{z}^{z},\\
Z_5\,=&\,h_{yz}+\frac{q}{\omega}h_{tz}, \\
Z_6\,=&\,h_{tz}+i\omega\frac{\rho}{8}\psi.
\end{split}
\end{align}
which are each gauge invariant under a gauge transformation of the form eq.\eqref{defgic}
with
\be
\label{gtpara2}
\xi_M=f_M(v) \,e^{i \, q \, y-i \,\omega \, t}, \qquad \text{for} \, \, M=t,x,y,z,v.
\ee

Each combination in eq.\eqref{gicmodeqz} satisfies an equation of the form in eq.\eqref{diffeqz}. The coefficient functions for each of the modes are given in appendix \ref{qnmxyap}. 
\begin{figure}
\begin{center}
\includegraphics[width=0.8\textwidth]{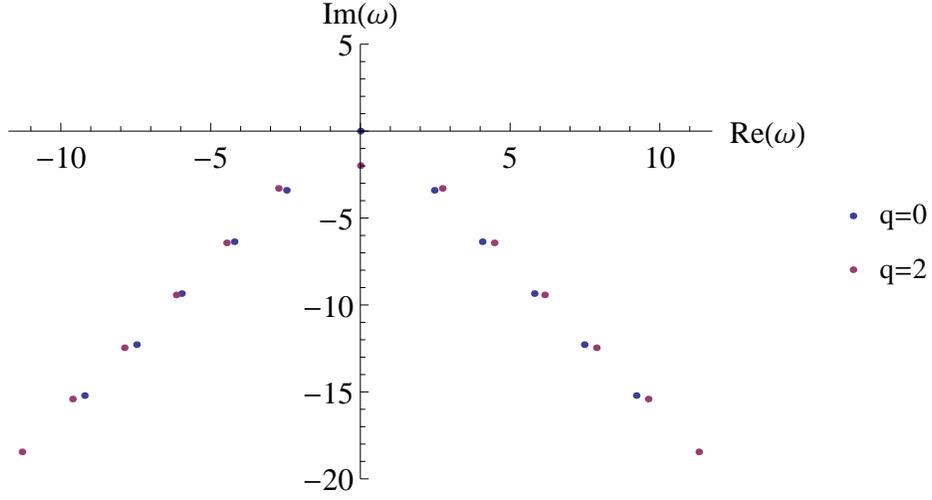}
\caption{QNM plot for $Z_1$ mode with $q=0,\,2$ with $\vec{q}$ along $y$}\label{sp2y}
\end{center}
\end{figure}

\begin{figure}
\begin{center}
\includegraphics[width=0.8\textwidth]{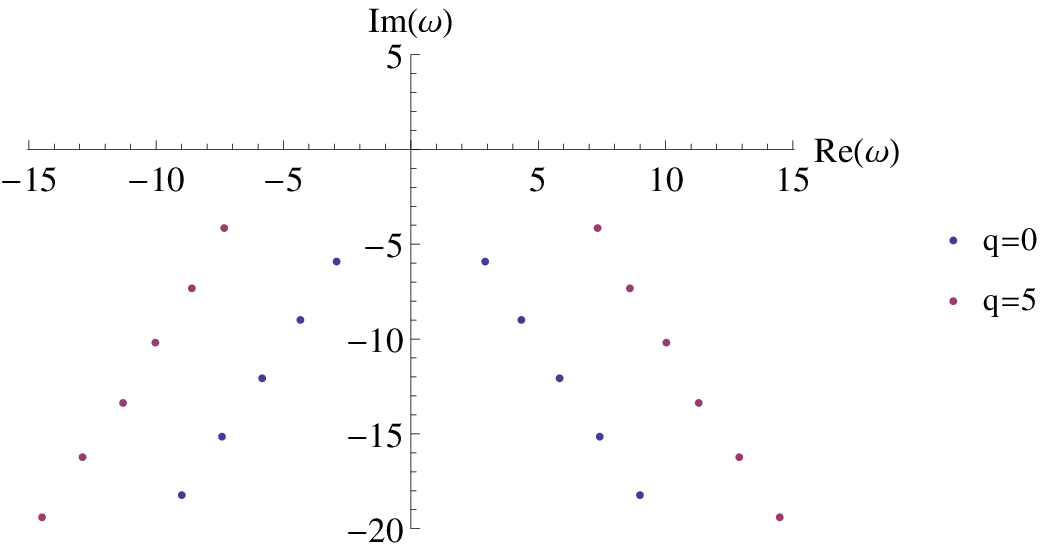}
\caption{QNM plot for $Z_2$ mode with $q=0,\,5$  with $\vec{q}$ along $y$}\label{sp1y}
\end{center}
\end{figure}

\begin{figure}
\begin{center}
\includegraphics[width=0.8\textwidth]{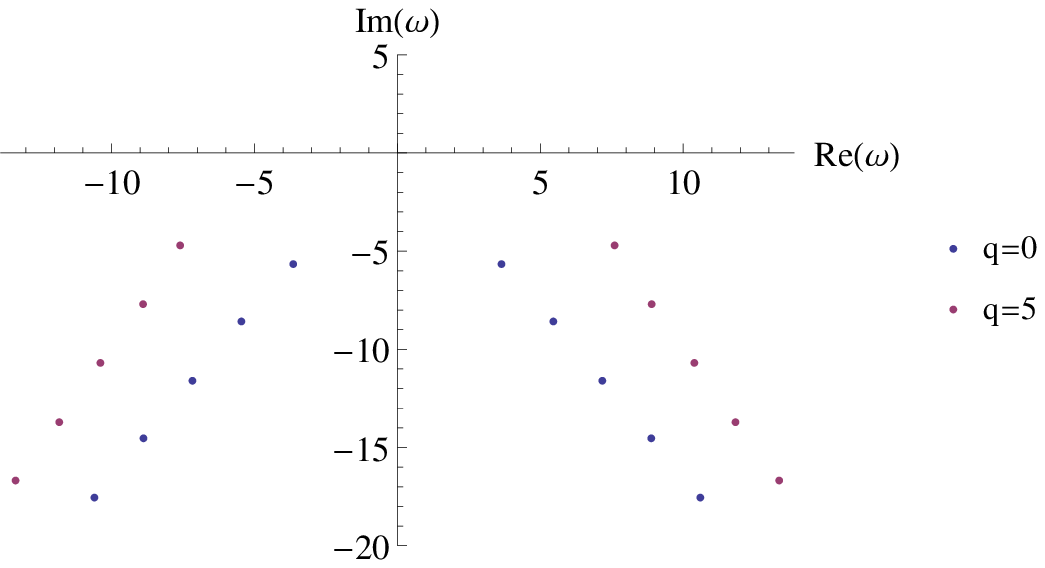}
\caption{QNM plot for $Z_3$ mode with $q=0,\,5$ with $\vec{q}$ along $y$}\label{sp0yZ}
\end{center}
\end{figure}

\begin{figure}
\begin{center}
\includegraphics[width=0.8\textwidth]{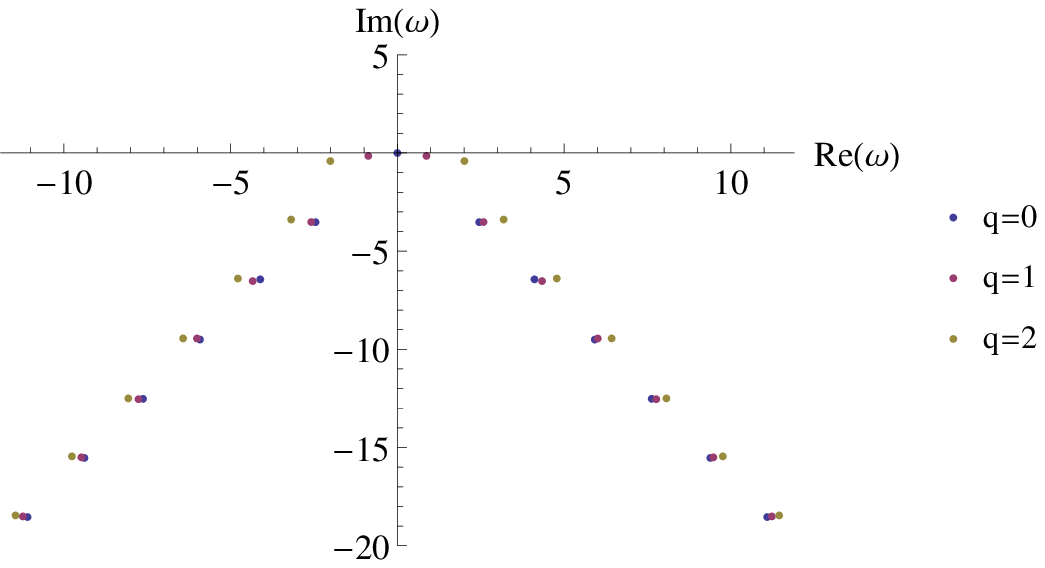}
\caption{QNM plot for $Z_4$ mode with $q=0,\,1,\,2$ with $\vec{q}$ along $y$}\label{sp0yZt}
\end{center}
\end{figure}

\begin{figure}
\begin{center}
\includegraphics[width=0.8\textwidth]{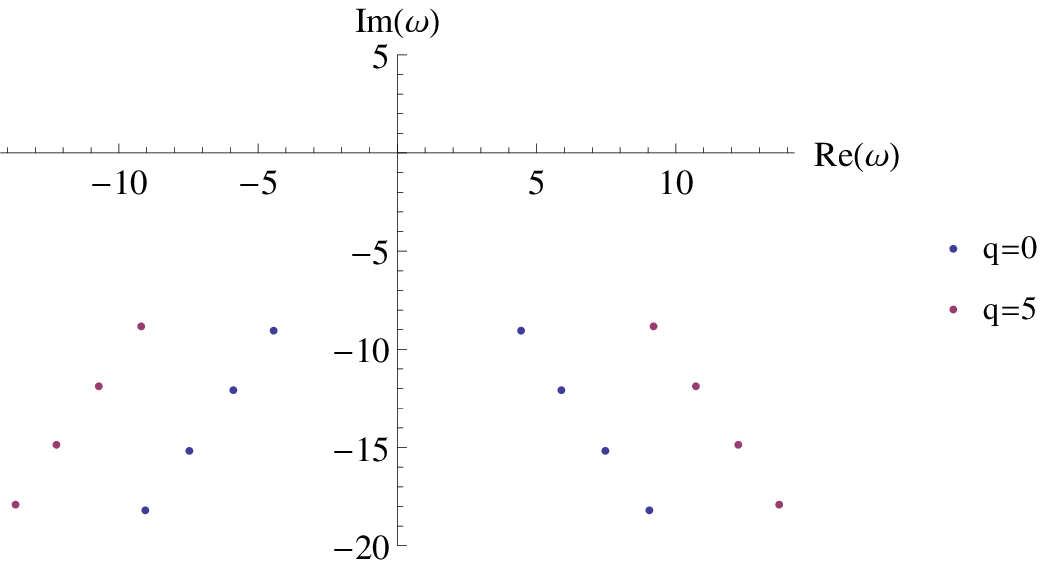}
\caption{QNM plot for $Z_5$ mode with $q=0,\,5$ with $\vec{q}$ along $y$}\label{ax_Z1}
\end{center}
\end{figure}

\begin{figure}
\begin{center}
\includegraphics[width=0.8\textwidth]{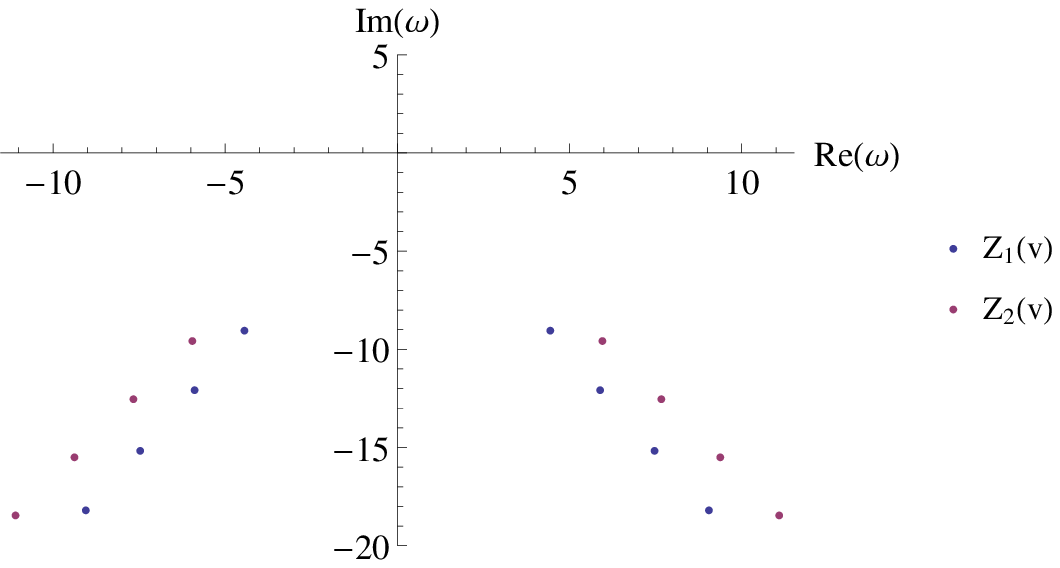}
\caption{QNM plot comparing both the $Z_5$ and $Z_6$ modes for $q=1$ with $\vec{q}$ along $y$.}\label{ax_Z1Z2}
\end{center}
\end{figure}
The resulting spectrum of quasi normal modes is given in Fig \ref{sp2y} to Fig \ref{ax_Z1Z2}. 
We observe that, mode $Z_1$ has a hydrodynamic mode in its spectrum. At, $q=0$ the hydrodynamic mode 
sits at the origin where as at finite momenta, this mode shifts in the imaginary axis as can be seen from
the plot for momenta $q=2.$ At smaller values of momenta the hydrodynamic mode takes the form (see appendix \ref{qnmyapsp2})
\be
\label{disy1}
\omega=- i\frac{q^2}{3\,\pi\,T}.
\ee

From the analysis of the hydrodynamic modes (see eq.\eqref{disp111} in appendix \ref{hydmoap}) we 
obtain 
\be\label{etas}
\omega=-i\frac{\eta_\parallel}{s T} \hat{q}^2,
\ee
where ${\hat q}$ is the momenta defined at the boundary of $AdS_5$.
To compare eq.\eqref{disy1}, eq.\eqref{etas}, 
we need to rescale the coordinates of spatial directions ($x$ and $y$) of $AdS_4$ and accordingly the momenta $q$, see eq.\eqref{defhatq}
 appendix \ref{qnmxyap}. 
Expressing eq.\eqref{disy1} in terms of the rescaled momenta ${\hat q}$ and using eq.\eqref{etas} we finally obtain (see appendix \ref{qnmyapsp2} for details)
 \be\label{etash}
\frac{\eta_{\parallel}}{s}=\frac{1}{4\pi}.
\ee
This matches with result in eq.\ref{etaprll1}. Note that, $Z_1$ in eq.\eqref{gicmodeqz} suggests that in the fluid dynamics side 
the fluid flow has nontrivial solutions for fluctuation for 
velocity $\delta u_{x}(t,y)$ 
 but the temperature is fixed to its equilibrium value. This is precisely what we found in subsection \ref{hydmoap1} 
eq.\eqref{hydmoap2}.

We also observe from Fig.\ref{sp0yZt} that, the mode $Z_4$ has a hydrodynamic mode and the dispersion relation is given by
\be
\label{valom}
\omega=\pm\sqrt{\frac{2}{3}}q-i\frac{q^2}{6\, \pi\, T },
\ee
which when expressed in terms of ${\hat q}$ gives 
\be
\label{valhatom}
\hat{\o}=\pm\frac{1}{\sqrt{2}}\hat{q}-i\frac{\hat{q}^2}{8\,\pi\, T }.
\ee
As explained in more detail in appendix E.2.3,  we can compare this dispersion relation to 
that of a suitably identified mode (see eq.(\ref{etaso})) in the  linearized analysis in fluid mechanics. 
Comparing, we find that the coefficient of the
 linear term in ${\hat q}$ corresponds to  a speed of sound,
\be
\label{valcs}
c_s=\frac{1}{\sqrt{2}}, 
\ee
which agreed with the result from fluid mechanics, while the coefficient of the quadratic term in ${\hat q}$, upon 
using the fact that \be{\eta_{\parallel} \over 2(\epsilon+ P)}={1 \over 8\,\pi\,T},\ee
we obtain a relation, eq.(\ref{defcs2}) between two transport coefficients $\zeta_2^a, \zeta_1^2$ defined in eq.(\ref{tmnrel4}). 
Note that, $Z_4$ in eq.\eqref{gicmodeqz} suggests that in the fluid dynamics side 
the fluid flow has nontrivial solutions for fluctuation for 
velocity $\delta u_{y}(t,y)$ and temperature fluctuation $\delta T(t,y)$. This is precisely what we found in subsection \ref{hydmoap1} 
eq.\eqref{hydmoap3}.

The quasinormal mode for $Z_5$ is plotted in Fig. \ref{ax_Z1}. The plot for the other mode, \emph{i.e.} $Z_6$, will be similar and we are not plotting quasi normal modes for it separately. In Fig. \ref{ax_Z1Z2} we plot the quasi normal modes corresponding to both, $Z_5$ and $Z_6$, for comparison. 

\subsection{Modes with general momentum turned on}\label{qyqz}
Finally we turn to modes where all components of the momentum, $q_x, q_y, q_z$ are turned on. 
Using the rotational symmetry we can set $q_x=0$. 
We have not been able to analyze the spectrum of quasi normal modes in full detail in this case. 
Some partial results are as follows.

\begin{figure}
\begin{center}
\includegraphics[width=0.8\textwidth]{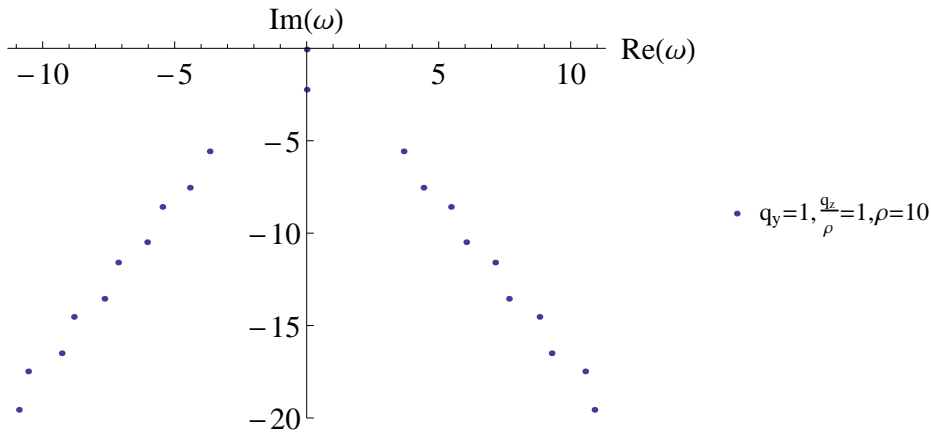}
\caption{QNM plot for $Z(v)$ and $\tilde{Z}(v)$ mode with both momentum turned on, $q_y= 1,\,\frac{q_z}{\rho}=1,\, \rho=10$}\label{bomom_qz1}
\end{center}
\end{figure}

\begin{figure}
\begin{center}
\includegraphics[width=0.8\textwidth]{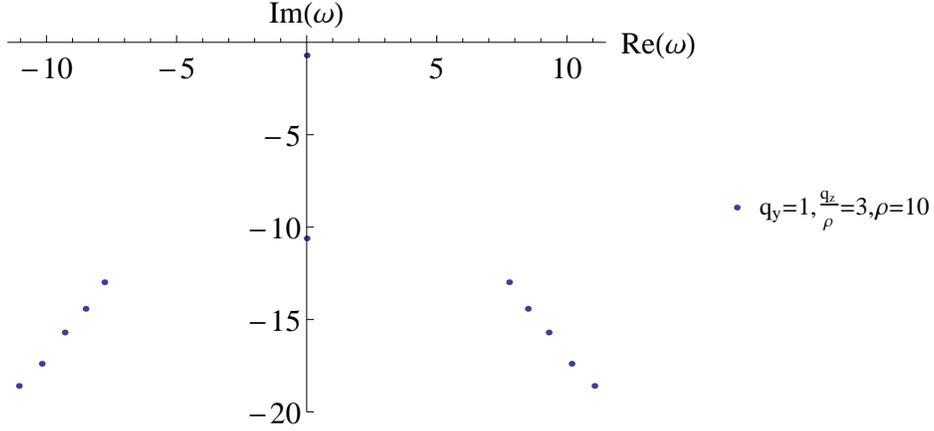}
\caption{QNM plot for $Z(v)$ and $\tilde{Z}(v)$ mode with both momentum turned on, $q_y= 1,\,\frac{q_z}{\rho}=3,\, \rho=10$}\label{bomom_qz3}
\end{center}
\end{figure}

Similar to the analysis for the previous two subsections \ref{momqz} and \ref{momqxy} in this case also there are $6$ 
independent perturbations. Four of the six perturbations couple among themselves and we do not analyze their quasi normal spectrum since it 
is very complicated. The rest of the two perturbations are also coupled among themselves and we carry out a similar analysis as in the 
previous subsections, to obtain the spectrum of their quasi normal modes. These two modes are 
\be
\begin{split}\label{Ztzxy}
 Z(v)=&h^x_y(v) + {q_y \over \omega} h^x_t(v), \\
 \tilde{Z}(v)=&h^x_z(v) + {q_z \over \omega} h^x_t(v).
\end{split}
\ee
The equation of motion for them is given in appendix \ref{apqyqz}.
The spectrum of the quasinormal modes for them are plotted in Fig.\ref{bomom_qz1} and Fig.\ref{bomom_qz3}.

We see from the figures, Fig.\ref{bomom_qz1} and Fig.\ref{bomom_qz3} that there are two hydrodynamic poles, which shifts accordingly along the negative imaginary axis for increasing momenta.

\section{Fluid mechanics in anisotropic phases}
\label{fluidmech}

\subsection{Anisotropic fluid mechanics: general discussion}
We begin this section by first describing how to set up the general equations of forced 
fluid mechanics  for an anisotropic phase 
 where the breaking of 
rotation invariance is characterized by a vector ${\bf \xi}$ which is constant in space-time. In the present context 
${\bf \xi}$ is given  by eq.(\ref{defxi}).

Fluid mechanics is an effective theory which  describes slowly varying situations. More precisely,
it is an  effective theory that is valid when the four velocity, temperature, and any other relevant parameters,
 are varying slowly compared to the mean free path of the quasi particle excitations in the system. 
In a conformal field theory, where 
there are no well defined quasi particles, the temperature plays the role of the mean free path, since it is the only scale which characterizes the equilibrium configuration.  

For systems we are interested in the fundamental variables of fluid mechanics will be  the four velocity $u^\mu$ and temperature $T$.  
To get the equations of fluid mechanics one expands the stress energy tensor in terms of these variables in a  derivative  expansion
and obtains  the constitutive relation. Then using the equations of momentum-energy conservation,
\be
\label{consstress1}
\partial_\mu T^{\mu\nu}=0,
\ee
or their generalization in the presence of forcing functions, 
leads to  the equations of fluid mechanics.

In the anisotropic case,  the equilibrium configuration is characterized by two scales, the 
temperature, $T$, and the length of the vector 
${\bf \xi}$. From eq.\eqref{defxi} and eq.\eqref{formdil} we see that 
\be
\label{lengthxi}
|{\bf \xi}| = \rho. 
\ee
We will take  the spatial-temporal gradients to be  small compared to both $T, \rho$. 
In such situations the stress energy tensor can be written in a derivative expansion,
\be
\label{stressteng}
T_{\mu\nu}=T^{(0)}_{\mu\nu}+ T^{(1)}_{\mu\nu} + \cdots,
\ee
where the superscript $0,1$ indicate terms with no and one derivative respectively and the ellipses stand for higher derivative terms. 

At zeroth order the constitutive relation takes the form, 
\be
\label{zerostress}
T^{(0) \mu\nu}=\epsilon \, u^\mu\, u^\nu + P (\eta^{\mu \nu}+ u^\mu\, u^\nu) + f\, {\hat \xi}^\mu\, {\hat \xi}^\nu,
\ee
where 
\be
\label{defxihat}
{\hat \xi}^\mu\equiv \xi^\mu+ (u \cdot \xi) \, u^\mu,
\ee
and $u^\mu$ is the four velocity of the fluid at any point in spacetime. 
Note that the three terms on the RHS of eq.(\ref{zerostress}) are the most general we can write down, keeping in mind the fact that the 
rotational anisotropy  of the equilibrium situation is characterized by a single vector,  ${\bf \xi}$. We have also imposed that stress tensor satisfies the condition
\be
\label{condstressev}
T^\mu_\nu\, u^\nu=\epsilon\, u^\mu.
\ee
This can be taken as the definition of $u^\mu$, namely that it is the timelike  eigenvector of $T^\mu_\nu$, with norm, $u^2=-1$, 
 and $\epsilon$ is the corresponding eigenvalue. 
The three coefficients, $\epsilon, P, f$ in eq.(\ref{zerostress}) depend in general  
 on the three independent scalars in the problem, $T, \rho$, and also, $u \cdot \xi$.
On imposing the symmetry
that $T^{\mu \nu}$ is invariant under
${\bf \xi } \rightarrow -{\bf \xi}$, we learn that $\epsilon,\, P ,\, f$ are even functions of $u \cdot \xi$.

We see  that  the form in eq.(\ref{zerostress}) agrees with eq.(\ref{lto}) after noting that in section \ref{thermoft} we considered situations where
 \be
\label{condvela}
u \cdot \xi =0,
\ee 
and therefore, 
${\hat \xi}^\mu= \xi^\mu$.
Due to eq.(\ref{condvela}),  $\epsilon, P, T,$    were    found to be functions of $T, \rho$ alone in  section \ref{thermoft}.

Going beyond leading order is now conceptually straightforward. For example, to obtain $T^{(1) \mu\nu}$ one writes down all the independent terms of the appropriate tensorial type involving $T,\, \xi^\mu,\, u^\mu$ and one derivative.
The coefficients of these terms, which are the analogue of $\epsilon,\, P,\, f$ above are functions of $T,\, \rho,\, u\cdot \xi$. 
In practice though this is quite complicated because in the absence of rotational invariance the number of independent terms proliferate. 
As discussed in appendix \ref{FMdetail} there are 10 independent terms\footnote{It turns out that 
 for the specific system at hand, the number of independent terms is smaller,
 as discussed in appendix(\ref{FMdetail}). The  trace conditions as discussed in 
eq.\eqref{fotracea}, eq.\eqref{fotraceb}, cuts down the number of independent terms to $8.$ Further equilibrium consideration, reduces the  number of transport coefficients to even smaller number. See discussion around  eq.\eqref{indeq}.} which can appear in $T^{(1) \mu \nu}$, eq.\eqref{tmnrel4}.

In a translationally invariant system
the stress-energy would be conserved and the equations of fluid mechanics would follow from demanding that
\be
\label{consmom}
\partial_\mu T^{\mu\nu}=0,
\ee
where we would substitute for $T^{\mu\nu}$ in terms of $u^\mu,\, T,\, \xi^\mu$ using the constitutive relation. 

For a forced system the forcing terms would appear on the right hand side of the equation above. 
For example with the dilaton turned on one gets 
\be
\label{consstress}
\partial_\mu T^{\mu\nu}=\langle O \rangle\, \partial^\nu \phi,
\ee
where $\langle O \rangle$ is the expectation value of the operator dual to the dilaton.

Due to this complication,  for the forced system at hand, we must also consider the behavior of $\langle O \rangle$ when studying the stress tensor. We can write, in a similar fashion to eq.(\ref{stressteng})  in a gradient expansion
\be
\label{expdila}
\langle O \rangle= \langle O \rangle^{(0)} + \langle O \rangle^{(1)} + \langle O \rangle^{(2)} + \cdots,
\ee
where the superscripts, $0,1,2,$ indicate the order of the derivative expansion. Next, we need to expand $\langle O \rangle^{(n)}$ in terms of
scalar terms involving  $u^\mu, T$ and  the required number of derivatives. Inserting this expansion on the RHS of eq.(\ref{consstress})
would give the required equations of fluid mechanics for this dilaton system. 
  Note that one needs to go upto $n=2$ on the RHS of eq.(\ref{expdila}) if $T^{\mu\nu}$ is being expanded upto 1st order 
for consistency in eq.(\ref{consstress}). 

In appendix \ref{FMdetail} we carry out this procedure of expanding both $T^{(1) \mu\nu}$ and $\langle O \rangle$ upto $\mathcal{O}(\langle O \rangle^{(2)})$.
As mentioned above $10$ independent terms appear in the expansion of $T^{(1) \mu\nu}$ which takes the form,
\be
 \label{relt4m}
 \begin{split}
 T^{(1)}_{\mu\nu} = \sum_{i=1}^3 \zeta^a_i \, \tilde P_{\mu\nu} \, S^{(1)}_i + \sum_{i=1}^3 \zeta^b_i \, \hxi_{\mu} \hxi_{\nu}\, S^{(1)}_i + \sum_{i=1}^3 {v}_i \, (\hxi_{\mu}\, {V_i}^{(1)}_{\nu}+\hxi_{\nu}\, {V_i}^{(1)}_{\mu})+ \eta \,t^{(1)}_{\mu\nu},
 \end{split}
\ee
where $S_i, V_{i\nu}^{(1)}, t_{\mu\nu}^{(1)}$ refer to scalar, vector and  tensor terms
 which are defined in eq.(\ref{defscal}), eq.(\ref{defvecs}, eq.(\ref{deftens}). 
Similarly,  $5$ terms appear in the expansion of $\langle O \rangle^{(1)}$.
 And $20$ independent terms appear in $\langle O \rangle^{(2)}$.   
Let us also mention that both the coefficients which appear in $T^{(1) \mu \nu}$ and in the expansion of $\langle O \rangle$ are in general functions 
of $T,\, \rho,\, u \cdot \xi$.
This completes our discussion of the fluid mechanics in anisotropic driven systems in general. 

Let us end this subsection with one comment. The constitutive relation for the stress tensor we have 
obtained in the anisotropic case is analogous to  that for a superfluid, \cite{Bhattacharyya:2012xi}, with the role of ${\hat \xi}$
 being played by the component of the superfluid velocity orthogonal to the velocity of the normal fluid.

\subsection{The Dilaton Driven System}
We are now ready to apply these general considerations to the dilaton driven system of interest
 here. In principle, in this case, the dependence of the coefficients which appear in 
$T^{(1)}$ and also in $\langle O \rangle$ can be obtained by carrying out a complete analysis on the gravity 
side. But we will not carry out this analysis in detail  here since it is quite complicated.  
 Important constraints on various constraints arise, for example, close to extremality. These will be discussed
further towards the end of this section and also in 
 in appendix \ref{FMdetail}.  

To begin, in the next subsection  we will focus on a very simple situation, free from many of the complications,
  which allows us to 
understand some of the physical consequences of the  small viscosity we had found close to extremality above. 

\subsection{Flow between two plates}
\label{thoughtex}
\begin{figure}
\begin{center}
\includegraphics[width=0.6\textwidth]{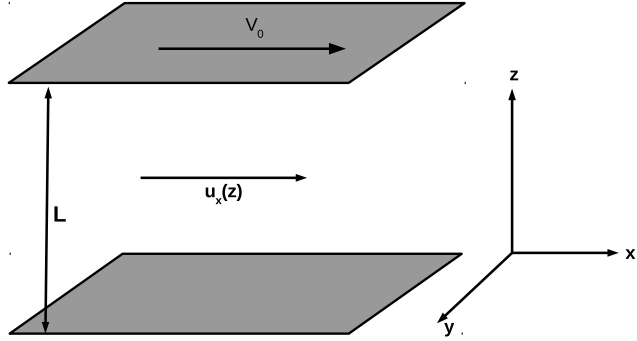}
\caption{Fluid flow between two parallel plates separated in the $z$-direction by a distance $L$. The upper plate has velocity $v_0$ along the $x$-direction and the lower plate is at rest. The fluid generates a flow along $x$-direction with velocity $u_x(z)$.}\label{plates}
\end{center}
\end{figure}

This situation arises when we consider the  fluid to be   enclosed between two parallel planes
which are moving with a relative velocity to each other. This is a standard situation which is studied to understand the role of viscosity, see for example, section 17 in \cite{Landau1987Fluid} and section 12.3 in \cite{REI65}. We take the two plates to be separated along the $z$ direction, this is the direction along which the dilaton varies in the gravity description, and it is therefore the direction along which rotational invariance is broken. The two plates are in the $x-y$ plane and taken to be infinite in extent, with the bottom plate, at $z=0$ being at rest, and the top plate, at $z=L$ moving with a velocity $v_0$ along the $x$ direction. See Fig.\ref{plates}. 
Usually, in such a situation  a velocity gradient develops  for the fluid, which in turn generates an off 
diagonal stress tensor proportional to the  viscosity. As a result the fluid exerts a frictional force per unit area for example on the upper  plate,
 and an opposing external force must be exerted on it to keep it moving   at a constant velocity. This opposing force must  then be also proportional to the viscosity and can be in fact used to measure it. 

We would like to understand if the anisotropic fluid we are studying  behaves similarly. In particular,
whether the required force becomes anomalously small close to extremality, where as we saw in section \ref{grvis} eq.\eqref{etaperpint} the component of 
viscosity $\eta_\perp$ vanishes.

In the situation discussed above it is reasonable to assume at steady state that the only gradients are along the $z$ direction. 
To simplify the analysis  we will also assume that the velocity $v_0$ is small (compared to the speed of light), so that the flow is non-relativistic, and we will work to first order in the spatial velocities, $u^i, i=x,y,z$. 
We will find that the Navier Stokes equations upto second order in the derivative expansion, are then all self consistently satisfied 
if 
\begin{eqnarray}
u^y & = & 0, \label{condux} \\
u^z & = & 0, \label{conduz} 
\end{eqnarray}
and if the temperature $T$ is a constant independent of $z$. 
The Navier Stokes equations upto second order are 
\be
\label{soNS}
\partial_\mu \,(T^{(0)\mu\nu} + T^{(1) \mu\nu}) = (\langle O \rangle^{(1)}+ \langle O \rangle^{(2)})\, \xi^\nu.
\ee
With the ansatz made above for $u^y, u^z, T$ one finds all the terms contributing to $\langle O \rangle^{(1)}, \langle O \rangle^{(2)}$ vanish. 
This is a tremendous simplification and essentially occurs because the flow we are considering has a shear nature making all the 
scalars $S_i$ which appear in eq.(\ref{relt4m}) and in eq.(\ref{forO}), eq.(\ref{tmrel3}) vanish. 
Also  eq.(\ref{soNS}) with $\nu=t,y,z$ are identically met to the required order.
This leaves only the $\nu=x$ component of eq.(\ref{soNS}). The only contribution to it  comes from $T^{(1) xz}$, for which in turn we 
find that only the ${V_1}_{\alpha}$ term is activated, eq.(\ref{relt4m}),  see also
 appendix \ref{FMdetail} eq.\eqref{st2ex}, resulting in the equation, 
\be
\label{nsvx}
\partial_z^2 u_x=0,
\ee
with solution
\be
\label{solvxso}
u_x=a \, z+ b.
\ee
Due to the viscosity the fluid will acquire the velocity of the plates at the two ends in the steady state solution we are describing here. 
Imposing these two boundary conditions  gives
\be
\label{fsolvx}
u_x=v_0 {z \over L}.
\ee

In fact this solution is the same as obtained for a more conventional isotropic fluid described in  section 17 of \cite{Landau1987Fluid}. 
Thankfully all the additional complications in the anisotropic case have dropped out for this simple fluid flow! 

We can finally now evaluate the stress tensor for this solution. It is easy to see that the relevant component to determine the force 
on the plates is $T^{xz}$. One finds for the solution above, to the order we are  working in,  that $T^{(0) xz}=0$ and thus 
\be
\label{txzsol}
T^{\mu\nu} = T^{(1) xz}= v_1 \, \rho^2 \, { v_0 \over L}.
\ee
where we have used eq.(\ref{relt4m}). 
This, upto a possible sign, is the force per unit area exerted by the fluid on the plates.
The force is a friction force, and should act in a direction 
to retard the relative  motion of the plates.
Using eq. (\ref{appetarel}) in  appendix \ref{kuboapp} to relate   $v_1$  to $\eta_\perp$ gives,
\be
\label{txzsol2}
T^{(1) xz}= \eta_\perp {v_0 \over L}.
\ee

We see that analogous to the rotationally invariant fluid  the force required to sustain the gradient flow is a proportional to 
the viscosity. In particular,  for the plates separated along the anisotropy direction $z$ the relevant
 component of the viscosity is $\eta_\perp$. As noted in section \ref{grvis} eq.\eqref{etaperpint} close to extremality this scales like 
$\eta_\perp \sim T^4/\rho$, and is much smaller than the entropy density, $s$. As a result, close to extremality the required force does become very small\footnote{
Note since gradients must be small compared to $T$, for fluid mechanics to be valid,  this requires $v_0/L$ to be sufficiently small.}.

In the set up analyzed above the plates were separated along the $z$ direction with a relative velocity along the $x$ direction.
Due to the breaking of rotational invariance other situations related to the one above by a rotation will behave differently.
For example, taking the plates to be separated say along $x$ with a relative velocity along $y$ leads to  a solution analogous to the one
 above. However, now the force is determined by $\eta_\parallel$, which close to extremality goes like 
$\eta_\parallel \sim T^2 \rho$, and is of order $s$  and  much bigger. 
A third configuration is where the plates are separated say along $x$ direction but the plates move in the $z$ direction.
Analyzing this situation is more complicated. 

\subsection{More On Fluid Mechanics Close to Extremality}
Having studied the simple situation above, let us now return to analyzing  the fluid mechanics
 which arises close to extremality
in our system in more generality.  

 For Lorentz invariant systems 
fluid mechanics can be thought of as an effective theory  that
 governs the dynamics of Goldstone modes which arise by carrying out local boost transformations or local changes in temperature. A local boost gives rise to a 
locally varying four velocity, $u^\mu$, starting from an equilibrium configuration. Combining this
 with a locally varying temperature gives rise to the $4$  variables of fluid mechanics, $u^\mu,T$ with   $u^\mu$ 
meeting
 the condition, $u^2=-1$.  Now for the  
the system at hand, when it is close to extremality, we see from the gravity description that
the near horizon geometry is a black brane in $AdS_4 \times R$. Thus the boost symmetry along the $z$ direction is 
 broken, due to the linearly varying dilaton. 
This shows that the degrees of freedom  in the near extremal   fluid motion are only   three,
$u^0,u^x,u^y$, meeting the condition, $u^2=-1$,  and $T$. In particular, the fluid cannot move in the $z$ direction,
at least in the fluid mechanics approximation, and  $u^z=0$. Note that 
$u^0,u^x,u^y, T$ are  functions
of $t,x,y$ and $z$  in general, since gradients along the $z$ direction are allowed.

As a result of this  observation
\be
\label{condce}
u \cdot  \xi =0
\ee
 for the fluid close to extremality and this leads to a some  simplification in the constitutive relation 
and in the resulting equations of motion.
It is worth emphasizing that even though  the degrees of freedom have been reduced  to $3$ in number, the  equations
that the fluid must meet are still given by eq.(\ref{consstress}),   with $\nu=t,x,y,z$, and are $4$ in number. 
Thus the  fluid must satisfy one non-trivial  constraint close to extremality. Ensuring that this additional equation 
is met would serve as an important check in a more detailed analysis of the coefficient functions etc. 
 
Some of the consequences of these comments are worth mentioning in more detail. 
Eq.(\ref{condce}) and eq.(\ref{defxihat}) imply that ${\hat \xi}=\xi$. 
As a result,  since $\xi$ is  a constant, $V_3^{(1)}$ in eq.(\ref{defvin}) vanishes leading to the corresponding 
term in the 
constitutive relation eq.(\ref{relt4m}), and also in $\langle O \rangle^{(2)}$, eq.(\ref{tmrel3}), being absent.
Also, $S_4^{(1)}, S_5^{(1)}$ vanish, so the first order relations eq.(\ref{foca}) and eq.(\ref{focb}) imply important constraints on the fluid.

Let us end with two more observations \footnote{We thank S. Minwalla for discussions related to these observations.}.
First, an analysis of the hydrodynamic modes was carried out in  detail in section \ref{qnmspec}, see also appendix \ref{appqnm}. Both for the case where $q_z$ is turned on and $q_y$ is turned on, one finds exactly $3$ Goldstone modes in the limit when the momentum vanishes. Also, the functional forms of the  hydrodynamic modes in this limit agree with that for the Goldstone modes
obtained by changing $T$ and carrying out boost along the $x,y$ directions of the near extremal brane, in agreement 
with the discussion above. 
Second, far from extremality when, $T/\rho \rightarrow \infty$, the black brane is approximately the $AdS_5$ 
Schwarzschild black brane and the fluid can clearly have a velocity along the $z$ direction, and  
 $4$ degrees of freedom, including $u^z\ne 0$,  must enter the fluid mechanics description. This is because there is an approximate Goldstone mode, with the boost along $z$ being broken only mildly by the linearly varying dilaton in this limit. As $T/\rho$ increases this breaking became more important until eventually when $T/\rho \rightarrow \infty$, close to extremality, the additional mode corresponding to boosts along $z$ is no longer light and should not be kept in the 
low-energy fluid dynamics. More physically, one expects a  drag force to bring the velocity along $z$ direction to
 zero on a time scale faster than $T$, close to extremality. It would be interesting to investigate 
 the change in behavior as $T/\rho$ is varied in more detail. 

\section{String embeddings}
\label{stringemb}
The gravity-dilaton system we consider in this paper, with a negative cosmological constant,
is well known to arise as a consistent truncation in string theory. 
The most famous example being that of IIB string theory in the  $AdS_5 \times S^5$ background,
 which is dual to the ${\cal N}=4$ SYM theory. 
Other examples are when the $S^5$ is replaced by some other  Einstein Manifold. 
For example $T^{1,1}$,
 which is the base of the conifold, or orbifolds of $S^5$. 
Thus it is easy to embed this system in string theory and/or supergravity. 

However once this embedding is done one must also analyze the additional modes which arise in the full theory in addressing the question of stability, and this poses a non-trivial constraint, as we now discuss. In particular, for IIB theory on  $AdS_5 \times S^5$,
 starting from the ten dim. SUGRA theory, one gets fields in the $5$ dim. theory which arise after 
KK reduction on the $S^5$ and which carry non-trivial $SO(6)$ charge. These fields do not mix with 
the dilaton and $5$ dim. graviton which are excited in the background black brane geometry we have 
analyzed above. If a mode of this type lies below the BF bound of the near horizon
 $AdS_4$ geometry present in the extremal or near extremal black brane, then   the 
corresponding brane solution is unstable, as  we show in appendix \ref{instads}.  
 Now, actually a scalar field  of this type
does arise in the $AdS_5 \times S^5$ theory. It has a mass
which in units of the  $AdS_5$ radius is 
\be
\label{bf5a}
m^2 L^2=-4
\ee
 and therefore it 
saturates  the BF bound for $AdS_5$. In $AdS_4$ the condition for a scalar to lie above the BF bound is 
\be
\label{bf4}
m^2 L_4^2 \ge -{9/4}.
\ee
Using the relation eq.(\ref{rad4})  between the $AdS_4$ and the $AdS_5$ radii we see that this scalar violates
 the BF bound of $AdS_4$.  
This scalar field  arises from the four-form and the metric with legs along the $S^5$, and transforms as a $20$ of $SO(6)$. It is the $k=2$ mode in the series described in the first equation in
eq.(2.34) of \cite{Kim:1985ez}.  In the gauge theory this mode corresponds to the bilinear made from two 
scalars, $Tr(\phi^i\phi^j) -1/6 \delta^{ij} Tr(\phi^i\phi^i)$. 

The extremal or near extremal black branes, with $T/\rho \ll 1$, are therefore unstable in the 
IIB theory compactified on  $S^5$.  As the temperature increases, the $AdS_4$ region 
disappears, when $T \sim \rho$, and the geometry of the black brane becomes more akin to 
an $AdS_5$ Schwarzschild. One expects that the  instability will therefore disappear
when $T \sim \rho$.  

Let us note that the  instability we have found above is akin to what was also found in \cite{Azeyanagi:2009pr} where extremal black branes with a linearly varying axion field was considered. 

It is easy to check that an   unstable mode close to extremality   
  is also present when  the $S^5$ is replaced by an $S^5/Z_N$ orbifold, and cannot be completely projected out. 
An  instability also arises in the $T^{1,1}$ case where for e.g., 
there is a scalar with a mass meeting eq.(\ref{bf5a}), which is dual to the operator, $|A|^2-|B|^2$, of dimension $2$.
The $T^{(1,1)}$ theory also has a scalar with mass $m^2L^2=-15/4$ which is  dual to the scalar component of the chiral  operator $Tr(AB)$ of dimension $3/2$. 
For the spectrum of IIB on $T^{(1,1)}$ see \cite{Ceresole:1999zs}, we are using the notation of \cite{Klebanov:1998hh} for the operators in the field theory,
see also \cite{Klebanov:1999tb}.  

It would be interesting to ask if there are compactifications where the $S^5$ is replaced by another suitable Einstein manifold for which there is no such instability. We leave this important 
question for the future.

\section{Conclusions}

In this paper we have analyzed a simple example of an anisotropic phase which arises when 
we consider a system of gravity coupled to the dilaton, with linearly varying boundary conditions for the dilaton. 
The dual description is that of  a CFT which is subjected to linearly varying source term. 
 The system actually breaks both rotational 
and translational invariance but it turns out that its behavior, as far as equilibrium thermodynamics is 
concerned, is in fact translationally invariant. We learn this from the  gravity theory where 
 the equilibrium configuration corresponds to a black brane at finite temperature. The back reaction of the 
dilaton breaks  rotational invariance, but not translational invariance, 
 since this back reaction depends on the gradient 
of the dilaton field which is   constant for a linearly varying profile.

There is one dimensionless parameter, $T/\rho$, where $T$ is the temperature and $\rho$ the dilaton gradient,
eq.(\ref{formdil}), which characterizes this phase. The behavior in the highly anisotropic case, 
 when $T/\rho \rightarrow0 $, is especially interesting. In this limit the near-horizon geometry is given by
an  $AdS_4 \times R$  attractor, which manifestly breaks rotational invariance, and many properties of the 
phase   can be understood from this near horizon geometry.  We find in fact that some novel features arise in the transport coefficients on account of the
 anisotropy. The viscosity  should now be thought of as a four index tensor, and 
has $5$ independent coefficients in this case. One component, $\eta_\perp$, with spin $1$ with 
respect to the unbroken $U(1)$
 rotation subgroup,
 satisfies the relation, eq.\eqref{etaperpint} 
and its ratio $\eta_{\perp}/s$ vanishes as $T/ \rho \rightarrow 0$.

We carried out a fairly detailed, but not fully exhaustive, analysis of the quasi normal mode spectrum near extremality. 
All the modes we study  are stable, with frequencies that have an imaginary part  in the lower 
complex plane. This suggests that the system is in fact stable. 

The system we analyze   
is quite analogous to the one studied in \cite{Mateos:2011ix}, \cite{Mateos:2011tv}, in 
which a linearly varying axion was considered in detail, and the subsequent paper \cite{Rebhan:2011vd} 
where it was was also  found that $\eta_\perp$ violates the KSS bound.

In the latter part of the paper we turn to an analysis of the resulting fluid mechanics  in some detail. 
We first show how to set up the equations of fluid mechanics, order by order in a derivative expansion, in a systematic manner. These provide us with the analogue of the  
Navier Stokes equations for an anisotropic driven 
system. 
In the absence of rotational symmetry the equations of fluid mechanics are much more complicated. 
We find that  many more terms can appear in the constitutive relation for the 
stress tensor. For example, there are  ten terms at first order  in the derivative expansion, while in 
the rotationally invariant case only two terms are allowed out of which one, proportional to the bulk viscosity, vanishes for the conformally invariant case. 
  Even more   parameters,  which originate from the forcing term due to the dilaton, 
enter in the  resulting equations of forced fluid mechanics.  

For the dilaton system at hand,  on carrying out the
 fluid mechanics analysis 
for linearized perturbations as discussed  in 
appendix \ref{hydmoap}, and also by coupling the fluid to  small metric perturbations, appendix \ref{kuboapp}, we can relate
some of  the coefficients appearing in the fluid mechanics to components of the viscosity calculated from
 gravity; we also find various additional  conditions that these coefficients must satisfy. 
In principle all coefficients etc which determine the fluid mechanics  can 
be obtained by carrying out a more systematic derivative expansion on the gravity side as discussed in \cite{Bhattacharyya:2008jc}, \cite{Bhattacharyya:2008xc},\cite{BLMNTW},\cite{Bhattacharya:2011tra}, \cite{Hubeny:2011hd}. 
We leave such a complete
 analysis  for the future. 

Having formulated the fluid mechanics in general, we then apply   it  to study 
a simple situation where the fluid in the dilaton system is enclosed  
between two parallel plates which are separated by some distance and are  moving with a 
relative velocity $v_0$
 which is small (non relativistic).
The equations simplify a lot in this situation and the dependence on  the many unknown coefficients 
drops out.   
Depending on how the plates are oriented we find that the frictional force exerted by the fluid on the plates is proportional to a different component of the viscosity.  In particular,  
  for an appropriate orientation  this force can be made
proportional to $\eta_\perp$ and therefore very small close to extremality. As a result the 
fluid can slip past the plates with very little friction.  

We have not found a string theory embedding of this system in which the black brane solution close to extremality is free from instabilities. 
In the case of IIB string theory on $AdS_5\times S^5$, we found a mode which lies below the BF bound in the $AdS_4$ near horizon region thereby 
signaling an  instability for the extremal and near-extremal cases. An instability is also present when the $S^5$ is replaced by an $S^5/Z_N$ orbifold
or by the base of the conifold, $T^{1,1}$.  This instability should disappear as the temperature increases, and $T$ becomes
bigger than $\rho$.  More generally, since we did not find any instabilities in the $AdS_5$ dilaton-gravity system, any instability would have to arise from the extra modes in the string embedding which could arise from KK reduction on an internal manifold.   It would seem surprising, at least at first glance, if in the vast string landscape 
the  masses of such modes cannot be made to lie about the BF bound of the $AdS_4$ near horizon spacetime.  
But of course  one will only 
 be sure  with a concrete example in hand \footnote{We should note that in \cite{Polchinski:2012nh} similar behavior was found for the viscosity in the susy NS5-F1 system 
which is stable.}. As noted above, if the instability turns out to be a general feature close to extremality, it should still go away
for $T\sim \rho$. It will not then be possible  to make the  viscosity to entropy ratio, 
for some components of the viscosity, arbitrarily small, but one can make it violate the 
KSS bound, eq.(\ref{kssbound}) by a fraction of order unity, analogous to \cite{Brigante:2007nu}, \cite{Brigante:2008gz}, \cite{Myers:2010jv}, \cite{Sen:2014nfa}.
 
A  natural expectation is that in strongly coupled systems the viscosity meets a bound       
\be
\label{boundvis}
\eta/s \sim \order (1).
\ee
As explained  in  \cite{Kovtun:2004de}, \cite{Son:2007vk} this expectation arises because for systems with weakly coupled quasiparticles
\be
\label{viscb}
{\eta\over s} \sim {l_{mfp}\over \lambda_{deBroglie}},
\ee
where $l_{mfp}, \lambda_{deBroglie}$ are the mean free path and the De Broglie wavelength of the quasi particles respectively.
When the system becomes strongly coupled one expect that $l_{mfp}\sim \lambda_{deBroglie}$ leading to eq.(\ref{boundvis}).
The violation we see in the anisotropic case, where the viscosity to entropy ratio can become parametrically small is
striking in view of this expectation and begs for a better explanation.
It is particularly important in light of this intuitive argument   to try and  find stable string embeddings for this system. 

It is worth pointing out, in this context, that the diffusion lengths for the different hydrodynamic modes, whose dissipation is governed by  different components of the viscosity, also become very different in the highly anisotropic limit. 
From subsection \ref{sp2momzsp1}  we see that the diffusion length $D_\perp$ of the mode described by eq.(\ref{gica1}), eq.(\ref{gica2}), eq.(\ref{sp1disz}), and also appendix \ref{apfltz},  eq.(\ref{disp110}),  and appendix \ref{qnmzapsp1} is given by 
\be
\label{defdperp}
D_{\perp} =  {\eta_\perp\over \epsilon + P}.
\ee
While the diffusion length of the mode discussed in section \ref{momqxy}, by 
 eq.(\ref{gicmodeqz}),  eq.(\ref{disy1}), eq.(\ref{etas}),  and also  appendix \ref{hydmoap1}, eq.(\ref{disp111}), and appendix \ref{qnmyapsp2},
\be
\label{defdpara}
D_{||} =  {\eta_{||} \over \epsilon + P}.
\ee
Here $\epsilon$ and $P$ are the energy density and pressure in the $x-y$ plane, eq.(\ref{relep}) and eq.(\ref{stxyz}). 
In the highly anisotropic limit we find that 
\begin{eqnarray}
D_{\perp}& = & {8 \pi T \over 3 \rho^2} \label{valdperpha}, \\
D_{||}& = & {1\over 4 \pi T} \label{valdpara}.
\end{eqnarray}
The modes with diffusion lengths $D_{\perp}, D_{||}$ carry momentum along the $z$ direction and the $x-y$ directions respectively. Thus, we see that  the externally applied dilaton gradient impedes the diffusion along the $z$ direction which is singled out by  the anisotropy,   as opposed to the $x-y$ directions where rotational invariance is preserved. It would be worth understanding  this behavior better in  the field theory description itself.  

Independent of  further theoretical investigations along these lines, it will be interesting to ask whether there are strongly coupled anisotropic systems in nature, in which some components of the viscosity, when compared to entropy density, violate the KSS bound, eq.\eqref{kssbound}, as we find here. 
In the system we study the gravity hologram makes this possibility geometrically very clear. 
The near horizon geometry close to extremality is
$AdS_4\times R$, with $z$ being the coordinate along $R$,  and  obviously breaks rotational invariance in the  spatial directions. The $5$ dimensional 
metric after KK reduction along the $z$ direction  gives rise to a $4$ dim. metric, a gauge field and a scalar, 
and the different components of the viscosity correspond to the viscosity, related to metric perturbations
in the lower dim. theory, the conductivity related to the gauge field, etc. These turn out to scale 
differently with temperature. This lesson is much more general  and one  expects it to be borne out 
  in other situations as well where the near horizon geometry breaks rotational invariance.
In fact, it is  known to be true in various  cases already   studied in the literature,
including,  \cite{Mateos:2011ix}, \cite{Mateos:2011tv}, \cite{Rebhan:2011vd}, where the near 
horizon geometry close to extremality is of Lifshitz type, in \cite{Polchinski:2012nh} where it is $AdS_3 \times R^4\times S^3$,
 and in \cite{Erdmenger:2010xm}, \cite{Basu:2011tt}, \cite{Erdmenger:2011tj}
which is dual to the $p$ wave superfluids \footnote{We note though  that the stability analysis has not been carried out in all these cases.}.  
This general behavior from the gravity side provides good motivation to  ask about whether a small component of viscosity can arise   in  nature as well.    

Another direction, would be to explore the behavior of viscosity, and more generally transport phenomenon,
in other phases which are homogeneous but anisotropic. Such phases have been recently discussed in \cite{Iizuka:2012iv}, \cite{Iizuka:2012pn}, \cite{Kachru:2013voa}. 
They  do not preserve the usual translation symmetries in general, but a sort of 
generalized version of them in which the symmetry generators do not commute and their algebra can be classified 
using the Bianchi classification. The behavior of  transport properties and the formulation 
of  fluid mechanics in such phases  are interesting open questions.  

Finally, as was mentioned in the introduction, the study here extends  the 
  work in \cite{BLMNTW} where the forced fluid with a  slowly varying dilaton  was studied. 
We see that the rapidly varying situation, 
$\rho/T \rightarrow \infty$,  is quite different and  corresponds to an $AdS_4 \times R$ near horizon geometry 
which is   quite distinct from that of the $AdS_5$ Schwarzschild geometry. It would be interesting to ask 
what happens if the system is similarly placed in a rapidly varying background metric.
 In some cases, for example when the spatial geometry is $S^3$, one knows that the $T\rightarrow 0$ limit is quite 
different and the system undergoes a Hawking Page transition. It will be fascinating to study this
 question in more generality as well.

\acknowledgments
It is a pleasure to thank   Sayantani Bhattacharyya, Sumit Das, Shouvik Datta, Nori Iizuka, Gautam Mandal, David Mateos, Krishna Rajagopal, Rajdeep Sensarma, Tadashi Takayanagi, Diego Trancanelli and Spenta Wadia for discussion. We especially thank Prithvi 
Narayan for discussions and collaboration for some of the work reported here, Jyotirmoy Bhattacharya for asking a key question about stability,
and Shiraz Minwalla for several very illuminating discussions. We also thank Sean Hartnoll for sharing his mathematica notebook which helped us in our quasinormal mode analysis. AS acknowledges support from a Ramanujan fellowship, Government of India. SPT acknowledges support from the DAE and the  J. C. Bose fellowship of the Government of India. Most of all we thank the people of India for generously supporting research in string theory. 
  
\appendix
\section{Details of computing thermodynamic quantities using FG expansion}
\label{thermofg}
In this appendix we provide the details for the calculation of various thermodynamic quantities like energy, pressure and entropy using the Fefferman-Graham (FG) expansion.
\subsection{Low anisotropy regime: $\rho/T\ll 1$}
\label{apthloan}
We expand the metric coefficients in eq.\eqref{lowansol} near the boundary upto quartic order in $\rho / T$.
 \begin{align}
 \label{hiansfg}
\begin{split}
A(u)|_{u\rightarrow\infty}&=u^{2}-\frac{\rho^{2}}{12}-\frac{1}{u^{2}}(u_{H}^{2}-\frac{u_{H}^{2}\rho^{2}}{12}-\frac{\rho^{4}}{216}-\frac{\pi \rho^{4}}{144})-\frac{\rho^{4}}{72u^{2}}\log[\frac{8u_{H}}{u}],\\ 
B(u)|_{u\rightarrow\infty}&=u^{2}-\frac{\rho^{2}}{12}+\frac{1}{u^{2}}(\frac{u_{H}^{2}\rho^{2}}{24}+\frac{7\rho^{4}}{864})-\frac{\rho^{4}}{72u^{2}}\log[\frac{2u_{H}}{u}],\\
C(u)|_{u\rightarrow\infty}&=u^{2}+\frac{\rho^{2}}{6}-\frac{1}{u^{2}}(\frac{u_{H}^{2}\rho^{2}}{12}+\frac{\rho^{4}}{432})+\frac{\rho^{4}}{36u^{2}}\log[\frac{2u_{H}}{u}],
\end{split}
\end{align}
 We move to the standard FG coordinates using the coordinate transformation 
\be u=\frac{1}{v}+\frac{\rho^{2}}{48}v+\frac{864u_{H}^{4}-72u_{H}^{2}\rho^{2}-(7+6\pi-36\log[2])\rho^{4}}{6912}v^{3}+\frac{\rho^{4}}{576}v^{3}\log[v u_{H}],\ee 
to obtain
\begin{align}
\label{hiansfg1}
\begin{split}
g_{tt}&=-1+\frac{\rho^{2}}{24}v^{2}+(\frac{3u_{H}^{4}}{4}-\frac{7\rho^{4}}{2304}-\frac{\rho^{2}u_{H}^{2}}{16}-\frac{\pi\rho^{4}}{192}+\frac{\rho^{4}\log[2]}{32})v^{4}+\frac{\rho^{4}}{96}v^{4}\log[v u_{H}],\\ 
g_{xx}&=g_{yy}=1-\frac{\rho^{2}}{24}v^{2}+(\frac{u_{H}^{4}}{4}+\frac{u_{H}^{2}\rho^{2}}{48}+\frac{5\rho^{4}}{768}-\frac{\pi\rho^{4}}{576}-\frac{\rho^{4}\log[2]}{288})v^{4}-\frac{\rho^{4}}{96}v^{4}\log[v u_{H}],\\
g_{zz}&=1+\frac{5\rho^{2}}{24}v^{2}+(\frac{u_{H}^{4}}{4}-\frac{5u_{H}^{2}\rho^{2}}{48}-\frac{\rho^{4}}{256}-\frac{\pi\rho^{4}}{576}+\frac{11\rho^{4}\log[2]}{288})v^{4}+\frac{\rho^{4}}{32}v^{4}\log[v u_{H}],
\end{split}
\end{align}
Using eq.\eqref{hiansfg1} in eq.\eqref{STnsor} we obtain
\begin{align}
\label{hiansth}
\begin{split}
\langle T_{tt} \rangle&=\epsilon=\frac{N_c^2}{1536\, \pi^2}(576u_{H}^{4}-48u_{H}^{2}\rho^{2}+(-3-4\pi+24\log[2])\rho^{4}),\\
\langle T_{xx} \rangle&=P_{x}=P_{y}=P=\frac{N_c^2}{4608\, \pi^2}(576u_{H}^{4}+48u_{H}^{2}\rho^{2}+(17-4\pi-8\log[2])\rho^{4}),\\
\langle T_{zz} \rangle&=P_{z}=\frac{N_c^2}{9216\, \pi^2}(576u_{H}^{4}-240u_{H}^{2}\rho^{2}+(-19-4\pi+88\log[2])\rho^{4}),
\end{split}
\end{align}
The near horizon form of the metric in eq.\eqref{lowansol} is
\begin{align}
\label{nhloan}
\begin{split}
A(u)=&(u-u_H) \left(\frac{\rho ^4 (2-\pi +\log (16))}{144 u_H^3}-\frac{\rho ^2}{6 u_H}+4 u_H\right)+{\cal O}(u-u_H)^2 \\
B(u)=&\frac{\rho ^4 \left(6 \left(1+7\log^2(2)+\log(4)\right)-\pi(6+\pi )\right)+1728 u_H^4-144 \rho ^2 u_H^2 \log(2)}{1728 u_H^2} \\ &~~+{\cal O}(u-u_H)^1\\
C(u)=&\frac{\rho ^4 \left((\pi -3) \pi -6 (\log (4)-1)^2\right)+864 u_H^4+144 \rho ^2 u_H^2 \log (2)}{864 u_H^2}+{\cal O}(u-u_H)^1
\end{split}
\end{align}
The temperature, defined in eq.\eqref{defT0}, is then given by 
\be T=\frac{1}{4\pi}\left(\frac{\rho ^4 (2-\pi +\log (16))}{144 u_H^3}-\frac{\rho ^2}{6 u_H}+4 u_H\right),\ee
Inverting this relation we can express 
\be \label{reluht} 
u_{H}=\pi T+\frac{\rho^{2}}{24\pi T}+\frac{-3+\pi-4\log[2]}{576\pi^{3}T^{3}}\rho^{4}+O(\rho^{6}),\ee
Using eq.\eqref{reluht} in eq.\eqref{hiansth} we obtain the thermodynamic quantities as given in eq.\eqref{Eloan}, eq.\eqref{Pxloan} and eq.\eqref{Pzloan}.

\section{Details of derivative expansion in anisotropic fluid dynamics}
\label{FMdetail}
In section \ref{fluidmech} we studied the fluid mechanics for an anisotropic phase, characterized by the vector $\xi_{\mu}$, eq.\eqref{defxi}, upto second order in derivative expansion. As it was mentioned, for a consistent analysis, following eq.\eqref{consstress}, we need to consider $T_{\mu\nu}$ upto first order in derivative expansion, eq.\eqref{stressteng}, and $\langle O \rangle$ upto second order in derivative expansion, eq.\eqref{expdila}. In this appendix we will carry out a more detailed analysis of this derivative expansion for both $T_{\mu\nu}$ and $\langle O \rangle$.

The main strategy for carrying out the derivative expansion is to consider the basic fluid dynamic variables at our disposal, \emph{e.g.}  $T,\, \hxi_{\mu}$ and $u_{\mu}$ along with their derivatives to construct all possible scalar, vector and tensorial quantities at higher orders

It will be useful for our discussion below to define the quantity 
\be
 \label{defproj}
 \tilde P^{\mu\nu}=\eta^{\mu\nu}+u^{\mu}u^{\nu}-\frac{1}{\hat{\xi}^2}\hat{\xi}^{\mu}\hxi^{\nu}, 
 \ee
which is the projector onto a plane perpendicular to both $u_{\mu}$  and $\hxi_{\mu}$, \emph{i.e.},
\be
\tilde P^{\mu\nu}\,u_{\mu}=\tilde P^{\mu\nu}\,\hxi_{\mu}=0.
\ee

As it has been already mentioned, at zeroth order of derivative expansion, the possible scalars are\footnote{Note that we are considering $\hxi^2$ as the scalar quantity in place of $u\cdot\xi$, but from eq.\eqref{defxihat} it is easy to see that they are related
\begin{equation*}
\hxi^2 =\rho^2 + (u\cdot\xi)^2.
\end{equation*}
} $T,\, \rho$  and $\hxi^2$. Similarly the possible vectors are $u_{\mu}$ and $\hxi_{\mu}$ and the possible tensors are $\eta_{\mu\nu},\, u_{\mu} u_{\nu}$ and $\hxi_{\mu} \hxi_{\nu}$.

At first order in derivatives we see that there are five possible scalars made out of combining zeroth order quantities and their derivatives. We   denote them as $S_i$, for $i=1,\cdots,5$,
 \be
 \label{defscal}
 \begin{split}
 S^{(1)}_{1} =& \partial.u, \qquad S^{(1)}_{2} = (u.\partial)T, \qquad S^{(1)}_{3} = (\hat{\xi}.\partial)T, \\ &  S^{(1)}_{4} = (u.\partial)(\hat{\xi}.\hat{\xi}), \qquad S^{(1)}_{5} = (\hat{\xi}.\partial)(\hat{\xi}.\hat{\xi}).
 \end{split}
 \ee

The vectors, denoted as ${V_i}_{\alpha}$ for $i=1,2,3,4$, are
\begin{equation}
\label{defvecs}
\begin{split}
&{V_1}^{(1)}_{\alpha}= \tilde P_{\alpha\nu}(\hxi.\partial)u^{\nu}, \qquad {V_2}^{(1)}_{\alpha}=\tilde P_{\alpha\nu}\partial^{\nu}T, \\ &{V_3}^{(1)}_{\alpha}=\tilde P_{\alpha\nu}\partial^{\nu} (\hxi . \hxi), \qquad {V_4}^{(1)}_{\alpha}= \tilde P_{\alpha\nu}(u.\partial)u^{\nu}.
\end{split}
\end{equation}
Similarly the tensors at first order in derivatives are,
\begin{equation}
\label{deftens}
 t^{(1)}_{\mu\nu}=\tilde P^{\alpha}_{\mu}\tilde P^{\beta}_{\nu}( \partial_\alpha u_{\beta}+\partial_\beta u_{\alpha}-\eta_{\alpha\beta}\tilde P^{\lambda\gamma}\partial_{\lambda}u_{\gamma})
\end{equation}
Therefore upto first order in derivatives $T^{(1)}_{\mu\nu}$ and $\langle O \rangle^{(1)}$  can be written as,
 \be
 \label{tmnrel2}
 \begin{split}
 T^{(1)}_{\mu\nu} = \sum_{i=1}^5 \zeta^a_i \, \tilde P_{\mu\nu} \, S^{(1)}_i +& \sum_{i=1}^5 \zeta^b_i \, \hxi_{\mu} \hxi_{\nu}\, S^{(1)}_i + \sum_{i=1}^4 {v}_i \, (\hxi_{\mu}\, {V_i}^{(1)}_{\nu}+\hxi_{\nu}\, {V_i}^{(1)}_{\mu})+ \eta \,t^{(1)}_{\mu\nu}, \\
 \langle O \rangle^{(1)} =& \sum_{i=1}^5 \zeta^c_i \,  S^{(1)}_i
 \end{split}
 \ee
 where $\zeta^a_i,\,\zeta^b_i,\, {v}_i, \, \eta $ are various transport coefficients and $\zeta^c_i$'s are arbitrary forcing coefficients which in principle can be determined from gravity.

{\bf Independent Data upto first order in derivatives: }
 
 In eq.\eqref{defscal} we have written all the possible scalars that can be obtained from zeroth order quantities and their derivatives. We can further make use of the equation of motion in eq.\eqref{consstress} to relate some of them in terms of others and obtain the independent datas. Upto first order in derivative expansion, using $T_{\mu \nu}$ upto zeroth order from eq.\eqref{zerostress} and $\langle O \rangle^{(1)}$  upto first order in derivatives from eq.\eqref{tmnrel2}, we get the the scalar relation by projecting eq.\eqref{consstress}\footnote{Notice that in eq.\eqref{consstress} on the RHS $\partial_{\nu} \phi=\xi_{\nu}$ and not $\hxi_{\nu}$.}
 along $\hxi_{\mu}$
 \be
\label{foca}
 \begin{split}
 (\epsilon+P)\hxi_{\nu}(u.\partial)u^{\nu} +(\hxi.\partial)P+\hxi^2(\hxi.\partial)f  +f\hxi^2(\partial.\hxi)+\frac{1}{2}f(\hxi.\partial)\hxi^2 = \hxi.\xi \sum_{i=1}^5 \zeta^c_i \,S^{(1)}_i
 \end{split}
 \ee
 where $(\epsilon+P)\hxi_{\nu}(u.\partial)u^{\nu}$ and $f(\hxi.\partial)\hxi^2$ can be expressed in terms of the scalars mentioned in eq.\eqref{defscal}.
 
 Similarly projecting eq.\eqref{zerostress} along $u_{\mu}$ we obtain the second scalar relation,
 \be
\label{focb}
 \begin{split}
  -(\epsilon+P)(\partial.u)-(u.\partial)(\epsilon+P)+(u.\partial)P+f u_{\nu}(\hxi.\partial)\hxi^{\nu} = u.\xi \sum_{i=1}^5 \zeta^c_i \,S^{(1)}_i
 \end{split}
 \ee
Thus we have obtained two scalar relation between all the five scalars given in eq.\eqref{defscal}. Hence there will be three independent scalars at the first order in derivatives. We can make a choice to work with the independent scalars as
 \be
 \label{defscin}
 \begin{split}
S^{(1)}_{1} = \partial.u, \qquad S^{(1)}_{2} = (u.\partial)T, \qquad S^{(1)}_{3} = (\hat{\xi}.\partial)T
 \end{split}
 \ee
 
 Similarly projecting eq.\eqref{consstress} with $\tilde P_{\mu \nu}$ along the directions perpendicular to both $\xi_{\mu}, \, u_{\mu}$, we obtain a vector relation between ${V_2}^{\mu}$ and ${V_4}^{\mu}$ 
 \be
 (\epsilon+P) \tilde{P}^{\alpha}_{\nu}(u.\partial)u^{\nu}+\tilde{P}^{\alpha}_{\nu}\partial^{\nu}P  +f \tilde{P}^{\alpha}_{\nu}(\hxi.\partial)\hxi^{\nu} =0
 \ee 
 Therefore we make a choice to work with the three independent vectors as
 \begin{equation}
\label{defvin}
\begin{split}
{V_1}^{(1)}_{\alpha}= \tilde P_{\alpha\nu}(\hxi.\partial)u^{\nu}, \qquad {V_2}^{(1)}_{\alpha}=\tilde P_{\alpha\nu}\partial^{\nu}T,  \qquad {V_3}^{(1)}_{\alpha}=\tilde P_{\alpha\nu}\partial^{\nu} (\hxi . \hxi).
\end{split}
\end{equation}
Upto first order in derivatives then we write $T^{(1)}_{\mu\nu}$  in terms of independent scalars, vectors and tensors as,
\be
 \label{tmnrel4}
 \begin{split}
 T^{(1)}_{\mu\nu} = \sum_{i=1}^3 \zeta^a_i \, \tilde P_{\mu\nu} \, S^{(1)}_i + \sum_{i=1}^3 \zeta^b_i \, \hxi_{\mu} \hxi_{\nu}\, S^{(1)}_i + \sum_{i=1}^3 {v}_i \, (\hxi_{\mu}\, {V_i}^{(1)}_{\nu}+\hxi_{\nu}\, {V_i}^{(1)}_{\mu})+ \eta \,t^{(1)}_{\mu\nu},
 \end{split}
\ee
and also for $\langle O \rangle^{(1)}$ as
\be 
\label{forO}
 \langle O \rangle^{(1)} = \sum_{i=1}^3 \zeta^c_i \,  S^{(1)}_i.
\ee
Note that there are 10 independent terms in eq.\eqref{tmnrel4}.

Actually there are important relations which reduce the number of independent coefficients among the ten terms that appear in eq.(\ref{tmnrel4}) and also  among the coefficients that appear in eq.(\ref{forO}). We discuss these in some detail now.  
As far as the coefficients in eq.(\ref{tmnrel4}) are concerned,  
the first order stress tensor should meet the condition
\be
\label{fotracea}
{T^{(1)\mu}}_\mu=0.
\ee
This follows from the fact that the full stress tensor must meet the  trace anomaly condition  given in eq.\eqref{traceano},
which is already satisfied by the  zeroth order stress tensor. 
In addition, if one is close to extremality,  $T/\rho\ll 1 $, we can think of the system as a state in $AdS_4$ and 
${T^{(1)}}_\mu^\mu$ must also meet the condition 
\be
\label{fotraceb}
{T^{(1)}}^t_t+ {T^{(1)}}^x_x+ {T^{(1)}}^y_y=0.
\ee
This is analogous to the conditions that $\delta \epsilon, \delta P, \delta P_z$ were found to satisfy in 
section \ref{mohianth}. 
These two conditions will result in two relations between the $10$ parameters in eq.(\ref{tmnrel4}), so that the
 number of 
independent transport coefficients close to extremality will be  $8$.

In addition, it further turns out that,  equilibrium consideration also impose conditions reducing the number of independent coefficients in eq.(\ref{tmnrel4}).  At equilibrium, following \cite{Bhattacharyya:2012xi}, one can write down three terms in partition function at first order
\be\label{indeq}
\xi^{i}\partial_{i} {\hat T},~~\nabla_{i}\xi^{i},~~ \xi_{i}\partial^{i}\left( {\xi\cdot\xi}\right),
\ee
where ${\hat T}$ is given in eq.\eqref{zopf}, $\xi_{\mu}=\{0,0,0,\rho\}$ as defined in eq.\eqref{defxi} and index $i$ denote spatial directions $x,y,z$. Note that $\xi_{\mu}$ is a constant vector. This implies that the third term in eq.\eqref{indeq} is zero. The second term $\nabla\cdot\xi$ in eq.\eqref{indeq} can be related to first term $\xi\cdot\partial {\hat T}$ by integration by parts. So, there is only one 
non-dissipative term. While we  do not go into details here, these consideration
further  reduces the number of independent transport coefficients. 

As far as the coefficients in eq.(\ref{forO}) are concerned, it turns out that   $\zeta_1^c, \zeta_2^c $,
 must vanish.
This can be argued as follows. From the gravity theory it can be easily seen 
 that there must be solutions  for which 
which $u^z=0$ and the remaining components of the velocity,  $u^0, u^x, u^y$ and $T$ are
 independent  of $z$. It is easy to see that  such a solution must satisfy the equation
\be
\label{solspo}
\partial^\mu T^{(0)}_{\mu z} =0.
\ee
One the other hand, the equations of motion, eq.(\ref{consstress}),  to this order gives 
\be
\label{cons1}
\partial^\mu T^{(0)}_{\mu z} =\langle O \rangle^{(1)} \rho.
\ee
On comparing with eq.(\ref{solspo}) we see that $\langle O \rangle^{(1)}$ must vanish for any such solution. 
Since $S_1^{(1)}, S^{(1)}_2$ do not vanish in general for such solutions it follows 
that
\be
\label{condzetac}
\zeta_1^c=\zeta_2^c=0.
\ee
 
Having discussed the constraints on the coefficients in eq.(\ref{tmnrel4}) and eq.(\ref{forO}) we are ready to proceed with our  discussion of the equations of  fluid mechanics to second order in the derivative expansion. This requires information about $\langle O \rangle^{(2)}$, for which in turn we   need to know the possible scalars upto second order in derivatives. They can be obtained using the independent first order quantities and their derivatives as follows 
 \be
 \label{defscal1}
 \begin{split}
 S^{(2)}_{i} \Rightarrow &\qquad t_{\mu\nu}t^{\mu\nu}, \\&\qquad {V_i}_{\mu}^{(1)}{V_j}^{(1)\mu},  \, \text{for} \,\, i, \, j =1, \cdots , 3, \\&\qquad S^{(1)}_{i}S^{(1)}_{j}, \,  \text{for} \,\, i, \, j=1, \cdots , 3, \\ &\qquad \partial_{\mu} {V_i}^{(1)\mu}, \,  \text{for} \,\, i=1, \cdots , 3, \\&\qquad (\hxi . \partial)S^{(1)}_{i}, \, \text{for} \,\, i=1, \cdots , 3, \\ &\qquad (u . \partial)S^{(1)}_{i}, \, \text{for} \,\, i=1, \cdots , 3,
 \end{split}
 \ee
Upto second order in derivatives we then obtain
\be
\label{tmrel3}
\begin{split}
 \langle O \rangle^{(2)} =& n^{(2)}_1\, t_{\mu\nu}t^{\mu\nu}+\sum_{i,\,j=1}^3  {n_2}^{(2)}_{ij}\, {V_i}_{\mu}^{(1)}{V_j}^{(1)\mu} + \sum_{i,\,j=1}^3  {n_3}^{(2)}_{ij}\, S^{(1)}_iS^{(1)}_j \\& + \sum_{i=1}^3  {n_4}^{(2)}_{i}\, \partial_{\mu} {V_i}^{(1)\mu}   + \sum_{i=1}^3  {n_5}^{(2)}_{i}\, (\hxi . \partial)S^{(1)}_{i} + \sum_{i=1}^3  {n_6}^{(2)}_{i}\, (u . \partial)S^{(1)}_{i}
\end{split}
\ee
with $n^{(2)}_i$'s being arbitrary coefficients and they are 22 in number.

Note that following our discussion above for the quantities at first order in derivatives we can also use the equation of motion to show that the scalars at second order in derivative listed in eq.\eqref{tmrel3} are not all independent. As it turns out, at second order in derivatives projecting eq.\eqref{consstress} along both $u_{\mu}$ and $\xi_{\mu}$, 2 of them can be expressed in terms of the other 20. So, there are 20 independent scalars at second order in derivatives.

Once we have obtained $T^{(1)}_{\mu\nu}, \, \langle O \rangle^{(1)}$ and  $\langle O \rangle^{(2)}$, eq.\eqref{consstress} is also known to us explicitly upto second order in derivatives.
\subsection{Consistent fluid configuration for the flow between two plates}\label{thtexap}
 In subsection \ref{thoughtex} we considered a specific example of a fluid flowing between two plates separated along the $z-$direction. We considered non-relativistic fluid flow along the $x-$direction with velocity $u_x(z)$ and having no components along the other two spatial directions, \emph{i.e} $y,\, z$. In this subsection we argue in some more detail that the specific fluid configuration we considered is a consistent solution of the equation of hydrodynamics eq.\eqref{consstress}, working to first order in the fluid velocity $u_x(z)$.
 Let us write the fluid velocity in the form
 \be
 u_{\mu}=\{-1,0,0,0\}+\lambda \, \{0,u_x(z),0,0 \}
 \ee
 where $\lambda \ll 1$ is a small parameter signifying that the fluid velocity is small and consistent with non-relativistic fluid flow.
 As already mentioned we will work upto the linear order in $\lambda$, \emph{i.e.} $\mathcal{O}(\lambda)$. 
 
 The anisotropy vector $\xi_{\mu}$ is given in eq.\eqref{defxi}. We also consider the situation where the temperature $T$ is a constant.
 
 For this fluid configuration the stress energy tensor in zeroth order of derivatives, in eq.\eqref{zerostress}, becomes upto $\mathcal{O}(\lambda)$,
 \be
 \label{st1ex}
 T^{(0)}_{\mu\nu} =\left(
\begin{array}{cccc}
 \epsilon  & -\lambda\,(P+\epsilon )   u_x(z) & 0 & 0 \\
 -\lambda\,(P+\epsilon )  u_x(z) & P& 0 & 0 \\
 0 & 0 & P& 0 \\
 0 & 0 & 0 & f \rho ^2+P\\
\end{array}
\right)
 \ee
 
At first order in derivatives all the independent scalars $S_i$, for $i=1,2,3,4,5$, defined in eq.\eqref{defscin}, vanishes upto $\mathcal{O}(\lambda)$ for this particular fluid configuration. Similarly, the tensor $t_{\mu\nu}$, defined in eq.\eqref{deftens} also vanishes upto $\mathcal{O}(\lambda)$. Among the vectors ${V_i}_{\alpha}$ for $i=1,2,3,$ defined in eq.\eqref{defvin}, only ${V_1}_{\alpha}$ contributes,
\be
\label{vec4f}
{V_1}_{\alpha} = \left\{0,\lambda \, \rho \, u_x'(z),0,0\right\}
\ee

Therefore the stress energy tensor in first order in derivatives becomes upto $\mathcal{O}(\lambda)$,
\be
\label{st2ex}
T^{(1)}_{\mu\nu} = \left(
\begin{array}{cccc}
 0 & 0 & 0 & 0 \\
 0 & 0 & 0 &   \lambda  \, v_1 \,  \rho ^2 \, u_x'(z) \\
 0 & 0 & 0 & 0 \\
 0 &  \lambda  \, v_1 \,  \rho ^2 \, u_x'(z) & 0 & 0 \\
\end{array}
\right)
\ee
and also 
\be
\label{relo1}
\langle O \rangle ^{(1)}=0
\ee

It turns out that the scalars at second order in derivative, given in eq.\eqref{defscal1}, vanish upto $\mathcal{O}(\lambda)$. Therefore,
\be
\label{relo2}
\langle O \rangle ^{(2)}=0
\ee

Therefore eq.\eqref{consstress} upon using eq.\eqref{st1ex}, eq.\eqref{st2ex}, eq.\eqref{relo1} and eq.\eqref{relo2} reduces to eq.\eqref{nsvx}.

\section{Kubo analysis}
\label{kuboapp}
In the previous subsection \ref{thtexap}, we have seen how the system behaves under the perturbation of velocity configuration of the fluid. The aim of this section is to study response of the system under background metric fluctuations. At the end of this analysis, we will 
be able to identify various two point functions of stress tensors which in the limit of small frequency gives various transport coefficients.
This is known by the name of the Kubo formula.

Let us consider the perturbed metric to be given by
\be
g_{\mu\nu}=\eta_{\mu\nu}+h_{\mu\nu}(t)
\ee
where we will work upto linear order in metric fluctuation $h_{\mu\nu}$. 
 Under this perturbation, the zeroth order stress tensor eq.\eqref{zerostress} is given by
 {\footnotesize 
 \be 
  T_{\mu\nu}^{(0)} =
 \left(
 \begin{array}{cccc}
 2 \epsilon + P\, h_{tt}(t) & P\, h_{xt}(t) & P\, h_{yt}(t) & P\, h_{zt}(t) \\
  P\, h_{xt}(t) &   2P\,(h_{xx}(t)+1) & P\, h_{xy}(t) & P\, h_{xz}(t) \\
  P\, h_{yt}(t) & P\, h_{xy}(t) & 2P\,(h_{yy}(t)+1) & P\, h_{yz}(t) \\
  P\, h_{zt}(t) & P\, h_{xz}(t) & P\, h_{yz}(t) & 2 \left(f \rho ^2+P\right)+P\, h_{zz}(t) \\
 \end{array}
 \right)
  \ee
}
The first order stress tensor eq.\eqref{tmnrel4} takes a more complicated form. In fact additional terms, 
with new coefficients, can now appear in its constitutive relation which vanish in the flat space limit. An example is the  term
\be
\label{exterm}
\delta T^{(1)}_{\mu\nu}= \zeta_6^a {\tilde P}_{\mu\nu} (\nabla.{\hat \xi}).
\ee 

In the fourier space with $h_{\mu\nu}(t)=\int\frac{dt}{2\pi} h_{\mu\nu}(\omega)e^{-i\omega t}$ the 
first order stress tensor is given by
\be
\label{resstrs}
\begin{split}
T^{(1)}_{xy}(\omega)&=-\frac{1}{2}\,i\,\omega \, \eta \, h_{xy}(\omega)\\
T^{(1)}_{xz}(\omega)&=-\frac{1}{2}\,i\,\omega\, v_1\, \rho^2\, h_{xz}(\omega)\\
T^{(1)}_{yz}(\omega)&=-\frac{1}{2}\,i\,\omega\, v_1\, \rho^2\, h_{yz}(\omega)\\
T^{(1)}_{tz}(\omega)&=2i\,\omega\, v_3\, \rho^3\, h_{zz}(\omega)\\
T^{(1)}_{zz}(\omega)&=-\frac{1}{2}\,i\, \omega \, \zeta^b_1\, \rho^2\left(h_{tt}(\omega)+h_{xx}(\omega)+h_{yy}(\omega)+h_{zz}(\omega)\right)+\cdots\\
trT^{(1)}&=\frac{T^{(1)}_{xx}(\omega)+T^{(1)}_{yy}(\omega)}{2} \\
&=-\frac{1}{2}\, i\,\omega\, \zeta_{1}^{a} \left(h_{tt}(\omega)+h_{xx}(\omega)+h_{yy}(\omega)+h_{zz}(\omega)\right)+\cdots
\end{split}
\ee
where the ellipses  denote contributions coming from the additional terms 
in the constitutive relation mentioned above which arise in curved space, e.g., eq.(\ref{exterm}).
We will not study the added complications due to such terms  further in this paper and leave them for future.

We can now use eq.\eqref{resstrs} to arrive at the Kubo's formula for various transport coefficients using eq.\eqref{stdef}. 
This gives

\be
\label{viskubo}
\begin{split}
\lim_{\omega\rightarrow 0}\langle T_{xz}(\omega) T_{xz}(\omega) \rangle &= i\,  \omega\,v_1 \rho^2  \\
\lim_{\omega\rightarrow 0}\langle T_{xy}(\omega) T_{xy}(\omega) \rangle &= i\, \omega \, \eta \\
\lim_{\omega\rightarrow 0}\langle T_{tz}(\omega) T_{zz}(\omega) \rangle &= -4i\,  \omega\, v_3\, \rho^3 \\
\lim_{\omega\rightarrow 0}\langle trT(\omega) trT(\omega) \rangle &= i\,  \omega \,\zeta_{1}^{a}+\cdots \\
\lim_{\omega\rightarrow 0}\langle T_{zz}(\omega) T_{zz}(\omega) \rangle &= i\,  \omega\, \zeta_{1}^{b}+\cdots.
\end{split}
\ee 
It might seem strange at first 
 that the coefficient $\zeta_2^a$, eq.(\ref{tmnrel4}) does not appear on the RHS above. 
However,   as discussed in appendix \ref{FMdetail}, after eq.\eqref{forO}, this coefficient is  in fact not independent of the others. 

Following our definition for $\eta_{\perp}$, $\eta_{\parallel}$ in eq.\eqref{defetperp1}, eq.\eqref{defetprl1} respectively, and comparing with first two relations in eq.\eqref{viskubo} we obtain
\be
\label{appetarel}
\eta_{\perp} = v_1 \rho^2  \qquad \text{and}  \qquad  \eta_{\parallel} =\eta.
\ee
\section{More on hydrodynamic modes}
\label{hydmoap}
In this appendix we discuss the hydrodynamic modes in detail through a field theory analysis. For that we need to study the normal modes of the linearized hydrodynamic equations. These solutions behave as $e^{i(- \omega t+\vec{k}.\vec{x})}$ and we will focus on the situation when $\omega \rightarrow 0$ as $k \rightarrow 0$.

We consider fluid mechanics in flat space with a metric 
\begin{equation}
 \eta_{\mu \nu} = \{ -1,1,1,1\}.
\end{equation}
The equation of hydrodynamics is given in eq.\eqref{consstress}. We consider the stress energy tensor upto first order in derivative expansion,$T^{\mu\nu}=T^{\mu\nu}_{0}+T^{\mu\nu}_{1}$, where 
$T^{\mu\nu}_{0}$ is given in eq.\eqref{lto} and $T^{\mu\nu}_{1}$ in \eqref{tmnrel4}. 
In what follows we consider various possible solutions for velocity field $u^{\mu}$ and temperature $T$, which are the dynamical variables of the fluid mechanics.
We can divide the modes into two sector. Fluid fluctuations which depends on time coordinate $t$ and one of the spatial direction say $y.$
Another case that we might consider is that fluid fluctuations which depends on time coordinate $t$ and $z$ direction. 
\subsection{Fluid fluctuations as a function of $t,y$}\label{hydmoap1}
\begin{itemize}
 \item First we consider fluid configuration of the form
 \be\begin{split}\label{hydmoap2}
 u_{\mu}&=u_{\mu}^0+\{0,\delta u_{x}(t,y),0,0\}\\
T&=T_0.
 \end{split}\ee
 with temperature and the background fluid velocity given by
 \be
 \label{fltu}
 \begin{split}
 T&=T_0\\
 u_{\mu}^0&=\{-1,0,0,0\}
\end{split} \ee where $T_0$ is constant. We expand $\delta u(t,y)$  in fourier modes as
\be
\delta u(t,y)=\int dk_y\, d\omega\, \delta u(\omega,k_y)\,e^{-i\,\omega\, t+i\, k_y\,y}
\ee
Demanding that there exists a nontrivial solution of eq.\eqref{consstress}, one obtains
\be\begin{split}\label{disp111}
 \omega&=-i k_y^2 \frac{\eta_{\parallel}}{P_0+\epsilon_0}\\   
   \end{split}
\ee 
where $\epsilon_0$ and $P_0$ are values of energy density and pressure respectively at temperature $T=T_0.$
Note that, in deriving eq.\eqref{disp111}, we have used the fact that $\delta u$ is very very small compared to background velocity and temperature ($u_{\mu}^0$ and $T=T_0$). 

\item We also consider fluid configuration of the form
 \be\begin{split}\label{hydmoap3}
 u_{\mu}&=u_{\mu}^0+\{0,0,\delta u_{y}(t,y),0\}.\\
T&=T_{0}+\delta T(t,y)
\end{split} \ee 
For this case, we obtain two modes,
\begin{enumerate}
 \item The first one as,
 \be\begin{split}\label{disp112}  
 \omega&=\pm k_y c_s+ i\frac{\zeta^a_{2}}{2( \partial_{T}\epsilon)_0} k_y^2-i k_y^2 \frac{\zeta^a_{1}+\eta_{\parallel}}{2(P_0+\epsilon_0)}  
   \end{split}
\ee
\item The other hydrodynamic mode is given by
 \be\label{disp1121}
 \begin{split}
i \omega  \left(\zeta^c_2 \left(P_0+\epsilon _0\right)-\zeta^c_1( \partial_{T}\epsilon)_0\right)&+k_y^2 \left( n_{4_2}+v_2\right) \left(P_0+\epsilon _0\right)\\
&+\omega ^2 \left(n_{6_2} \left(P_0+\epsilon _0\right)-n_{6_1}( \partial_{T}\epsilon)_0\right)=0,
   \end{split}
\ee where $\epsilon_0,\,P_0,\,(\partial_{T}\epsilon)_0$ are evaluated at temperature $T=T_0.$
From the gravity analysis of section \ref{momqxy}, we have seen that we have got only two hydrodynamic modes with momenta along $y$ direction. 
 As will be argued below eq.\eqref{etaso}, we see that while eq.\eqref{disp112} has a interpretation from gravity analysis, there seem to exist no mode
of the form presented in eq.\eqref{disp1121}. 
From eq.(\ref{condzetac})  we see that $\zeta_1^c, \zeta_2^c$ vanish. The remaining conditions for this mode to not be present are then
\be\begin{split}\label{rlads4}
&v_2+ n_{4_2}=0\\
&n_{6_2}=\frac{n_{6_1} ( \partial_{T}\epsilon)_0}{ \left(P_0+\epsilon _0\right)}.
   \end{split}
\ee
\end{enumerate}
\end{itemize}
\subsection{Fluid fluctuations as a function of $t,z$}
\label{apfltz}
\begin{itemize}
\item Let us consider fluid configuration of the form
 \be\begin{split}\label{apfltz1}
 u_{\mu}&=u_{\mu}^0+\{0,\delta u_{x}(t,z),0,0\}\\
T&=T_0.
\end{split} \ee where $T_0$ and $u_{\mu}^0$  given in eq.\eqref{fltu}. Expanding $\delta u(t,z)$ similarly in fourier modes we get
\be
\delta u_x(t,z)=\int dk_z d\omega \delta u_x(\omega,k_z)e^{-i\omega~ t+i k_z~z}
\ee
Further we demand that there exists a nontrivial solution of eq.\eqref{consstress} to obtain
\be\begin{split}\label{disp110}
 \omega&=-i k_z^2 \frac{v_1 \rho^2}{P_0+\epsilon_0}=-i k_z^2 \frac{\eta_{\perp}}{P_0+\epsilon_0},
    \end{split}
   \ee
where we have used eq.\eqref{appetarel}. 
\item Let us consider fluid configuration of the form
 \be\begin{split}\label{apfltz2}
 u_{\mu}&=u_{\mu}^0+~\{0,0,0,\delta u_{z}(t,z)\}\\
T&=T_0+\delta T(t,z)
\end{split}
\ee
The hydrodynamic mode in this case is given by 
\be\begin{split}\label{hydrozz1}
&\rho ^2 k_z^2 \left(f \rho ^2+P_0+\epsilon _0\right) \left(\rho  k_z \left(\zeta^b_3+n_{5_3}\right)-i \zeta^c_3\right)\\
&-i \omega ^2( \partial_{T}\epsilon)_0 \left(f \rho ^2+P_0+\epsilon _0\right)-i\omega k_z \rho  \left(\zeta^c_1( \partial_{T}\epsilon)_0-\zeta^c_2 \left(f \rho ^2+P_0+\epsilon _0\right)\right)\\
&-\omega k_z^2\rho ^2 \left(\left(\zeta^b_2+n_{5_2}+n_{6_3}\right) \left(f \rho ^2+P_0+\epsilon _0\right)-( \partial_{T}\epsilon)_0 \left(\zeta^b_1+n_{5_1}\right)\right)\\
&+\omega^2 k_z \rho  \left(n_{6_2} \left(f \rho ^2+P_0+\epsilon _0\right)-n_{6_1}( \partial_{T}\epsilon)_0\right)\\
&+\rho ^2 \omega  k_z^2 \left(\left(\zeta^b_2+n_{5_2}+n_{6_3}\right) \left(f \rho ^2+P_0+\epsilon _0\right)-( \partial_{T}\epsilon)_0 \left(\zeta^b_1+n_{5_1}\right)\right)=0.
\end{split}\ee
Turning on momenta along $z$ direction gives in the gravity side two hydrodynamic modes as discussed in section \ref{momqz}. We have identified
eq.\eqref{disp110} with eq.\eqref{sp1disz}. So, eq.\eqref{hydrozz1} should be identified with eq.\eqref{sp0disz}. 
In order to do that we need to assume that 
\be
\zeta^c_1=\zeta^c_2=\zeta^c_3=0
\ee which is consistent with eq.\eqref{condzetac}.
Using this we obtain
\be\label{rela123}
\begin{split}
\omega=-i \rho^2 k_z^2 \bigg({n_{5_1} +\zeta^b_1 \over P_0+\epsilon_0+f \rho ^2} - {n_{5_2}+n_{6_3}+\zeta^b_2 \over( \partial_{T}\epsilon)_0 } \bigg)
\end{split}
\ee
In order to match with eq.\eqref{sp0disz}, we need to compute various transport coefficients appearing in eq.\eqref{rela123} in $AdS_4$ background.
We leave this as a future exercise.
\end{itemize}

\section{More on quasi normal modes}
\label{appqnm}
In this appendix we discuss in more detail the analysis for the spectrum of the quasi normal modes. We will provide the intermediate steps for the analysis in section \ref{qnmspec}.

\subsection{Modes with $q_x,\, q_y=0$}\label{qnmzap}
First we consider the situation when the modes have momentum turned on along the $z$-direction.
 \subsubsection{Spin 2}\label{qnmzapsp2}

It was mentioned in subsection \ref{sp2momz} that each of the modes $h_{xy}$ and $(h_{xx}-h_{yy})/2$ satisfy decoupled equation of the form
 eq.\eqref{diffeqz}. We considered the mode $h_{xy}$ and for that the coefficient functions are given as
\begin{align}
\label{abcsp2qz}
\begin{split}
a(v)&=-\frac{16(1-v^3)^2 \rho^2}{v^3},\\
b(v)&=\frac{16(2-v^3-v^6)\rho^2}{v^4},\\
c(v)&=\frac{96q^2 (1-v^3)-9 v^2 \rho^2 \omega^2}{v^5},\\
\end{split}
\end{align}
We further decompose this mode as in eq.\eqref{expZser} and obtain
\be
\gamma=\frac{3+\sqrt{24 \left({q \over \rho}\right)^2+9}}{2}
\ee
\subsubsection{Spin 1}\label{qnmzapsp1}

The mode $Z_1^{(1)}$ defined in eq.\eqref{gica1}, also satisfies an equation of the form eq.\eqref{diffeqz} with the coefficient functions given as 
\begin{align}
\label{abcsp1qz}
\begin{split}
a(v)=&16 \rho ^2 v^2 \left(v^2-1\right)^2 \left(32 q^2 \left(v^2-1\right)+3 \rho ^2 v^2 \omega ^2\right)\\
b(v)=&-32 \rho ^2 v \left(v^2-1\right) \left(32 q^2 \left(v^2-1\right)^2+3 \rho ^2 v^2 \left(v^2-2\right) \omega ^2\right)\\
c(v)=&96 \rho ^2 v^2 \left(v^2-1\right) \omega ^2 \left(6 q^2+\rho ^2 v^2\right)+1024 q^2 \left(v^2-1\right)^2 \left(3 q^2+\rho ^2 
v^2\right) \\ &+27 \rho ^4 v^4 \omega ^4.
\end{split}
\end{align}
Further decomposing this mode according to eq.\eqref{expZser} will give us
\be
\gamma=\frac{3+\sqrt{24 \left({q \over \rho}\right)^2+9}}{2}
\ee

{\bf Hydrodynamic Mode:}

We now proceed for the calculation of the hydrodynamic modes of the shear channel in the z direction. 
We will solve the equation perturbatively in the limit $q,\omega \rightarrow 0$ and expand only upto the linear order in
 $q,\omega$. For this we introduce $\epsilon$ such that $q\rightarrow \epsilon q,\,\, \omega\rightarrow \epsilon \omega$ and expand upto 
linear order in $\epsilon$ which at the end we will set to 1.
\be 
Z(v)=(1-v^3)^{-\frac{\imath\omega}{4}}{\mathcal{C}}_{1} [Z_{0}(v)+\epsilon Z_{1}(v)+O(q^2,\omega^2)]
\ee 
with ${\mathcal{C}}_1$ is a normalization constant. Solving eq.\eqref{diffeqz} with eq.\eqref{abcsp1qz} consistently upto linear order in $\epsilon$
we obtain 
\begin{align}
\begin{split}
Z_{0}(v)&=1,\\
Z_{1}(v)&=\frac{1}{24\rho^2 \omega}\bigg(64i q^2(1-v^3)+i\rho^2\omega^2\bigg(9+2\sqrt{3}\pi-9v^2\tan^{-1}\bigg(\frac{1+2v}{\sqrt{3}}\bigg) 
\\ & \qquad +3\log\bigg[\frac{3+ (1+2v)}{12}\bigg]\bigg)\bigg),
\end{split}
\end{align}
The hydrodynamic modes are found by putting the Dirichlet boundary condition at the boundary
\be \label{dirichcond}
 Z(v)|_{v\rightarrow0}=0,
\ee
Upto leading order in $q$ we obtain 
 \be 
 \omega=-i\frac{8\,q^2}{3\rho^2}.
 \ee
 Now inserting factor of $u_H$ by dimensional analysis gives
 \be
 \omega=-i\frac{8\,u_H\,q^2}{3\rho^2}.
 \ee
 Now using eq.\eqref{defT}, 
the dispersion relation corresponding to the hydrodynamic pole is same as given in eq.\eqref{sp1disz}.
Comparing eq.\eqref{sp1disz} with eq.\eqref{disp110}, we conclude that
\be\label{appetaperp}
\frac{\eta_{\perp}}{s }=\frac{8\pi\,T^2}{3\rho^2},
\ee
where we have used $\epsilon+P=s T.$
Note that eq.\eqref{appetaperp} is in perfect agreement with eq.\eqref{etaperpint}.

\subsubsection{Spin 0}\label{qnmzapsp0}

The gauge invariant combinations for the spin $0$ modes $Z_0^{(1)}(v),\, Z_0^{(2)}(v)$ are defined in eq.\eqref{sp0defzt} and eq.\eqref{sp0defz}.
 The decoupled equations satisfied by them are given in eq.\eqref{sp0eqnZ} and eq.\eqref{sp0eqnZt} with the coefficients as follows,
\begin{align}
\label{abcsp0qz}
\begin{split}
a_{1}(v)&=16 \rho ^2 v^2 \left(v^3-1\right)^2 \bigg(192 q^4 \left(v^3-2\right)^2
+16 \rho ^2 q^2  (v^2  (3  (v^3-2 ) \omega ^2-16 v ) \\ & +16 )+\rho ^4 v^2 \omega ^2 \left(32 v^3+3 v^2
   \omega ^2-32\right)\bigg)\\
b_{1}(v)&=16 \rho ^2 v \left(v^3-1\right) \bigg(192 q^4  (v^9-8 v^6+8 v^3+8)\\
&+16 \rho ^2 q^2 \left(32 \left(-2 v^6+v^3+1\right)+3 v^2 \left(2 v^6-3 v^3-8\right) \omega ^2\right)\\
&+\rho ^4 v^2
   \left(v^3+2\right) \omega ^2 \left(64 v^3+9 v^2 \omega ^2-64\right)\bigg)\\
c_{1}(v)&=3 \bigg(16 \rho ^4 v^4 \omega ^4 \left(3 q^2 \left(5 v^3-8\right)+\rho ^2 \left(v^6+10 v^3-2\right)\right)\\
&+64 q^2 \rho ^2 v^2 \omega ^2 \left(3 q^2 \left(11 v^6-36 v^3+28\right)+2 \rho ^2 \left(v^9+3
   v^6-36 v^3+14\right)\right)\\
&+1024 q^4 \left(6 q^2 \left(v^3-1\right) \left(v^3-2\right)^2+\rho ^2 \left(-3 v^9+4 v^6+16 v^3-8\right)\right) \\&+9 \rho ^6 v^6 \omega ^6\bigg)
\end{split}
\end{align}
and
\begin{align}
\label{abcsp0qz1}
\begin{split}
a_{2}(v)&=16 \rho ^2 v^2 \left(v^6+v^3-2\right)^2,\\
b_{2}(v)&=-16 \rho ^2 v \left(v^{12}-19 v^9-30 v^6+32 v^3+16\right),\\
c_{2}(v)&=96 q^2  (v^9+3 v^6-4 )+\rho ^2  \bigg(9  (v^4+2 v )^2 \omega ^2\\&+16  (v^{12}-40 v^9+96 v^6+32 v^3-8 )\bigg),
\end{split}
\end{align}

{\bf Hydrodynamic limit:}

For studying the hydrodynamic mode we have seen that only the gauge invariant combination $Z_0^{(1)}(v)$ has a hydrodynamic pole. Following the same
 procedure as for the spin $1$ mode we write upto linear order in $\omega,\, q$ 
\be
 Z_0^{(1)}(v)=(1-v^3)^{-\frac{i\omega}{4}-1}{\mathcal{C}}(\tilde{Z}_{0}(v)+\tilde{Z}_{1}(v)+O(\omega^2,q^2))
\ee
Further solving eq.\eqref{sp0eqnZt} consistently upto linear order in $\omega,\, q$ we obtain
\begin{align}
\begin{split}
\tilde{Z}_{0}(v)&=(1-v^3),\\
\tilde{Z}_{1}(v)&=\frac{1}{24\rho^2 \omega}(72iq^2-i\rho^2\omega^2(-2\sqrt{3}\pi(1-v^3)+3(2+3v^2+\log[1728]\\
&-v^3(2+\log[1728])))-3(1-v^3)\rho^2\omega^2(-(3i+\sqrt{3})\log[\sqrt{3}-i(1+2v)]\\
&+(\sqrt{3}-3i)\log[\sqrt{3}+i(1+2v)]))
\end{split}
\end{align}
Now imposing the Dirichlet boundary condition at the boundary eq.\eqref{dirichcond} we obtain the dispersion relation upto quadratic order in the momenta $q$
as
\be
 \omega=-3i \frac{q^2}{\rho^2}
\ee  
which upon inserting the appropriate factor of $u_H$ gives
as
\be
 \omega=-3i \frac{u_H\,q^2}{\rho^2}=-3i \frac{\pi\,T\,q^2}{\rho^2}
\ee
 Note that the hydrodynamic mode does not have a real part. This implies that the velocity of the sound mode in the z-direction is $0$. 
 
\subsection{Modes with $q_z=0$}
\label{qnmxyap}
 It was mentioned earlier that while considering modes with momentum along $x-y$ direction we have to rescale the coordinates 
\be 
(x,y)\rightarrow(\sqrt{\b}x,\sqrt{\b}y)
\ee
 to ensure that the asymptotic form of the metric at the boundary becomes of the form
 eq.\eqref{ads5}. The value of $\b$ was obtained from numerical interpolation in eq.\eqref{betaval}. This enforces us to rescale 
 the momentum along the $(x,y)$ direction as 
\be
\label{defhatq}
q\rightarrow \frac{\hat{q}}{\sqrt{\b}}
\ee where $\hat{q}$ corresponds to momenta in $AdS_5.$

\subsubsection{Mode $Z_1$}\label{qnmyapsp2}
The gauge invariant combination for the spin 2 mode is given by 
\be 
Z_1(v)=q \, h^{x}_{t}+\omega \, h^{x}_{y}.
\ee
It satisfies an equation of the form eq.\eqref{diffeqz} 
where the coefficients are given by, 
\begin{align}
\begin{split}
a(v)&=-\frac{16(1-v^3)^2(3\omega^2-4q^2 (1-v^3))}{v^6},\\
b(v)&=\frac{16(1-v^3)(3(2+v^3)\omega^2-8q^2 (1-v^3)^2)}{v^7},\\
c(v)&=-\frac{3(3\omega^2-4q^2(1-v^3))^2}{v^6},\\
\end{split}
\end{align}

\textbf{Hydrodynamic limit:}

The $Z_1$ mode admits hydrodynamic pole in the limit $q,\omega \rightarrow0$ as can be seen from Fig.\ref{sp2y}. To analysis this further we do an expansion of the function $Z_1(v)$ in this limit.
 We introduce a book keeping parameter $\epsilon$ such that $q\rightarrow \epsilon \, q,\omega\rightarrow \epsilon \, \omega$ and do the expansion of 
$Z_1(v)$ in $\epsilon$ upto linear order. At the end we will set $\epsilon=1$. Thus
\be 
Z_1(v)=(1-v^3)^{-\frac{i \, \omega}{4}}{\mathcal{C}}_{1} [Z_{1}^{0}(v)+\epsilon Z^{1}_{1}(v)+\epsilon^2 O(q^2,\omega^2)]
\ee
 where 
\begin{align}
\begin{split}
Z_{1}^{0}(v)&=1,\\
Z_{1}^{1}(v)&=\frac{3\omega^2 \log[3]+4q^2(1-v^3)-3\omega^2\log[1+v+v^2]}{12 \omega},\\
\end{split}
\end{align}
${\mathcal{C}}_1$ is an overall constant fixed by normalization.
Note that in the above expansion $Z_1^{1}(v)\sim O(q,\omega)$.
The hydrodynamic modes are found by putting the Dirichlet boundary condition at the boundary
\be Z_1(v)|_{v\rightarrow 0}=0,\ee
Solving this upto leading order in $q$ we obtain the dispersion relation for the hydrodynamic pole as
 given in eq.\eqref{disy1}, which we write in terms of the rescaled momentum $\hat{q}$, defined in eq.\eqref{defhatq}, as
\be\label{etap1} 
\omega=-i\frac{\hat{q}^2}{4}.
\ee
In obtaining the above relation we used the value of $\b$ obtained from numerical interpolation as $\b=4/3$.
Upon inserting, appropriate factor of $u_H$ in eq.\eqref{etap1} we obtain
\be\label{etap} 
\omega=-i\frac{\hat{q}^2}{4u_H}=-i\frac{\hat{q}^2}{4\,\pi\,T}.
\ee
As discussed in appendix \ref{hydmoap} in eq.\eqref{disp111}, the relevant hydrodynamic mode is given by 
\be\label{etasap}
\omega=-i\frac{\eta_{\parallel}}{s T} \hat{q}^2,
\ee
where we have used eq.\eqref{threl1}.
Thus by comparing eq.(\ref{etap}) with eq.(\ref{etasap} we obtain
\be\label{etashap}
\frac{\eta_{\parallel}}{s}=\frac{1}{4\pi},
\ee
which reproduces eq.\eqref{etash}.

\subsubsection{Mode $Z_2$}\label{qnmyapsp1}
The gauge invariant combination for this channel is given by
\be Z_2(v)=h^{x}_{z}\ee
It satisfies the equation of the form eq.\eqref{diffeqz} with the coefficients are given by,
\begin{align}
\begin{split}
a(v)&=-16v(1-v^3)^2, \\
b(v)&=16(1-v)(1+v+v^2)(4-v^3),\\
c(v)&=3v(4q^2-4q^2v^2 +3\omega^2),
\end{split}
\end{align}

This mode is stable against the perturbations since there are no modes in the upper half plane of the spectrum of the quasi normal mode. It also
 does not admit any hydrodynamic mode since the Dirichlet boundary condition
\be \label{dirconZ2}
 Z_2(v)|_{v\rightarrow0}=0,
\ee at the boundary is not compatible with the condition $q,\omega\ll 1$.

\subsubsection{Modes $Z_3$, $Z_4$}\label{qnmyapsp0}
The two decoupled equations are 
\begin{align}
\begin{split}
a_{1}(v)Z_3''(v)+b_{1}(v)Z_3'(v)+c_{1}(v)Z_3(v)&=0,\\ 
a_{2}(v)Z_4''(v)+b_{2}(v)Z_4'(v)+c_{2}(v)Z_4(v)&=0,
\end{split}
\end{align}
where the coefficients are given as 
\begin{align}
\begin{split}
a_{1}(v)&=16v^4(1-v^3)^2\\
b_{1}(v)&=-16v^3(8-13v^3+5v^6)\\
c_{1}(v)&=3 v^2 \left(4 \left(v^3-1\right) \left(v^2 \left(q^2+12 v\right)-16\right)+3 v^2 \omega ^2\right)
\end{split}
\end{align}
and
\begin{align}
\begin{split}
a_{2}(v)&=16 v^4 \left(v^3-1\right)^2 \left(q^2 \left(v^3-4\right)+3 \omega ^2\right)\\
b_{2}(v)&=-16 v^3 \left(v^3-1\right) \left(q^2 \left(5 v^6-4 v^3+8\right)-3 \left(v^3+2\right) \omega ^2\right)\\
c_{2}(v)&=3 v^4 \left(3 q^2 \left(5 v^3-8\right) \omega ^2+4 q^2 \left(v^3-1\right) \left(q^2 \left(v^3-4\right)+12 v^4\right)+9 \omega ^4\right)
\end{split}
\end{align}
The equation for $Z_4(v)$ gives the hydrodynamic modes for the sound channel which we will discuss in the following 
section. Similar to the earlier section we can use the ingoing boundary conditions near the horizon and normalizability
 near the boundary. Thus both the modes admit the solutions of the form 
\be \begin{split}
Z_3(v)=&(1-v^3)^{-\frac{i\omega}{4}}v^{\frac{1}{2}(9+\sqrt{33})}{\tilde Z}_3(v),\\
Z_4(v)=&(1-v^3)^{-\frac{i\omega}{4}} v^{3}{\tilde Z}_4(v).
\end{split}\ee 
Figure \ref{sp0yZ} and \ref{sp0yZt} shows the quasi normal mode plot for the mode $Z_3(v),Z_4(v)$ respectively for various momenta. As we increase the momenta the QNMs get separated. 

\textbf{Hydrodynamic limit:}\\
The sound mode admits hydrodynamic pole for the mode $Z_4(v)$ which can be determined as follows.  We expand the equation perturbatively 
in $\omega,q \ll 1$ upto linear order. For convenience we introduce the bookkeeping parameter $\epsilon$ such that 
$q\rightarrow \epsilon q,\omega\rightarrow \epsilon \omega$ and expand in $\epsilon$ upto linear order and eventually put equal to 1.
 We can perturbatively expand the solution as 
\be 
Z_4(v)=(1-v^3)^{-\frac{i\omega}{4}}{\mathcal{C}}(Z_{4}^{0}(v)+\epsilon Z_{4}^1(v)+\epsilon^2 O(\o^2,q^2)),
\ee
where ${\mathcal{C}}$ is a normalization constant. The explicit form of the functions are 
\be Z_{4}^0(v)=\frac{q^2 (2+v^3)-3\omega^2}{10q^2-3\omega^2}, \  Z_{4}^1(v)=-\frac{i q^2 \omega (1-v^3)}{10q^2-3\omega^2}.\ee
The hydrodynamic modes are given as the solutions of the equation 
\be Z_4(0)=0,\ee 
which gives
\be q^2(2-i \omega)-3\omega^2=0.\ee
Solving this we obtain eq.\eqref{valom} which is the dispersion relation for the hydrodynamic modes for $Z_4(v)$. In terms of the rescaled momentum $q=\hat{q}/\sqrt{\b}$ we write the hydrodynamic modes as
\be \hat{\o}=\pm\frac{1}{\sqrt{2}}\hat{q}-i\frac{\hat{q}^2}{8}.\ee
Again inserting appropriate factor of $u_H$ we obtain
\be \hat{\o}=\pm\frac{1}{\sqrt{2}}\hat{q}-i\frac{\hat{q}^2}{8\,u_H}=\pm\frac{1}{\sqrt{2}}\hat{q}-i\frac{\hat{q}^2}{8\,\pi\,T}.\ee
Now comparing above equation with (eq.\eqref{disp112})
\be\label{etaso}
\omega=\pm \hat{q}\, c_s+ i\frac{\zeta^a_{2}}{2\partial_{T}\epsilon}\hat{q}^2-i \hat{q}^2 \frac{\zeta^a_{1}+\eta_{\parallel}}{2(P+\epsilon)},
\ee
we obtain 
\be\label{cseta}\begin{split}
c_s=\frac{1}{\sqrt{2}}, \qquad \text{and} \qquad
\frac{\zeta^a_{1}+\eta_{\parallel}}{2(P+\epsilon)}-\frac{\zeta^a_{2}}{2\partial_{T}\epsilon}=\frac{1}{8\,\pi\,T}.
\end{split}\ee
Note that value of sound velocity $c_s$ in eq.\eqref{cseta} is consistent with 
the definition
\be
\label{defcs}
c_s=\sqrt{\frac{\partial_T P}{\partial_T \epsilon}},
\ee
where $\epsilon$ and $P$ is given in eq.\eqref{reldele}, eq.\eqref{reldelp}.
Now using the fact that $\frac{\eta_{\parallel}}{2(P+\epsilon)}=\frac{1}{8\,\pi\, T}$ (see eq.\eqref{etash})
we immediately see 
\be
\label{defcs2}
-\frac{\zeta^a_{2}}{2\partial_{T}\epsilon}+\frac{\zeta^a_{1}}{2(P+\epsilon)}=0.
\ee

\subsubsection{Modes $Z_5$,~$Z_6$}\label{qnmyapax}
The modes $Z_5$ and $Z_6$ as defined in eq.\eqref{gicmodeqz}
 satisfy the set of coupled differential equations given by 
\begin{align}
\begin{split}
&Z_{5}''(v)+f_{1}(v)Z'_{5}(v)+f_{2}(v)Z_{5}(v)+f_{3}(v)Z'_{6}(v)=0\\
&Z_{6}''(v)+g_{1}(v)Z'_{6}(v)+g_{2}(v)Z_{6}(v)+g_{3}(v)Z'_{5}(v)=0
\end{split}
\end{align}
where the coefficients are given as
\begin{align}
\begin{split}
f_{1}(v)&=\frac{96 v^5+9 v^4 \omega ^2-96 v^2}{\left(v^3-1\right) \left(4 q^2 \left(v^3-1\right) v^2+32 v^3+3 v^2 \omega ^2-32\right)}\\
f_{2}(v)&=\frac{3 \left(4 q^2 \left(v^3-1\right) v^2+32 v^3+3 v^2 \omega ^2-32\right)}{16 v^2 \left(v^3-1\right)^2}\\
f_{3}(v)&=-\frac{96 q v^2}{\omega  \left(4 q^2 \left(v^3-1\right) v^2+32 v^3+3 v^2 \omega ^2-32\right)}
\end{split}
\end{align}
and
\begin{align}
\begin{split}
g_{1}(v)&=\frac{v \left(v^3+2\right) \left(4 q^2 \left(v^3-1\right)+3 \omega ^2\right)}{\left(v^3-1\right) \left(4 q^2 \left(v^3-1\right) v^2+
32 v^3+3 v^2 \omega ^2-32\right)}\\
g_{2}(v)&=\frac{3 \left(4 q^2 \left(v^3-1\right) v^2+32 v^3+3 v^2 \omega ^2-32\right)}{16 v^2 \left(v^3-1\right)^2}\\
g_{3}(v)&=-\frac{4 q v \left(v^3+2\right) \omega }{4 \left(v^3-1\right) \left(q^2 v^2+8\right)+3 v^2 \omega ^2}.
\end{split}
\end{align}
The boundary behavior of the modes are given by 
\begin{align}
\begin{split}
Z_{5}(v)&=(1-v^3)^{-\frac{i\o}{4}} v^3 \hat{Z}_{5}(v)\\
Z_{6}(v)&=(1-v^3)^{-\frac{i\o}{4}} v^3\hat{Z}_{6}(v).
\end{split}
\end{align}
The quasi normal mode plots are given in figure \ref{ax_Z1} and \ref{ax_Z1Z2}.
We notice that there are no poles on the upper half plane as well as there is no hydrodynamic mode. 
This we have verified from the explicit computation.
\subsection{Modes with both $q_y$ and $q_z$ turned on}\label{apqyqz}
As discussed in section \ref{qyqz}, there are total $6$ modes, out of which we have only analyzed two modes 
\be
\begin{split}
 Z(v)=&h^x_y(v) + {q_y \over \omega} h^x_t(v) \\
 \tilde{Z}(v)=&h^x_z(v) + {q_z \over \omega} h^x_t(v).
\end{split}
\ee
These two modes couple with each other and their equitations are given by
\be\begin{split}
&Z''(v)+a_1(v) Z'(v)+a_2(v)Z(v)+a_3(v){\tilde Z}'(v)=0\\
&{\tilde Z}''(v)+b_1(v) {\tilde Z}'(v)+b_2(v){\tilde Z}(v)+b_3(v)Z'(v)=0
\end{split}\ee
with
\be\begin{split}
&a_1(v)=-\frac{12 \rho ^2 v^4 q_y^2}{32 \left(v^3-1\right) q_z^2+\rho ^2 v^2 \left(4 \left(v^3-1\right) q_y^2+3 \omega ^2\right)}-\frac{v^3+2}{v-v^4}\\
&a_2(v)=\frac{3 \left(4 \left(v^3-1\right) \left(\rho ^2 v^2 q_y^2+8 q_z^2\right)+3 \rho ^2 v^2 \omega ^2\right)}{16 \rho ^2 v^2 \left(v^3-1\right)^2}\\
&a_3(v)=-\frac{96 v^2 q_y q_z}{32 \left(v^3-1\right) q_z^2+\rho ^2 v^2 \left(4 \left(v^3-1\right) q_y^2+3 \omega ^2\right)}
\end{split}\ee
and
\be\begin{split}
&b_1(v)=-\frac{64 \left(v^3-1\right)^2 q_z^2+\rho ^2 v^2 \left(v^3-4\right) \left(4 \left(v^3-1\right) q_y^2+3 \omega ^2\right)}{v \left(v^3-1\right) \left(32
   \left(v^3-1\right) q_z^2+\rho ^2 v^2 \left(4 \left(v^3-1\right) q_y^2+3 \omega ^2\right)\right)}\\
&b_2(v)=\frac{3 \left(4 \left(v^3-1\right) \left(\rho ^2 v^2 q_y^2+8 q_z^2\right)+3 \rho ^2 v^2 \omega ^2\right)}{16 \rho ^2 v^2 \left(v^3-1\right)^2}\\
&b_3(v)=-\frac{4 \rho ^2 v \left(v^3+2\right) q_y q_z}{32 \left(v^3-1\right) q_z^2+\rho ^2 v^2 \left(4 \left(v^3-1\right) q_y^2+3 \omega ^2\right)}.
\end{split}\ee
Details of quasinormal mode is plotted in Fig.\ref{bomom_qz1} and Fig.\ref{bomom_qz3}.
\section{Instability in AdS space}
\label{instads}
It is well known that for a field with  a mass lying below the BF bound in AdS space there 
is  a tachyonic mode which grows exponentially in time. If $R$ is the radius of $AdS_{d+1}$, the mass of a scalar field lies above the BF bound when  
\be
\label{condBF}
m^2R^2\ge d^2/4.
\ee

If the $AdS_{d+1}$ spacetime is not the full solution of interest
 but  arises only  in the far infrared one might wonder if the  instability is still present, 
or whether the process of gluing the IR geometry to another  one
in the UV (say an $AdS$ spacetime of higher dimension) can get rid of the instability sometimes? 
Here we will argue that this is not possible, and show that the instability in the IR  $ AdS_{d+1}$ 
space will continue to be present in the full  geometry as well. The essential point is that there are 
 unstable modes which are located sufficiently close to the horizon and this makes them  
 insensitive to the exact nature of the  UV geometry \footnote{We are grateful to Shiraz Minwalla for outlining  the proof given below.}.

Although our analysis is more general, for concreteness
we will restrict ourselves to the case at hand where the IR geometry is 
\footnote{ More correctly the IR is $AdS_{4}\times R$ but we will loosely refer to it as  $AdS_4$
since it is only the propagation in $AdS_4$ that will be of interest.} $AdS_{4}$
 and the UV geometry is $AdS_{5}$. 
And we will consider a scalar, $\phi$,  which lies below the BF bound in $AdS_4$  but not in $AdS_5$. 
The full geometry is given by eq.(\ref{metans1}). The mode we consider has $\phi \sim e^{-iwt}$ and is independent of 
$x,y,z$. It satisfies an equation which can be cast in the form of a one-dimensional Schroedinger equation:
\be
\label{oneds}
-\partial_x^2 \psi + V(x) \psi =\omega^2 \psi
\ee
where 
\begin{align}
x & =  -\int {du \over A(u)} \label{svar} \\
 \psi & =  \phi(u) (\text{det} (g))^{1/4} \label{defpsi} \\
\text{det}(g) & =  \beta^2 B(u)^2 C(u) \label{defdet} 
\end{align}
The Schroedinger variable $x\in [0, \infty]$ where $x=0$ is the boundary of $AdS_5$ space and $x=\infty$ is the 
horizon of $AdS_4$.  

We see from eq.(\ref{oneds})  that  $\omega^2$ is the energy in the Schroedinger problem. A bound state which is normalisable and meets the required boundary conditions would have negative energy and would correspond to imaginary $\omega$ and thus to  a tachyonic 
instability.

Let us first consider the behavior of bound states which  arises in pure $AdS_4$. For this case eq.(\ref{oneds})
 becomes
\be
\label{seads41}
-\partial_x^2 \psi + {\alpha \over x^2} \psi  = E \psi
\ee
This equation is  invariant under the scaling symmetry
\be
\label{scalsym}
\psi\rightarrow \psi, x \rightarrow \lambda x, E \rightarrow E/\lambda^2
\ee
which will play a crucial role in the discussion below. Note that  the energy $E$ is given by
\be
\label{connegE}
E=\omega^2
\ee
and  a negative energy state has $E<0$. Due to the scaling symmetry we can denote the solution to eq.(\ref{seads41}) by $\psi(\sqrt{|E|} x)$.  
Eq.(\ref{seads41}) can be cast as a (modified) Bessel equation
\be
\label{modbe}
 x^2 \partial_x^2{\tilde \psi} + x \partial_x {\tilde \psi} +(x^2 E {\tilde \psi} +(\alpha -1/4)) {\tilde \psi}=0
\ee
where
\be
\label{deftpsi}
{\tilde \psi}=x^{-1/2} \psi 
\ee
For $\alpha <1/4$ it is easy to see that 
there are negative energy  normalisable bound states which decay exponentially towards the horizon, $\psi \sim e^{
-\sqrt{|E|} x}$, and which vanish as 
\be
\label{decaybnd}
\psi \simeq C_1 (\sqrt{|E|} x)^{1/2} ((\sqrt{|E|}x)^{i \sqrt{\alpha-1/4} } + \text{Complex Conj.})
\ee
towards the boundary. 
In fact  such  a state exists for any value of $E$.  
Thus there are an infinite number of normalisable bound states and the system is unstable. 
 A bound state with energy $E$ is located roughly at $x \sim \sqrt{|E|}$ as also follows from eq.(\ref{scalsym}).  
Also, from eq.(\ref{modbe}) we see that the limiting behavior  in eq.(\ref{decaybnd}) is valid for
\be
\label{condxas}
x\ll {1\over \sqrt{|E|}}.
\ee
We will normalize $\psi$ so that the coefficient $C_1$ is of order unity.  A normalisable state can then  be written as 
\be
\label{nstate}
\psi_{N}= c |E|^{1/4}\psi(\sqrt{|E|}x)
\ee
and satisfies the condition
\be
\label{ncond}
\int d x |\psi_N|^2 = c^2 \int_{y=0}^{y=\infty}  d y |\psi(y)|^2  = 1
\ee
with $c \sim O(1)$.  
In eq.(\ref{ncond}) the coordinate $y$ is related to $x$ by 
\be
\label{defyc}
y = \sqrt{|E|} x
\ee

Now we turn to the case  where the $AdS_4$ geometry is connected to  the $AdS_5$ geometry in the UV. 
By rescaling the $u$ coordinate in eq.(\ref{metans1}) we can always set $\rho=1$. The geometry then has no scale left
 and  the region where the transition from the $AdS_4$ near horizon geometry to the asymptotic $AdS_5$ geometry occurs
is   around $u \sim O(1)$.  In the $x$ coordinate, eq.(\ref{svar})
 the transition occurs around $x \sim O(1)$. 

We will  use a variational argument to conclude  that the system continues
 to have negative energy states and thus an instability.    
We do this by demonstrating that a trial  normalisable wave function exists which meets the boundary conditions and which 
has a negative expectation value for the energy.  This is enough to show that there must be  a  negative eigen value for the energy and therefore
an instability. 

The trial wave function is obtained by  smoothly 
patching together the solution of the Bessel equation, eq.(\ref{modbe}) above,  for the case where
\be
\label{condE}
|E| \ll 1,
\ee  
with
the solution to the  Schroedinger equation also with energy $E$ obtained in the asymptotic $AdS_5$ region.
The details of the interpolating function will not be important so we are not very explicit about it here. 

In the $AdS_5$ region eq.(\ref{oneds}) also reduces to a Bessel function of the type, eq.(\ref{seads41}), but now with $\alpha\ge 1/4$
(since we are assuming no instability in $AdS_5$). 
We denote the corresponding solution, with the appropriate
 boundary conditions that are to be imposed as $x \rightarrow 0$, as
$\psi_5(\sqrt{|E|} x)$. The trial function, $\psi_T$  is then obtained by smoothly 
pasting together $\psi_N$, eq.(\ref{nstate}) with 
$\psi_5(\sqrt{|E|} x)$ around $x \sim O(1)$.  We write,
\begin{eqnarray}
x \gg 1 &\Rightarrow \ \psi_T = &  \psi_N \\
x \ll 1 & \Rightarrow \ \psi_T = &  C \,\psi_5 \\
x \sim O(1) &\Rightarrow \ \psi_T = &  \psi_{I}
\end{eqnarray}
where $\psi_N$ denotes the normalized function given in eq.(\ref{nstate}) and 
$\psi_I$ stands for an interpolating function.  

Note that in the patching region, $x \sim O(1)$,  and   $\sqrt{|E|} x  \ll 1$, so that  $\psi_5$ is given by 
\be
\label{asbe5}
\psi_5 \simeq B_1 x^{1/2 + \nu}
\ee
where $\nu$ is determined by the mass of the scalar in $AdS_5$ units.  
In passing we note that $\nu$ need not be positive, but square integrability  gives rise to the condition 
\be
\label{sqint}
\nu >-1
\ee
which agrees with the unitarity bound in $AdS_5$  \cite{Klebanov:1999tb}.
Equating $\psi_5$ and $\psi_N$ in the matching region $x \sim O(1)$ then gives 
\be
\label{valC}
|B_1|\sim  |E|^{1/2}
\ee  

The expectation value of the energy for $\psi_T$ is 
\be
\label{expE}
\langle E\rangle ={\int_0^\infty dx\, \psi_T^* \, (-\partial_x^2+ V(x) )\,  \psi_T \over \int_0^\infty dx\,  \psi_T^*\,  \psi_T}
\ee
The Denominator  on  RHS can be estimated as
\be
\label{den}
\text{denominator}= \int_0^\infty  dx\, |\psi_T|^2 \simeq \int_0^\infty dx\,|\psi_N|^2 - \int_0^1dx \,|\psi_N|^2 + \int_0^1dx\,|\psi_5|^2
\ee where we have not included the contribution of the interpolating region explicitly  since it will not change the estimate. 
The second term in eq.(\ref{den}) is $O(|E|)$, and the third term is  $O(B_1^2)$ and thus also of the same order. 
These are much smaller than the first term which is unity. So  we see that  $\psi_T$ is approximately of unit norm. 
The numerator can be similarly estimated as 
\be
\label{num}
\text{numerator}= \int_0^\infty dx\, \psi_N^* (H_4) \psi_N -\int_0^1 dx\,  \psi_N^* (H_4) \psi_N + \int_0^1dx\, \psi_5^* (H_5) \psi_5
\ee
with 
\be
\label{defh4}
H_4 = -\partial_x^2+ \alpha/x^2
\ee
being the Hamiltonian in the $AdS_4$  region and similarly for $H_5$. Just as for the denominator we find that 
 the last two terms in eq.(\ref{num})  are comparable, each  being $O(E |E|)$,  and    much smaller
than the first term which equals   $E$.
  
In summary we find that $\langle E\rangle  \simeq E$ for small $E$, upto small corrections, so that  $\psi_T$ has a negative 
 expectation value for energy. This proves that  the instability persists in the full geometry. 

We dealt with the extremal geometry, but it is clear now that a similar argument will also apply for a near-extremal situation.
It is also clear that a similar argument applies in other dimensions 
and also one expects to be able to generalize this to Lifshitz near horizon regions.
As the temperature is increased the horizon moves to the transition region where $x \sim O(1)$, and eventually, for large enough $T$
 the instability should disappear.

\section{QNM analysis for $AdS_3\times R$} 
\label{ads3rap}
In this appendix we discuss an independent check for the consistency of the numerical analysis we performed to find the QNM spectrum for the $AdS_4 \times R$. A detailed analysis of the QNM spectrum in $2+1$ dimensions can be found in \cite{Sachs:2008gt, Birmingham:2010gj,Birmingham:2002ph, Birmingham:2001pj, Cardoso:2001hn,Decanini:2009dn} and \cite{Son:2002sd}. For that we work in one lower spacetime dimension, \emph{i.e.} in (3+1) dimensions with coordinates $t,\,r,\,y,\,z$,  but with the same gravity set up of a dilaton coupled to gravity with negative cosmological constant given in eq.\eqref{action}. Working with the same units and conventions, eq.\eqref{cetlambda},eq.\eqref{valL} as in section \ref{gravitysol} and also with the same linearly varying profile for the dilaton in eq.\eqref{formdil}, we obtain a finite temperature solution of the form $AdS_3\times R$ in the highly anisotropy region with $\rho/T \gg1$, 
\be
\label{metads3r}
\begin{split}
ds^2&=\frac{du^2}{A(u)}-A(u)dt^2+B(u)dy^2+C(u)dz^2, \\
\text{with} \ A(u)&={3\over2} u^2\left(1-{u_H^2\over u^2}\right),\ B(u)=u^2, \ C(u)={\rho^2 \over 6}
\end{split}
\ee
The advantage of working with this anisotropic solution in one lower dimension is that the QNM spectrum can be obtained analytically 
 for the geometry given in eq.\eqref{metads3r}. This enables us to compare the QNM spectrum  obtained using the numerical technique  
developed in section \ref{qnmspec} for the geometry in eq.\eqref{metads3r} against the same obtained from analytical calculation directly. 
 We will finally show that they are in good agreement with each other. As already mentioned, this provides an independent check of the numerical  
technique we used for extracting the QNM spectrum for the $AdS_4 \times R$ geometry, where the analytical calculation is too complicated to be done. 
 
The perturbations in the metric and in the dilaton are defined in eq.\eqref{metperta} and eq.\eqref{dilperta}. Note that there is no $x-$coordinate now as we are working in one lower dimension. We also make a choice of the gauge same as given in eq.\eqref{gaugepa} and eq.\eqref{gaugepb}. Therefore the metric perturbations have $6$ components and the dilaton perturbation has one additional, making it $7$ in total. They can be listed as 
\be 
h_{tt}, \ h_{yy}, \ h_{zz}, \ h_{ty}, \ h_{tz},\  h_{yz},\ \text{and} \ \phi. 
\ee  
Furthermore following the same argument discussed in section \ref{qnmspec}, the $4$ constraint equations out of the  
$10$ components of the Einstein equations further reduces the number of independent perturbations down to $3$. We will identify suitable combinations of the perturbations such that the linearized equations decouple. 
 
The perturbations, denoted by the momentum $\vec{q} = (q_y, q_z)$ and frequency $\omega$ depending on $(y,z,t)$, as given in eq.\eqref{pertdep}. We will consider here two situations, firstly when the momentum is along $z$-direction with $q_y=0$ and secondly, when the momentum is along $y$-direction with $q_z=0$. We will not consider the more general situation when the momentum is turned on along both $y$ and $z$-direction because our aim is to give an example where both the analytic and numerical calculation produces same result for the QNM.  
 
\subsection{Modes with momentum along $z$ direction} 
We are considering the situation when the modes have momentum along the $z$-direction only and we will denote in this subsection the momentum by 
\be 
q_z=q 
\ee 
 
The gauge invariant combinations for this case are 
\begin{align} 
\label{momzmodes} 
\begin{split} 
\text{shear mode: } \ Z(u)=& q h_{ty}+\omega h_{yz},\\ 
\text{sound mode: } \  Z_1(u)=& h_{tt}+\frac{2\o}{q}h_{tz}+\frac{i \o^2\r}{3q}\phi+\frac{A'}{B'}h_{yy}, \\  
Z_2(u)=&h_{zz}-\frac{i q\rho}{3}\phi 
\end{split} 
\end{align} 
We will only consider the mode $Z(u)$ in eq.\eqref{momzmodes} corresponding to shear channel for the purpose of providing an example, leaving out the rests. It satisfies an linearized equation of the form 
\be 
a(u)Z''(u)+b(u)Z'(u)+c(u)Z(u)=0, 
\ee 
where, 
\begin{align} 
\begin{split} 
a(u)&=6 (-1 + u)^2 u^2 \r^2 (6 q^2 u + (-1 + u) \r^2 \o^2),\\ 
b(u)&=6 (-1 + u) u \r^2 (12 q^2 u^2 + (1 - 3 u + 2 u^2) \r^2 \o^2),\\ 
c(u)&=-36 q^4 u^2 - 12 q^2 (-1 + u) u \r^2 \o^2 - (-1 + u) \r^4 \o^2 (-\o^2 + u (6 + \o^2)), 
\end{split} 
\end{align} 
This equation can be solved analytically in the limit $q\rightarrow0$, as the solution for general $q$  
is very complicated to be solved analytically. The solution is given by, 
\be 
Z(u)\bigg|_{q\rightarrow0}=\frac{u^{-i\o/\sqrt{6}}}{1-u}\ \ _{2}F_1[-\frac{i\o}{\sqrt{6}},-1-\frac{i\o}{\sqrt{6}},1-i\sqrt{\frac{2}{3}}\o,u]. 
\ee 
Imposing the Dirichlet boundary condition \emph{i.e.} $Z(u)=0$ at the boundary $u=1$ we obtain the exact QNMs, at least in the $q=0$ limit, as 
\be\label{hz}  
\omega=-i\sqrt{6}(n+1), \ n=0,1,2,\cdots, 
\ee 
 
We would like to point out that it is insufficient to conclude from \eqref{hz} that there is no hydrodynamic mode because \eqref{hz} does not capture the hydrodynamic mode. The reason is that in \eqref{hz} we are setting $q=0$ and this kills of any $\rho$ dependence in the equation. Whereas for the hydrodynamic mode we should actually retain $\rho$ and set $q/\rho$ to be small. So the hydrodynamic mode is derived in some very specific limit.   
 
\begin{figure} 
\begin{center} 
\includegraphics[width=0.7\textwidth]{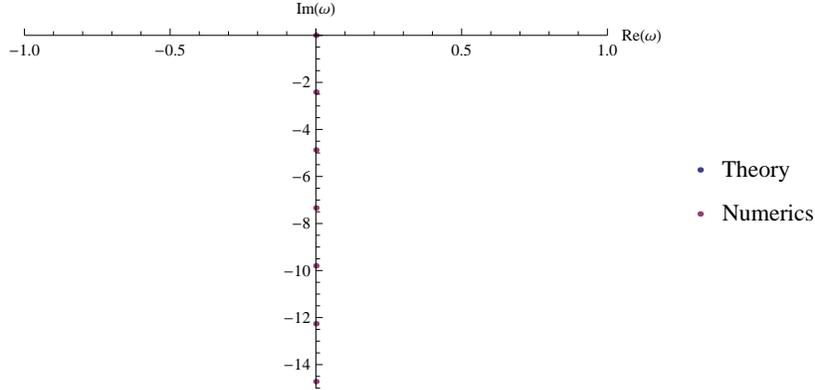} 
\caption{Comparison between the numerical estimates and the theoretical values of the QNMs for the shear mode with momentum along $z$ direction with $a=0$} 
\label{cmpshz} 
\end{center} 
\end{figure}

We also carry out a numerical analysis along the general lines discussed in section \ref{qnmspec} to obtain the QNM spectrum for this mode. In Fig. \ref{cmpshz} we plot the QNMs for this mode where we compare the analytical result against the numerical one and find that they are in good agreement with each other. We also note that the agreement is within $2\%$. From the plots it is not technically possible to cleanly isolate the hydrodynamic modes. Thus in the appendix as well as in the paper we analyze the hydrodynamic modes separately. 
  
\textbf{Hydrodynamic limit}: 
 
Since \eqref{hz} does not feature the hydrodynamic modes we verify them explicitly in this section by doing a small expansion around $\omega,q \ll 1$. The mode equation can then be solved perturbatively by assuming  
\be  
Z(u)=u^{-\frac{i\omega}{\sqrt{6}}}Z_{t}(u), 
\ee 
where $Z_t(u)=Z_0(u)+\epsilon Z_1(u)$ and at the end we put $\epsilon=1$. These functions are defined as 
\be\label{hydr} 
Z_t(u)={\mathcal{C}}\, \bigg(\frac{1}{1-u}+\frac{iu(6q^2+\rho^2\omega^2)}{(1-u)\sqrt{6}\rho^2\omega}\bigg), 
\ee 
 By imposing the Dirichlet boundary condition on the modes $Z(u)=0$ at the boundary $u=1$ we obtain the dispersion relation for $\omega$ and $q$ of the form, 
\be\label{hydr1} 
\sqrt{6}\rho^2\omega+i(6q^2+\rho^2\omega^2)=0. 
\ee 
Note that if we naively put $q=0$ in equation \eqref{hydr}, then we will wrongly conclude that there is a root at $\omega=0$. In fact the hydrodynamic mode exists only in the limit $q/\rho\rightarrow0$ and not at $q=0$. When $q\neq0$ we have equation \eqref{hydr1}. This equation has two roots given by considering $q/\rho=a$, 
\be 
\omega= -i\sqrt{\frac{3}{2}}\bigg(-1+\sqrt{1+4a^2}\bigg), \ \omega= i\sqrt{\frac{3}{2}}\bigg(1+\sqrt{1+4a^2}\bigg)\,, 
\ee 
The second root does not satisfy the hydrodynamic limit and the first one satisfies the conditions where both $\omega$ and $q$ are small. This is given by, 
\be 
\label{dispapp3} 
\omega=-i\sqrt{6}a^2, 
\ee 
 
\subsection{$\eta_\perp/s$ for $AdS_3\times R$} 
We can calculate the shear viscosity to entropy density ratio in the similar way as discussed in section \ref{grvissub1}. Using the prescription given in eq.\eqref{finaletaperp}.  for the geometry in eq.\eqref{metads3r} one obtains  
\be 
\label{etsapp} 
\frac{\eta_{\perp}}{s}=\frac{4\pi T^2}{\rho^2}. 
\ee 
 
Further, from eq.\eqref{metads3r} one can obtain a relation for the temperature, 
\be \label{reltads3} T=\frac{\sqrt{6}u_H}{4\pi}.\ee 
Using eq.\eqref{reltads3} along with the dispersion relation obtained via Kubo analysis that defines the viscosity $\eta_\perp$ as given in eq.\eqref{disp110}, one can obtain from eq.\eqref{dispapp3} that the ratio $\eta_{\perp}/s$ indeed agrees with eq.\eqref{etsapp}.

\subsection{Modes with momentum along $y$ direction} 
Next we consider the situation when the modes have momentum along the $y$-direction only and denote in this subsection the momentum by 
\be 
q_y=q 
\ee 
The gauge invariant combinations for this case are 
\begin{align} 
\label{modyapp} 
\begin{split} 
 Z(u)&=C'\left( h_{ty}+q^2 h_{tt}+\omega^2 h_{yy}\right)+(q^2 A'-\o^2 B') h_{zz},\\ Z_1(u)&=h_{yz}+\frac{q}{\o}h_{tz}, \ Z_2(u)=h_{tz}+i\frac{C}{\r}\o \phi, 
\end{split} 
\end{align} 
where $A,B,C$ are given in eq.\eqref{metads3r}. In order  
Note that from eq.\eqref{metads3r} $C'(u)=0$ and $Z(u)$ becomes  
 \be 
 Z(u)=(q^2A'-\o^2B')h_{zz}. 
 \ee 
By a coordinate transformation \cite{Son:2002sd} we can bring the $AdS_3$ part of the metric in eq.\eqref{metads3r} to BTZ form and the equation for the mode $Z(u)$ takes the form 
\be 
Z''(u)+\frac{Z'(u)}{u}+[\frac{\o^2}{6u^2(1-u)}-\frac{q^2}{6u(1-u)}-\frac{1}{u(1-u)^2}]Z(u)=0, 
\ee 
which is of the form of a massive scalar field in BTZ background given by 
\be  
f''_k(u)+\frac{f'_k(u)}{u}+(\frac{\omega^2 l^2}{4u^2(1-u)}-\frac{q^2 l^2}{4u(1-u)}-\frac{m^2 l^2}{4u(1-u)^2})f_k(u)=0. 
\ee  
From this we can identify $l=2/3$ and $m=\sqrt{6}$. This equation admits as exact solution of the type, 
\be  
Z(u)=e^{\sqrt{\frac{2}{3}}\pi \omega}(-1+u)^{\frac{1-\sqrt{5}}{2}}u^{-\frac{i\omega}{\sqrt{6}}}\,{\mathcal{C}} \ \  _2 F_1(a,b,c,u), 
\ee 
where, 
\be 
a= \frac{1}{6}(3(1-\sqrt{5})-i\sqrt{6}\alpha), \ b=\frac{1}{6}(3(1-\sqrt{5})+i\sqrt{6}\alpha-2i\sqrt{6}\omega), \ c=1-i\sqrt{\frac{2}{3}}\omega, 
\ee 
The exact solution above demonstrates the near horizon ingoing boundary conditions where horizon is at $u=0$ and in addition we put $Z(u)=0$ at the boundary $u=1$ and where $\alpha=q+\o$ for convenience. Using these conditions we get,  
\be  
\lim_{\epsilon\to 0}Z(1-\epsilon)=\frac{\pi \csc(\sqrt{5}\pi)\Gamma(c)}{\Gamma(\frac{1}{6}(3+3\sqrt{5}-i\sqrt{6}\alpha))\Gamma(\frac{1}{6}(3+3\sqrt{5}+i\sqrt{6}\alpha-2i\sqrt{6}\omega))}. 
\ee 
The exact solutions QNMs will correspond to the poles of $\Gamma$ functions in the denominator. These are given by  
\be 
\label{disapp3y} 
\omega=\pm q-i\sqrt{6}[n+\frac{1}{2}(1+\sqrt{5})], \  n=0,1,2,\cdots,  
\ee 
 
\begin{figure} 
\begin{center}  
\includegraphics[width=0.7\textwidth]{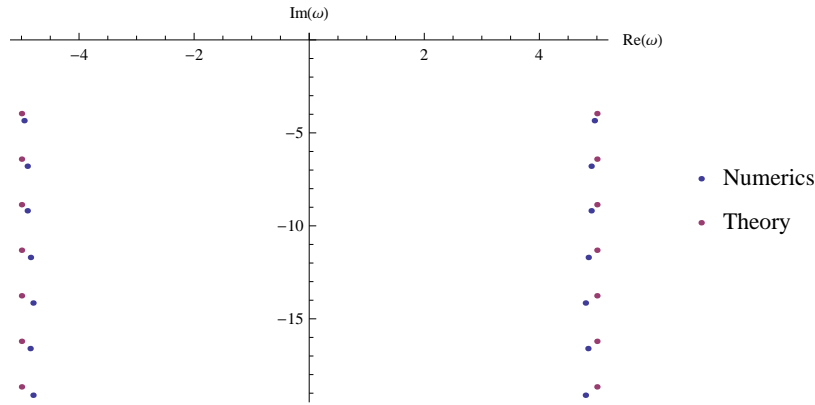} 
\caption{Comparison between the numerical estimates and the theoretical values of the QNMs for the sound mode with momentum along the $y$ direction for $q=5$} 
\label{compare} 
\end{center} 
\end{figure}

In Fig.\ref{compare} we plot the results for the QNM spectrum obtained both analytically and numerically.  It shows that they are in agreement  with each other to within $2\%$.  
 
We also carried out the procedure as done in the previous subsection and found no hydrodynamic mode for this case.

\bibliographystyle{jhepmod}
\bibliography{references}
\end{document}